\documentclass[11pt]{article}

\usepackage{fullpage}

\usepackage{graphicx}

\usepackage{hyperref}
\usepackage{amsmath,amssymb}

\newtheorem{thm}{Theorem}[section]

\newcommand{\cA}{\mathcal{A}}

\newcommand{\Df}{{\rm D}f}

\newcommand{\D}{{\mathbb D}}
\newcommand{\R}{{\mathbb R}}
\newcommand{\T}{{\mathbb T}}
\newcommand{\Z}{{\mathbb Z}}

\newcommand{\fix}{\operatorname{Fix}}

\renewcommand{\d}{{\rm d}}
\newcommand{\e}{{\rm e}}
\newcommand{\PD}[2]{\frac{\partial #1}{\partial #2}}
\newcommand{\FD}[2]{\frac{\d #1}{\d #2}}

\newcommand{\ID}[1]{{\d #1}/{\d t}}

\DeclareSymbolFont{AMSb}{U}{msb}{m}{n}
\DeclareMathSymbol{\NSet}{\mathalpha}{AMSb}{"4E}
\DeclareMathSymbol{\bbbn}{\mathalpha}{AMSb}{"4E}
\DeclareMathSymbol{\ZSet}{\mathalpha}{AMSb}{"5A}
\DeclareMathSymbol{\RSet}{\mathalpha}{AMSb}{"52}
\DeclareMathSymbol{\bbbq}{\mathalpha}{AMSb}{"51}
\DeclareMathSymbol{\CSet}{\mathalpha}{AMSb}{"43}

\begin{document}




\title{Mathematical frameworks for oscillatory network dynamics in neuroscience}

\author{Peter Ashwin\\Centre for Systems, Dynamics and Control\\
Harrison Building\\ University of Exeter\\
Exeter EX4 4QF, U.K.\and
Stephen Coombes\\
School of Mathematical Sciences\\ University of Nottingham\\
University Park\\
Nottingham NG7 2RD, U.K. 
\and
Rachel Nicks\\
School of Mathematical Sciences\\University of Birmingham\\
Watson Building\\
Birmingham B15 2TT, U.K. 
}






\maketitle

\begin{abstract} 

The tools of weakly coupled phase oscillator theory have had a profound impact on the neuroscience community, providing insight into a variety of network behaviours ranging from central pattern generation to synchronisation, as well as predicting novel network states such as chimeras.  However, there are many instances when this theory is expected to break down, say in the presence of strong coupling, or must be carefully interpreted, as in the presence of stochastic forcing. There are also surprises in the dynamical complexity of the attractors that can robustly appear - for example, heteroclinic network attractors. In this review we present a set of mathematical tools that are suitable for addressing the dynamics of oscillatory neural networks, broadening from a standard phase oscillator perspective to provide a practical framework for further successful applications of mathematics to understanding network dynamics in neuroscience.

%
\end{abstract}

\tableofcontents

\section{Introduction}
\label{sec:introduction}

Coupled oscillator theory is now a pervasive part of the theoretical neuroscientist's toolkit for studying the dynamics of models of biological neural networks.  Undoubtedly this technique originally arose in the broader scientific community through a fascination with understanding synchronisation in networks of interacting heterogeneous oscillators, and can be traced back to the work of Huygens  on ``an odd kind of sympathy" between coupled pendulum clocks" \cite{Huygens1893}.  Subsequently the theory has been developed and applied to the interaction between organ pipes \cite{Rayleigh1896}, phase-locking phenomena in electronic circuits \cite{VanDerPol1927}, the analysis of brain rhythms \cite{Wiener1948}, chemical oscillations \cite{Kuramoto84}, cardiac pacemakers \cite{Michaels1987}, circadian rhythms \cite{Liu1997}, flashing fireflies \cite{Ermentrout1991}, coupled Josephson junctions \cite{Wiesenfeld1998}, rhythmic applause \cite{Neda2000}, animal flocking \cite{Ha2010}, fish schooling \cite{Paley2007}, and behaviours in social networks \cite{Assenza2011}.
For a recent overview of the application of coupled phase oscillator theory to areas as diverse as vehicle coordination, electric power networks, and clock synchronisation in decentralised networks see the recent survey article by D\"orfler and Bullo \cite{Dorfler2014}.

Given the widespread nature of oscillations in neural systems it is no surprise that the science of oscillators has found such ready application in neuroscience \cite{Velazquez2006}.  This has proven especially fruitful for shedding light on the functional role that oscillations can play in feature binding \cite{Singer1993,Uhlhaas2009}, cognition \cite{Wang2010a}, memory processes \cite{Fell2011}, odour perception \cite{Wehr1996,Laurent2001}, information transfer mechanisms \cite{Buzsaki2004}, inter-limb coordination \cite{Haken1985,Kelso1995}, and the generation of rhythmic motor output \cite{Stein1999}.
Neural oscillations also play an important role in many neurological disorders, such as excessive synchronisation during seizure activity in epilepsy \cite{Milton2003,Coombes2012a}, tremor in patients with Parkinson's disease \cite{Titcombe2004} or disruption of cortical phase synchronisation in schizophrenia \cite{Bressler2003}.
As such it has proven highly beneficial to develop methods for the control of (de)synchronisation in oscillatory networks, as exemplified by the work of Tass \textit{et al}. \cite{Tass1999,Tass2006} for therapeutic brain stimulation techniques.  From a transformative technological perspective, oscillatory activity is increasingly being used to control external devices in brain-computer interfaces, in which subjects can control an external device by changing the amplitude of a particular brain rhythm \cite{Rao2013}.

Neural oscillations can emerge in a variety of ways, including intrinsic  mechanisms within individual neurons or by interactions between neurons.  At the single neuron level, sub-threshold oscillations can be seen in membrane voltage as well as rhythmic patterns of action potentials.  Both can be modelled using the Hodgkin-Huxley conductance formalism, and analysed mathematically with dynamical systems techniques to shed light on the mechanisms that underly various forms of rhythmic behaviour, including tonic spiking and bursting (see e.g. \cite{Coombes2005}).  The high dimensionality of biophysically realistic single neuron models has also encouraged the use of reduction techniques, such as the separation of time-scales recently reviewed in \cite{Rinzel2013,Kuehn2015},  or the use of phenomenological models, such as FitzHugh-Nagumo (FHN) \cite{FitzHugh1961}, to regain some level of mathematical tractability.  This has proven especially useful when studying the response of single neurons to forcing \cite{Coombes2000}, itself a precursor to understanding how networks of interacting neurons can behave.
When mediated by synaptic interactions, the repetitive firing of pre-synaptic neurons can cause oscillatory activation of post-synaptic neurons. At the level of neural ensembles, synchronised activity of large numbers of neurons gives rise to macroscopic oscillations, which can be recorded with a micro-electrode embedded within neuronal tissue as a voltage change referred to as a local field potential (LFP).
These oscillations were first observed outside the brain by Hans Berger in 1924 \cite{Berger1929} in electroencephalogram (EEG) recordings, and have given rise to the modern classification of brain rhythms into frequency bands for alpha activity (8-13 Hz) (recorded from the occipital lobe during relaxed wakefulness), delta (1-4 Hz), theta (4-8 Hz), beta (13-30 Hz) and gamma (30-70 Hz).  The latter rhythm is often associated with cognitive processing, and it is now common to link large scale neural oscillations with cognitive states, such as awareness and consciousness.  For example, from a practical perspective the monitoring of brain states via EEG is used to determine depth of anaesthesia \cite{Voss2007}.  Such macroscopic signals can also arise from interactions between different brain areas, the thalamo-cortical loop being a classic example \cite{Sherman2013}.
Neural mass models (describing the coarse grained activity of large populations of neurons and synapses) have proven especially useful in understanding EEG rhythms \cite{Wright1996}, as well as in augmenting the dynamic causal modelling framework (driven by large scale neuroimaging data) for understanding how event-related responses result from the dynamics of coupled neural populations \cite{David2003}.

One very influential mathematical technique for analysing networks of neural oscillators, whether they be built from single neuron or neural mass models, has been that of weakly coupled oscillator theory, as comprehensively described by Hoppensteadt and Izhikevich \cite{Hoppensteadt97}.  In the limit of weak coupling between limit cycle oscillators invariant manifold theory \cite{Fenichel1971} and averaging theory \cite{GuckenheimerHolmes1990} can be used to reduce the dynamics to a set of phase equations in which the relative phase between oscillators is the relevant dynamical variable.  This approach has been applied to neural behaviour ranging from that seen in small rhythmic networks \cite{Kopell1988} up to the whole brain \cite{Cabral2011}.
Despite the powerful tools and wide-spread use afforded by this formalism, it does have a number of limitations (such as assuming the persistence of the limit cycle under coupling) and it is well to remember that there are other tools from the mathematical sciences relevant to understanding network behaviour.  In this review we wrap the weakly coupled oscillator formalism in a variety of other techniques ranging from symmetric bifurcation theory and groupoid formalisms through to more ``physics-based'' approaches for obtaining reduced models of large networks.  This highlights the regimes where the standard formalism is applicable, and provides a set of complementary tools when it does not.  These are especially useful when investigating systems with strong coupling, or ones for which the rate of attraction to a limit cycle is slow.

In \S~\ref{sec:neuronoscillators} we review some of the key mathematical models of oscillators in neuroscience, ranging from single neuron to neural mass, as well as introduce the standard machinery for describing synaptic and gap junction coupling.  We then present in \S~\ref{sec:collectivedyns} an overview of some of the more powerful mathematical approaches to understanding the collective behaviour in coupled oscillator networks, mainly drawn from the theory of symmetric dynamics.  We touch upon the master stability function approach and the groupoid formalism for handling coupled cell systems.  In \S~\ref{sec:coupledlimitcycles} we review some special cases where it is either possible to say something about the stability of the globally synchronous state in a general setting, or that of phase-locked states for strongly coupled networks of integrate-and-fire neurons.  The challenge of the general case is laid out in
\S~\ref{sec:reduced}, where we advocate the use of phase-amplitude coordinates as a starting point for either direct network analysis or network reduction.  To highlight the importance of dynamics off cycle we discuss the phenomenon of shear-induced chaos.  In the same section we review the reduction to the standard phase-only description of an oscillator, covering the well known notions of isochrons and phase response curves. The construction of phase interaction functions for weakly coupled phase oscillator networks is covered in \S~\ref{sec:weakcoupling}, together with tools for analysing phase-locked states.  Moreover, we go beyond standard approaches and describe the emergence of turbulent states in continuum models with non-local coupling.  Another example of something more complicated than a periodic attractor is that of a heteroclinic attractor, and these are the subject of  \S~\ref{sec:heteroclinic}. The subtleties of phase reduction in the presence of stochastic forcing are outlined in \S~\ref{sec:stochastic}.  The search for reduced descriptions of very large networks is the topic of \S~\ref{sec:macroscopic}, where we cover recent results for Winfree networks that provide an exact mean-field description in terms of a complex order parameter.  This approach makes use of the Ott-Antonsen ansatz that has also found application to chimera states, and which we discuss in a neural context.  In \S~\ref{sec:Applications} we briefly review some examples where the mathematics of this review have been applied, and finally in \S~\ref{sec:discussion} we discuss some of the many open challenges in the field of neural network dynamics.

We will assume the reader has familiarity with the following:
\begin{itemize}
\item The basics of nonlinear differential equation descriptions of dynamical systems such as linear stability and phase-plane analysis.
\item Ideas from the qualitative theory of differential equations/dynamical systems such as asymptotic stability, attractors and limit cycles.
\item Generic codimension-one bifurcation theory of equilibria (saddle-node, Hopf) and of periodic orbits (saddle-node of limit cycles, heteroclinic, torus, flip).
\end{itemize}
There are a number of texts that cover this material very well in the context of neuroscience modelling, for example \cite{Izhikevich05,Ermentrout10}. At the end we include a glossary of some abbreviations that are introduced in the text.

\section{Neurons and neural populations as oscillators}
\label{sec:neuronoscillators}

Nonlinear ionic currents, mediated by voltage-gated ion channels, play a key role in generating membrane potential oscillations and action potentials.  There are many ordinary differential equation (ODE) models for voltage oscillations, ranging from biophysically detailed conductance-based models through to simple integrate-and-fire (IF) caricatures.  This style of modelling has also proved fruitful at the population level, for tracking the averaged activity of a near synchronous state.  In all these cases bifurcation analysis is especially useful for classifying the types of oscillatory (and possibly resonant) behaviour that are possible.   Here we give a brief overview of some of the key oscillator models encountered in computational neuroscience, as well as models for electrical and chemical coupling necessary to build networks.

\subsection{The Hodgkin-Huxley model and its planar reduction}

\label{sec:HH}

The work of Hodgkin and Huxley in elucidating the mechanism of action potentials in the squid giant axon
is one of the major breakthroughs of dynamical modelling in physiology \cite{Hodgkin52}, and see \cite{Rinzel90} for a review.  Their work underpins all modern electrophysiological models, exploiting the observation that cell membranes behave much like electrical circuits.  The basic circuit elements are 1) the phospholipid bilayer, which is analogous to a capacitor in that it accumulates ionic charge as the electrical potential
across the membrane changes; 2) the ionic permeabilities of the membrane, which are
analogous to resistors in an electronic circuit; and 3) the electrochemical driving forces,
which are analogous to batteries driving the ionic currents. These ionic currents are
arranged in a parallel circuit.   Thus the electrical behaviour of cells is based upon the transfer
and storage of ions such as K$^+$ and Na$^+$.

Our goal here is to illustrate, by exploiting specific models of excitable membrane, some of the concepts and techniques which can be used to understand, predict, and interpret the excitable and oscillatory behaviours that are commonly observed in single cell electrophysiological recordings.  We begin with the mathematical description of the Hodgkin-Huxley model.

The standard dynamical system for describing a neuron as a spatially isopotential cell
with constant membrane potential $V$ is based upon conservation of electric
charge, so that
\begin{equation}
C  \FD{}{t} V = I_\text{ion} +I ,
\nonumber
\end{equation}
where $C$ is the cell capacitance, $I$ the applied current and $I_\text{ion}$ represents the sum of individual ionic currents:
\begin{equation}
I_\text{ion} = -g_K (V-V_K) - g_{Na} (V-V_{Na}) - g_L (V-V_L).
\nonumber
\end{equation}
In the Hodgkin-Huxley (HH) model  the membrane current arises mainly through the conduction of sodium and
potassium ions through voltage dependent channels in the membrane.
The contribution from other ionic currents is assumed to obey Ohm's
law (and is called the leak current).   The $g_K$, $g_{Na}$ and $g_{L}$ are conductances (conductance=1/resistance).

\begin{figure}[htbp]
\begin{center}
\includegraphics[width=3in]{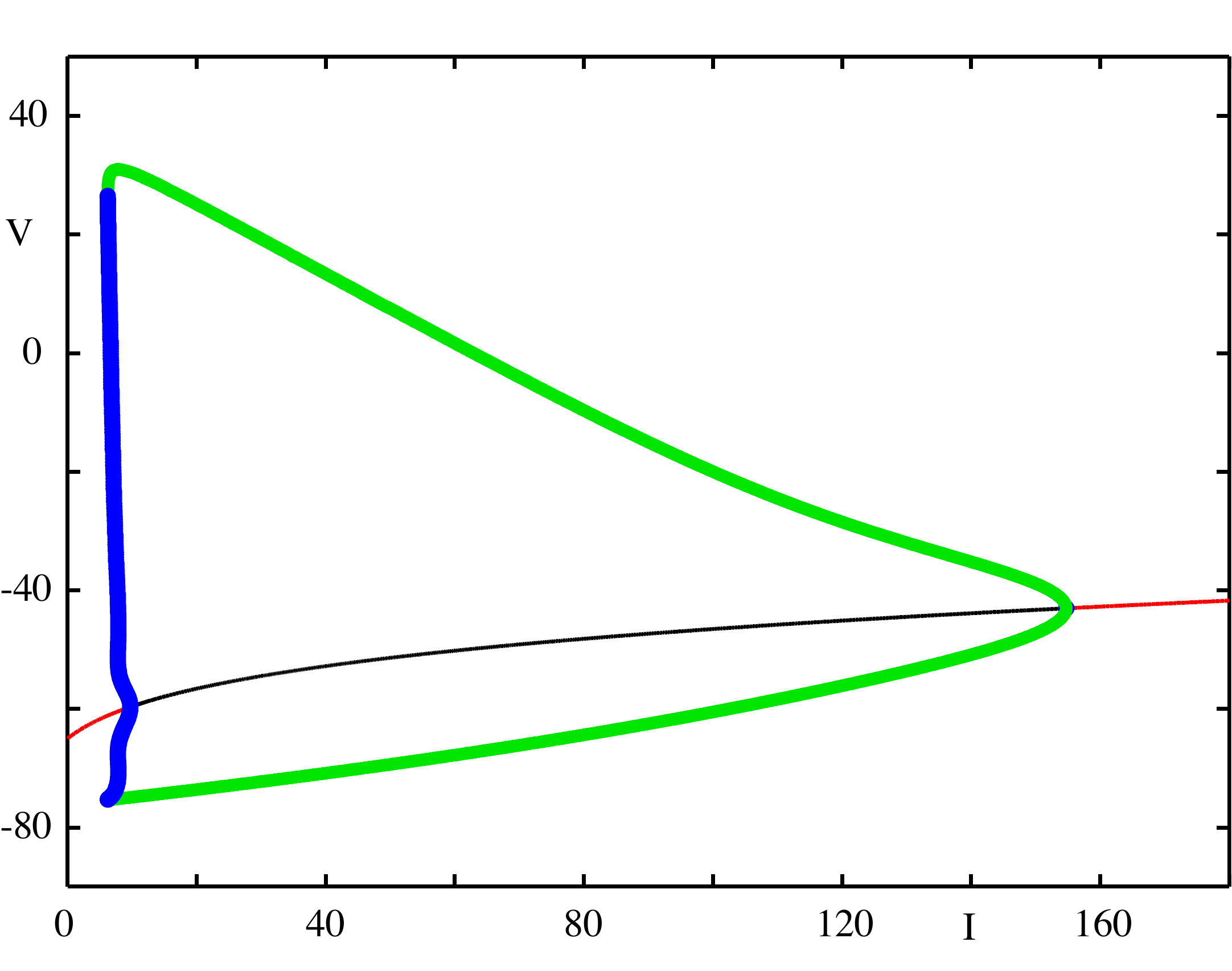}
\caption{
Bifurcation diagram of the Hodgkin-Huxley model as a function of the
external drive $I$.  The green circles show the amplitude of a stable limit cycle and the blue
circles indicate unstable limit cycle behaviour.  The solid red line shows the stable fixed
point and the black line shows the unstable fixed point.  Details of the model are in Appendix~A.
\label{Fig:HHbif}
}
\end{center}
\end{figure}

The great insight of Hodgkin and Huxley was to realise that $g_K$ depends upon four activation gates: $g_K = \overline{g}_K n^4$, whereas $g_{Na}$ depends upon three activation gates and one inactivation gate:
$g_{Na} = \overline{g}_{Na} m^3 h$.  Here the gating variables all obey equations of the form
\begin{equation}
\frac{\d}{\d t} {X} = \frac{X_\infty(V)-m}{\tau_X(V)}, \qquad X \in \{ m,n,h\}.
\nonumber
\end{equation}
The conductance variables described by $X$ take values
between $0$ and $1$ and approach the asymptotic values $X_\infty(V)$ with time constants $\tau_X(V)$.
These six functions are obtained from fits with experimental data.
It is common practice to write $\tau_X(V) = 1/(\alpha_X(V) + \beta_X(V))$, $X_\infty (V) =\alpha_X(V) \tau_X(V)$, where $\alpha $ and $\beta$ have the interpretation of opening and closing channel transition rates respectively.  The details of the HH model are provided in Appendix A for completeness.  A numerical bifurcation diagram of the model in response to constant current injection is shown in Fig.~\ref{Fig:HHbif}, illustrating that oscillations can emerge in a Hopf bifurcation with increasing drive.

\begin{figure}[htbp]
\begin{center}
\includegraphics[width=3in]{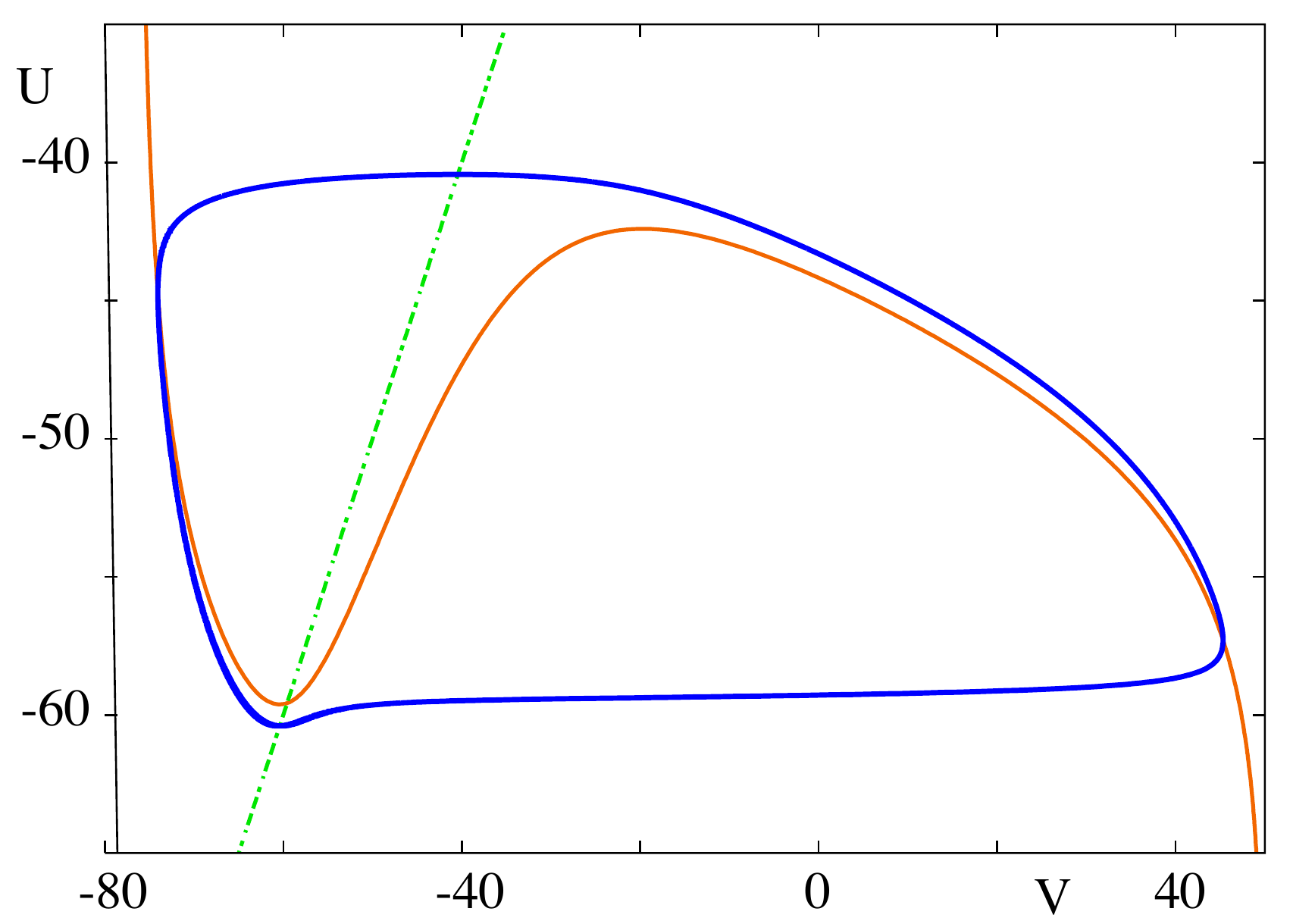}
\caption{
Nullclines (red for $V$ and green for $U$) of the reduced HH neuron mode, obtained using the reduction technique of Abbott and Kepler \cite{Abbott90}, for the oscillatory regime
($I=10$) capable of supporting a periodic train of spikes.  The periodic orbit is shown in blue.
\label{Fig:HHR}
}
\end{center}
\end{figure}

The mathematical forms chosen by Hodgkin and Huxley for the functions $\tau_{X}$ and $X_\infty$, $X \in \{m,n,h\}$, are all transcendental functions.  Both this and the high dimensionality of the model make analysis difficult.  However, considerable simplification is attained with the following observations:  (i)
$\tau_m(V)$ is small for all $V$ so that the variable $m$ rapidly approaches its equilibrium value $m_\infty(V)$, and (ii) the equations for $h$ and $n$ have similar time-courses and may be \textit{slaved} together.  This has been put on a more formal footing by Abbott and Kepler \cite{Abbott90}, to obtain a reduced planar model for $(V,U)$ obtained from the full Hodgkin-Huxley model under the replacement $m \rightarrow m_\infty(V)$ and $X = X_\infty (U)$ for $X \in \{ n , h \}$ with a prescribed choice of $U$.  The phase-plane and nullclines for this system are shown in Fig.~\ref{Fig:HHR}.

For zero external input the fixed point
is stable and the neuron is said to be
\textit{excitable}.  When a positive external current is applied the
low-voltage portion of the $V$ nullcline moves up
whilst the high-voltage part remains relatively unchanged.
For sufficiently large constant external input
the intersection of the two nullclines falls within the portion of the
$V$ nullcline with positive slope.  In this case the fixed point is unstable and
the system may support a limit cycle.  The system is said to be oscillatory as it may
produce a train of action potentials.  Action potentials may also be induced in the
absence of an external current for synaptic stimuli of sufficient
strength and duration.  This simple planar model captures all of the essential features of the original HH model yet is much easier to understand from a geometric perspective.  Indeed the model is highly reminiscent of the famous FHN model, in which the voltage nullcline is taken to be a cubic function.  Both models show
the onset of repetitive firing at a non-zero frequency as observed in
the HH model (when an excitable state loses stability via a subcritical Hopf bifurcation).  However, unlike real cortical neurons they cannot fire at arbitrarily low frequency.  This brings us to consider modifications of the original HH formalism to accommodate bifurcation mechanisms from excitable to
oscillatory behaviours that can respect this experimental observation.

\subsection{The cortical model of Wilson}

\begin{figure}[htbp]
\begin{center}
\includegraphics[width=4in]{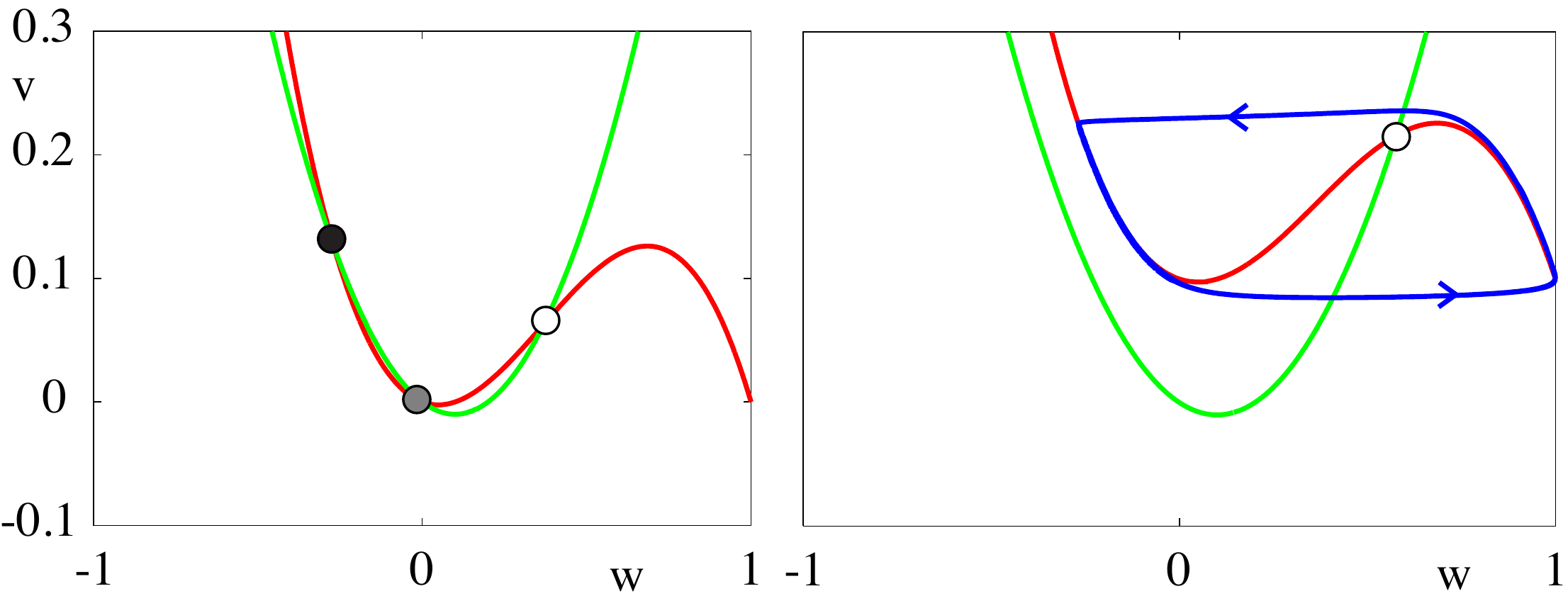}
\caption{Phase portrait for cortical neuron model with quadratic recovery variable, $a=0.1$, $\beta=\gamma=0.5$, $v_1=0$,$v_2=0.2$.  The voltage nullcline is shown in red and that of the recovery variable in green.
Left: $I=0$, showing a stable fixed point (black filled circle), a saddle (grey filled circle) and an unstable fixed point (white filled circle).  Right: $I=0.1$, where there is an unstable fixed point (white filled circle) with a stable limit cycle (in blue) for $C=0.01$.
\label{Fig:Cortical}
}
\end{center}
\end{figure}

Many of the properties of real cortical neurons can be captured by making the equation for the recovery variable of the FHN equations quadratic (instead of linear).  We are thus led to the cortical model of Wilson \cite{Wilson99}:
\begin{align}
C \frac{\d}{\d t} {v} & = f(v)  - w +I, \qquad f(v) = v(a-v)(v-1) , \nonumber  \\
\frac{\d}{\d t} {w} & = \beta (v-v_1)(v-v_2)-\gamma w \nonumber ,
\end{align}
where $0<a<1$ and $b,\gamma, \mu >0$.  Here $v$ is like the membrane potential $V$, and $w$ plays the role of a gating variable.
In addition to the single fixed point of the FHN model it is possible to have another  pair of fixed points, as shown in Fig.~\ref{Fig:Cortical} (right).  As $I$ increases two fixed points can annihilate at a {\em saddle node on an invariant curve} (SNIC) bifurcation at $I=I_c$ \cite{Izhikevich05}.  In the neighbourhood of this global bifurcation the firing frequency scales like $\sqrt{I-I_c}$.
For large enough $I$ there is only one fixed point on the middle branch of the cubic, as illustrated in Fig.~\ref{Fig:Cortical} (left).  In this instance an oscillatory solution occurs via the same mechanism as for the  FHN model.

\subsection{Morris-Lecar with homoclinic bifurcation}

\begin{figure}[htbp]
\begin{center}
\includegraphics[width=3in]{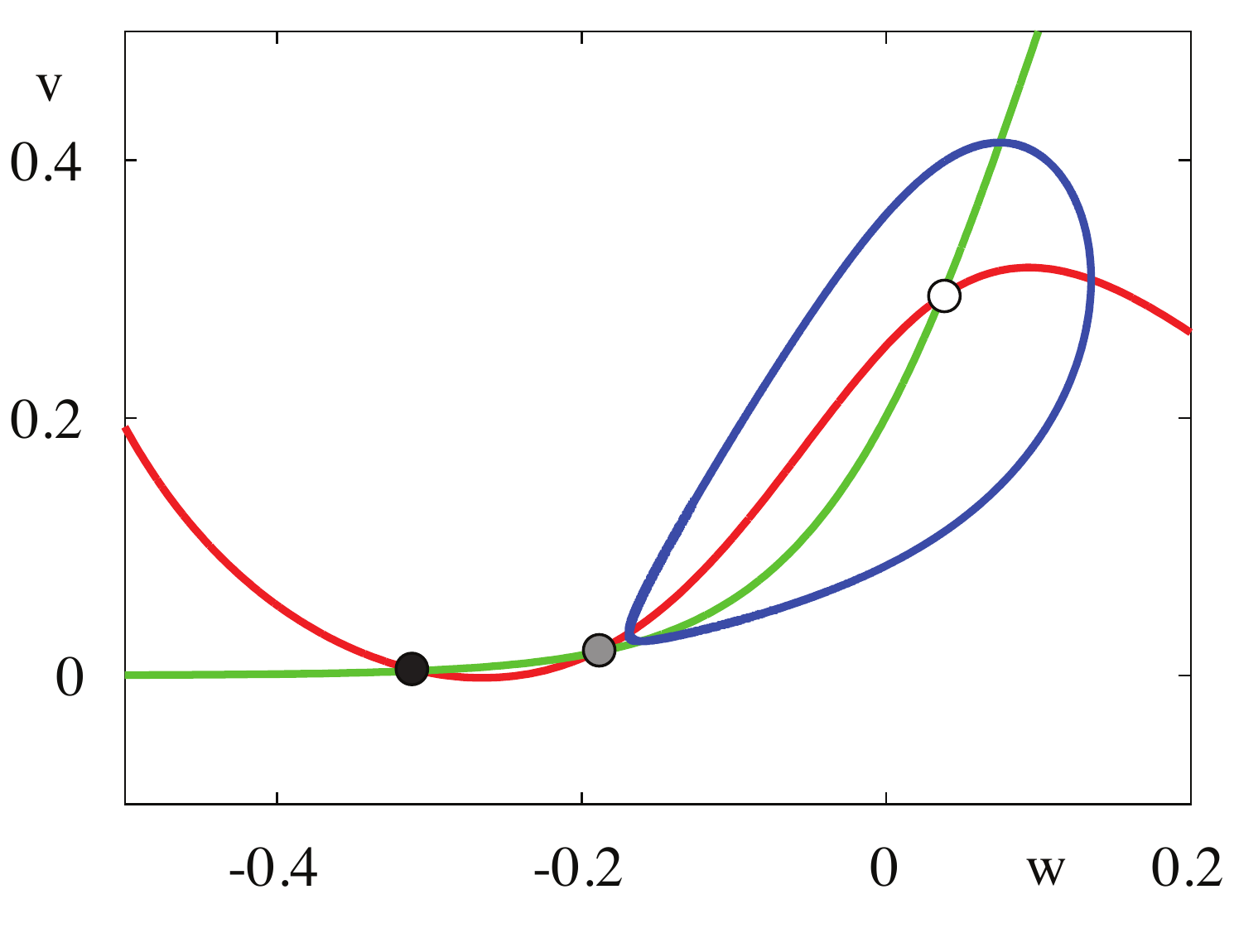}
\caption{Phase portrait of the Morris-Lecar model at $I=0.075$ with
$C=1$, $V_k=-0.7$, $V_L=-0.5$, $V_{Ca}=1$, $g_K=2$, $g_L=0.5$,
$V_1=-0.01$, $V_2=0.15$, $g_{Ca}=1.33$,
$V_3=0.1$, $V_4=0.145$ and $\phi=1/3$.
The voltage nullcline is shown in red and that of the gating variable in green.
The filled black circle indicates a stable fixed point, the grey filled circle a saddle and the filled white circle an unstable fixed point.  The periodic orbit is shown in blue.
\label{Fig:ML}
}
\end{center}
\end{figure}

A SNIC bifurcation is not the only way to achieve a low firing rate in a single neurone model.  It is also possible to achieve this via a homoclinic bifurcation, as is possible in the Morris-Lecar (ML) model \cite{Morris81}.  This was originally developed as a model for barnacle muscle fiber.  Morris and Lecar introduced a set of coupled ordinary differential equations (ODEs) incorporating two ionic currents: an outward going, non-inactivating potassium current and an inward going, non-inactivating calcium current.  Assuming that the calcium  currents operate on a
much faster time scale than the potassium current one they formulated the following two dimensional system:
\begin{align}
C\FD{}{t}V &= g_L(V_L-V) +g_K w (V_K-V) + g_{Ca}m_\infty(V)(V_{Ca}-V) + I ,\nonumber \\
\FD{}{t}w &= \lambda(V)(w_\infty(V)-w), \nonumber
\end{align}
with $m_\infty(V) = 0.5(1+\tanh[(V-V_1)/V_2])$, $w_\infty(V) = 0.5(1+ \tanh[(V-V_3)/V_4])$, and $\lambda(V) = \phi \cosh[(V-V_3)/(2V_4)]$.  Here $w$ represents the fraction of  K$^+$ channels open, and the Ca$^{2+}$ channels respond to $V$ so rapidly that we assume instantaneous activation.  Here $g_L$ is the leakage conductance, $g_K,g_{Ca}$ are the potassium and calcium conductances, $V_L,V_K,V_{Ca}$ are corresponding reversal potentials,
$m_{\infty}(V)$, $w_{\infty}(V)$ are voltage-dependent gating functions and $\lambda(V)$ is a
voltage-dependent rate.  Referring to Fig.~\ref{Fig:ML}, as $I$ decreases the periodic orbit grows in amplitude, it comes closer to a saddle point and slows down such that near the homoclinic bifurcation, where the orbit collides with the saddle at $I=I_c$, the frequency of oscillation scales as $-1/\log(I-I_c)$.

\subsection{Integrate-and-fire}

Although conductance-based models like that of Hodgkin and Huxley provide a level of detail that helps us to understand how the kinetics of channels (with averaged activation and inactivation variables) can underlie action-potential generation their high dimensionality is a barrier to studies at the network level. The goal of a network-level analysis is to predict emergent computational properties in populations and recurrent networks of neurons from the properties of their component cells.
Thus simpler (lower dimensional and hopefully mathematically tractable) models are more appealing - especially if they fit single neuron data.

A one-dimensional nonlinear IF model takes the form
\begin{equation}
\tau \FD{}{t} v = f(v) + I \quad \mbox{ if }v<v_{\text{th}} ,
\label{eq:one}
\end{equation}
such that $v(t)$ is reset to $v_{\text{R}}$ just after reaching the value $v_{\text{th}}$. In other words we seek a piecewise discontinuous solution $v(t)$ of (\ref{eq:one}) such that
$$
\lim_{t\rightarrow t_0-}v(t_0)=v_{\text{th}}~\mbox{ implies that }\lim_{t\rightarrow t_0+} v(t)=v_{\text{R}}.
$$
Firing times are defined iteratively according to
\begin{equation}
T_n = \inf \{ t ~| v(t) \geq v_{\text{th}}~;~t \geq T_{n-1} \}. \nonumber
\end{equation}
Real cortical data can be very accurately fit using:
\begin{equation}
f(v)= v_\text{L}-v + \kappa \e^{(v-v_{\kappa})/\kappa}, \nonumber
\end{equation}
with $v_\text{L}=-68.5$ mV, $\tau=3.3$ ms, $v_\kappa = -61.5$ mV and $\kappa=4$ mV \cite{Badel08}.
There are many varieties of nonlinear IF model, with the quadratic one \cite{Latham00} being well known as a precursor for the planar Izhikevich spiking model \cite{Izhikevich03}, itself capable of generating a wide variety of firing patterns, including bursting and chattering as well as regular spiking.  For a more thorough discussion of IF models and the challenges of analysing non-smooth dynamical systems we refer the reader to \cite{Coombes12}.

\subsection{Neuronal coupling}
\label{sec:coupling}

At a synapse presynaptic firing results in the release of
neurotransmitters that cause a change in the membrane conductance of
the postsynaptic neuron.  This postsynaptic current may be written $I_s(t) = g_s s(t) (V_s - V(t))$
where $V(t)$ is the voltage of the postsynaptic neuron, $V_s$ is the membrane reversal potential and $g_s$ is a constant conductance.  The variable $s$ corresponds to the probability that a synaptic receptor channel is in
an open conducting state.  This probability depends on the presence and concentration  of neurotransmitter released by the presynaptic neuron. The sign of $V_s$ relative to the resting potential (which without loss of generality we may set to zero) determines whether the synapse is excitatory ($V_s >0$) or inhibitory ($V_s < 0$).

The effect of some synapses can be described with a function that fits the shape of the postsynaptic response due to the arrival of action potential at the pre-synaptic release site.  A postsynaptic potential (PSP) $s(t)$
would then be given by $s(t) = \eta(t-T)$, $t \geq T$ where $T$ is the arrival time of a pre-synaptic action potential and $\eta(t)$ fits the shape of a realistic PSP (with $\eta(t)=0$ for $t<0$). A common (normalised) choice for $\eta(t)$ is a
difference of exponentials:
\begin{equation}
\eta(t) = \left(\frac{1}{\alpha} - \frac{1}{\beta}\right)^{-1}(\e^{-\alpha t} - \e^{-\beta t}) ,
\label{eq:doe}
\end{equation}
or the alpha function $\alpha^2 t \e^{-\alpha t}$ obtained from (\ref{eq:doe}) in the limit $\beta \rightarrow \alpha$.
The PSP arising from a train of action potentials is given by
\begin{equation}
s(t) = \sum_{m \in \ZSet} \eta (t-T_m),
\label{eqstt}
\end{equation}
where $T_m$ denotes the arrival time of the $m$th action potential at a synapse.

\begin{figure}[htbp]
\begin{center}
\includegraphics[width=3in]{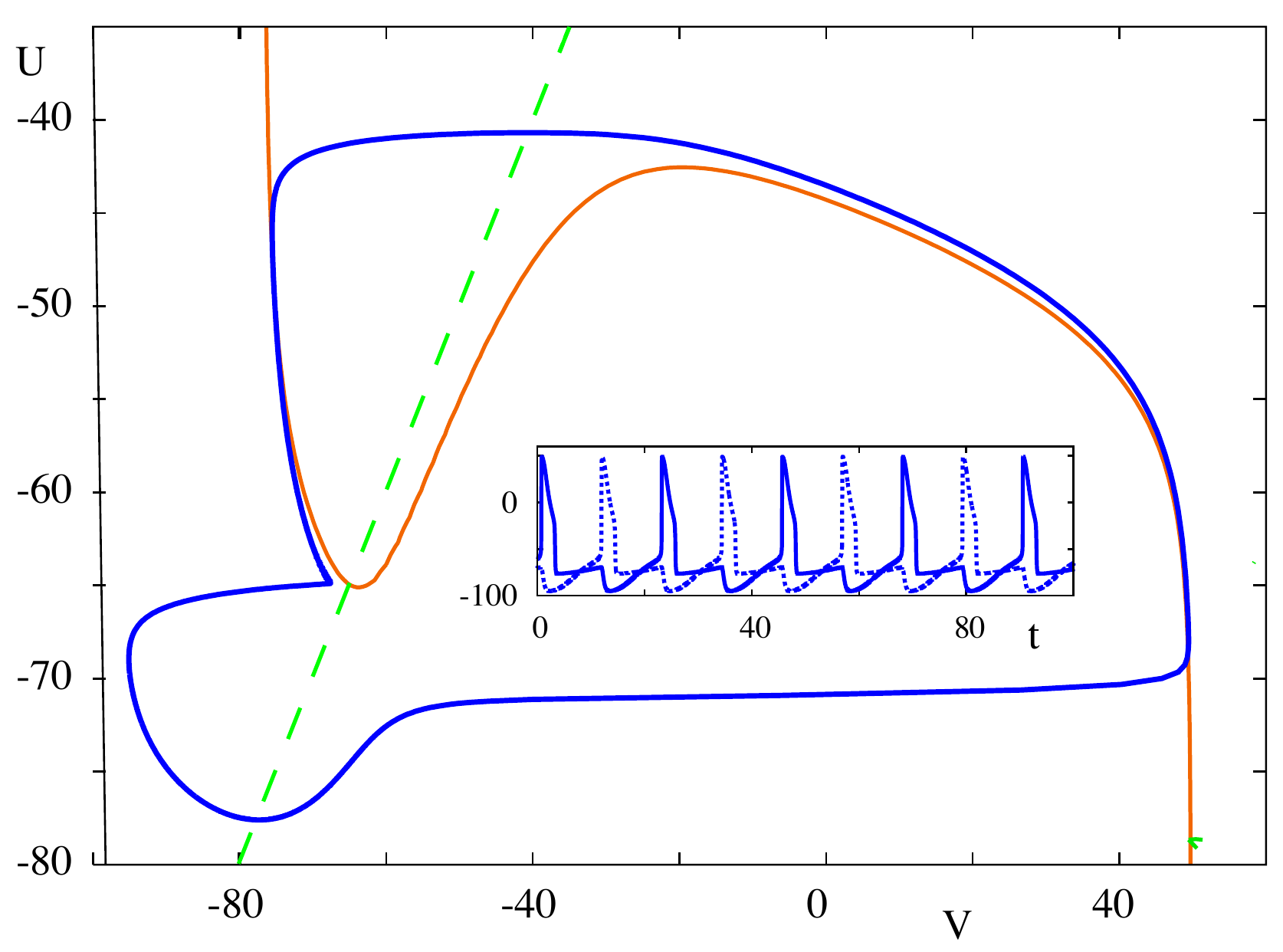}
\caption{A Half Center oscillator built from two (reduced) Hodgkin-Huxley neurons with mutual reciprocal inhibition
modelled by an $\alpha$-function synapse and a negative reversal potential.  The inset shows the voltage traces (solid and dashed lines) for the two neurons.  The solid blue line in the $(V,U)$ space shows the common orbit of the two neurons (though each neuron travels half a period out of phase with the other).
Parameters: $\alpha=1$ ms, $V_s=-100$ mV, $g_s=100$ mS cm$^{-2}$.  Spike times determined by a voltage threshold at $V=-35$ mV.
\label{Fig:HalfCenter}
}
\end{center}
\end{figure}

Interestingly even purely inhibitory synaptic interactions between non-oscillatory neurons
can create oscillations at the network level, and can give rise to central pattern generators of ``half-center" type \cite{Brown1914}. To see this we need only consider a pair of (reduced) Hodgkin-Huxley neurons
with mutual reciprocal inhibition mediated by an $\alpha$-function synapse with a negative reversal potential.
The phenomenon of anode break excitation (whereby a neuron fires an action potential in response to termination of a hyperpolarising current) can underlie a natural anti-phase rhythm, and is best understood in terms of the phase plane shown in Fig.~\ref{Fig:HHR}.  In this case inhibition will effectively move the voltage nullcline down, and the system will equilibrate to a new hyperpolarised state.  Upon release of inhibition the fixed point will move to a higher value, though to reach this new state the trajectory must jump to the right hand branch (since now $\ID{V}>0$). An example is shown in Fig.~\ref{Fig:HalfCenter}.

Gap junctions differ from chemical synapses in that they allow for direct communication between cells.
They are typically formed from the juxtaposition of two hemichannels (connexin proteins) and allow the free movement of  ions or molecules across the intercellular space separating the plasma membrane of one cell from another.
As well as being  found in the neocortex,  they occur in many other brain regions, including the hippocampus, inferior olivary nucleus in the brain stem,
the spinal cord, and the thalamus \cite{Connors1995}.  Without the need for receptors to recognise
chemical messengers, gap junctions are much faster than chemical synapses at relaying signals. The synaptic delay for a chemical synapse is typically in the range $1$--$100$ms, while the synaptic delay for an electrical synapse may be only about $0.2$ms.

It is common to view the gap-junction as nothing more than a channel that conducts current according to a simple ohmic model.  For two neurons with voltages $v_i$ and $v_j$ the current flowing into cell $i$ from cell $j$ is given by $I_\text{gap}(v_i,v_j)=g (v_j-v_i)$, where $g$ is the constant strength of the gap junction conductance.  They are believed to promote synchrony between oscillators (e.g. see \cite{Torben2012}), though the story is more subtle than this as we shall  discuss in \S~\ref{sec:coupledlimitcycles}.

\subsection{Neural mass models}

\label{subsec:neuralmass}

As well as supporting oscillations at the single neuron level, brain tissue can also generate oscillations at the tissue level.  Rather than model this using networks built from single neuron models, it is has proven especially useful to develop low dimensional models to mimic the collection of thousands of near identical interconnected neurons with a preference to operate in synchrony.  These are often referred to as neural mass models, with state variables that track coarse grained measures of the average membrane potential, firing rates or synaptic activity.
They have proven especially useful in the description of human EEG power spectra \cite{Liley02}, as well as resting brain state activity \cite{Deco11} and mesoscopic brain oscillations \cite{Young2009}.

\begin{figure}[htbp]
\begin{center}
\includegraphics[width=3in]{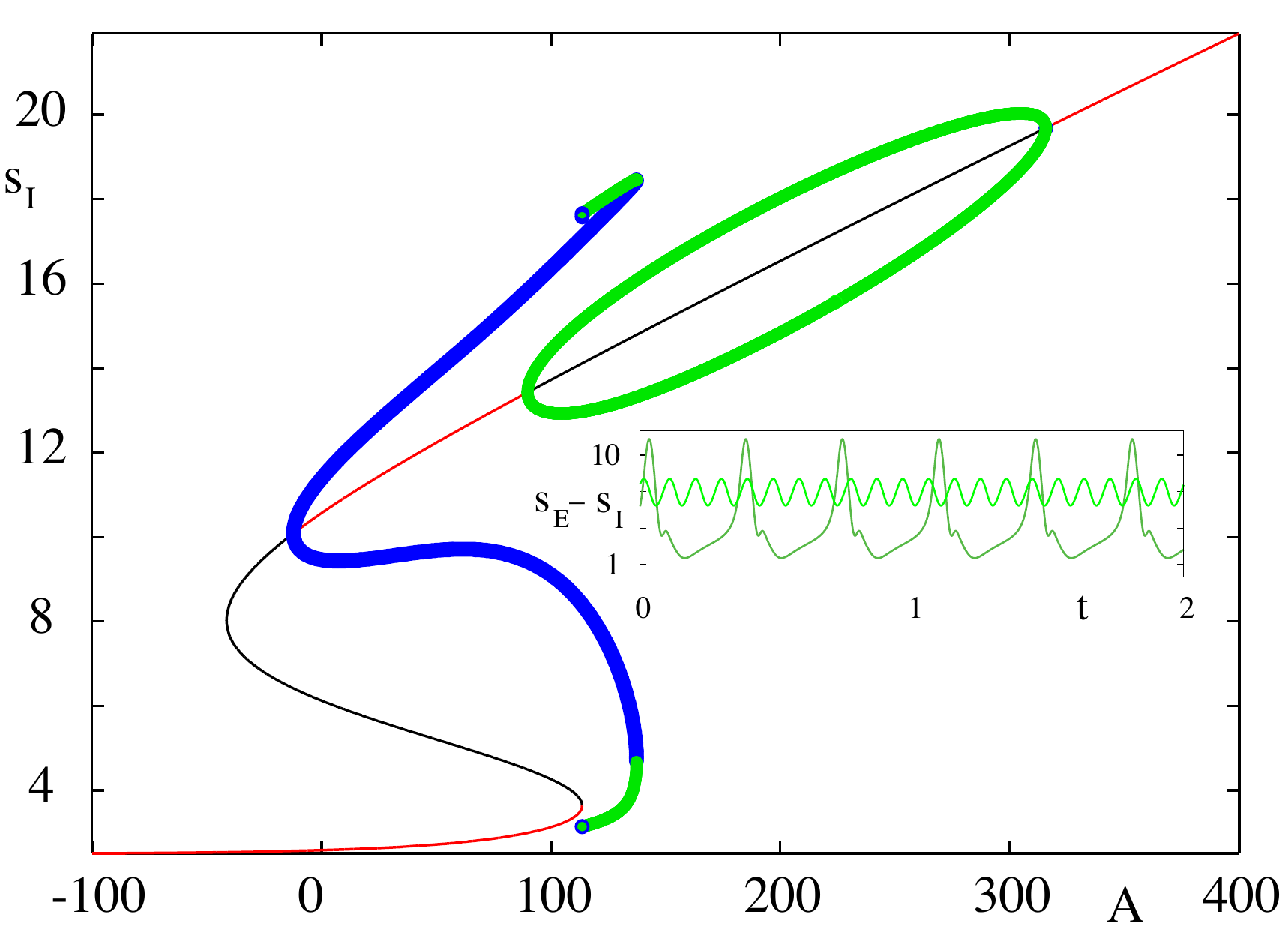}
\caption{Bifurcation diagram for the Jansen-Rit model with
$\beta_E=100$ s$^{-1}$, $\beta_I=50$ s$^{-1}$, $A_E=3.25$ mV, $A_I=22$ mV, $\nu = 5$ s$^{-1}$, $v_0=6$ mV, $r=0.56$ mV$^{-1}$, $C_1=135$, $C_2=0.8 C_1$, $C_3=0.25 C_1= C_4$.
Solid red (black) lines represent stable (unstable) fixed points.  Green (blue) points denote the amplitude of stable (unstable) periodic orbits that emerge via Hopf bifurcations.
The inset shows the co-existence of two stable periodic orbits at $A=125$ Hz.
\label{Fig:JR}
}
\end{center}
\end{figure}

In many neural population models, such as the well known Wilson-Cowan model \cite{Wilson72}, it is assumed that the interactions are mediated by firing rates rather than action potentials (spikes) \textit{per se}.
To see how this might arise we rewrite (\ref{eqstt}) in the equivalent form using a sum of Dirac $\delta$-functions
\begin{equation}
\left (1 + \frac{1}{\alpha} \FD{}{t} \right ) \left (1 + \frac{1}{\beta} \FD{}{t} \right ) s = \sum_{m \in \ZSet} \delta(t-T_m).
\label{eq:ss}
\end{equation}
Identifying the right hand side of (\ref{eq:ss}) as a train of pre-synaptic spikes motivates the form of a phenomenological rate model in the form
\begin{equation}
Q s = f ,
\label{eq:rate}
\end{equation}
with $f$ identified as a firing rate and $Q$ identified as the differential operator $(1+\alpha^{-1} \d /\d t)(1+\beta^{-1} \d /\d t)$.  At the network level it is then common practice to close this system of equations by specifying $f$ to be a function of pre-synaptic activity.  A classic example is the Jansen-Rit model \cite{Jansen95}, which describes a network of interacting pyramidal neurons (P), inhibitory interneurons (I) and excitatory interneurons (E), and has been used to model both normal and epileptic patterns of cortical activity \cite{Touboul2011}.  This can be written in the form
\begin{equation}
Q_{E} s_P = f(s_E -s_I) , \quad Q_{E} s_E = C_2f(C_1 s_P) +A , \quad Q_{I}s_I = C_4 f(C_3 s_P) ,
\nonumber
\end{equation}
which is a realisation of the structure suggested by (\ref{eq:rate}), with the choice
\begin{equation}
f(v) = \frac{\nu}{1+\e^{-r(v-v_0)}},
\nonumber
\end{equation}
and $Q_a = (1+\beta_a^{-1}\d /\d t)^2 \beta_a/A_a$ for $a \in \{E,I\}$.
Here $A$ is an external input.  When this is a constant we obtain the bifurcation diagram shown in Fig.~\ref{Fig:JR}.
Oscillations emerge via Hopf bifurcations and it is possible for a pair of stable periodic orbits to coexist.  One of these has a frequency in the alpha band and the other is characterised by a lower frequency and higher amplitude.  Recently a network of such modules, operating in the alpha range and with additive noise, has been used to investigate mechanisms of cross-frequency coupling between brain areas \cite{Jedynak2015}.  Neural mass models have also previously been used to model brain resonance phenomena \cite{Spiegler2011}, for modelling of epileptic seizures \cite{Breakspear2006,Terry2012,Chowdhury2014}, and are very popular in the Neuroimaging community for model driven EEG/fMRI fusion \cite{Valdes-Sosa2009}.

Now that we have introduced some oscillator models for neurons and neural populations it is appropriate to consider the set of tools for analysing their behaviour at the network level.

\section{Dynamical systems approaches to collective behaviour}
\label{sec:collectivedyns}

We give a brief overview of some dynamical systems approaches, concepts and techniques that can be used to understand collective behaviour that spontaneously appears in coupled dynamical system models used for neuroscience modelling. We do not give a complete review of this area but try to highlight some of the approaches and how they interact; some examples of applications of these to neural systems are given in later chapters.

\subsection{Synchrony and asynchrony}
\label{subsec:synchrony}

A set of $N$ coupled nonlinear systems represented by ODEs can be written
\begin{equation}
\frac{\d}{\d t}x_i = g_i(x_i;x_1,\ldots,x_N) ,
\label{eq:gen-coupledcells}
\end{equation}
where $x_i\in\R^d$ represent the state space of the individual systems whose evolution $g_i$ is affected both by the current state of the system and by the states of those coupled to that system. In the mathematics literature, the $x_i$ are often called ``cells'' though we note that the $x_i$ may include degrees of freedom in the coupling as well as variables such as membrane potential that reflect the state of the ``biological cell''. Note there is potential for notational confusion here: to clarify this we write
$$
x\in\R^{Nd},~~~~x_i\in\R^d,~~~~x_i^{(j)}\in \R.
$$

One of the most important observations concerning the collective dynamics of coupled nonlinear systems relates to whether the collection behaves as one or not - whether there is an attracting synchronous state, or whether more complex spatio-temporal patterns such as clustering appear. There is a very large literature, even restricting to the case of applications of synchrony, and one where we cannot hope to do the whole area justice. We refer in particular to \cite{Arenas2008,Pikovsky01}. Various applications of synchrony of neural models are discussed, for example, in \cite{Brette2012,Brunel2008,Casagrande2005,Deville2012,Nowotny2008,Parga2007,Pinsky1995,Popovych2006,Rubin2002} while there is a large literature (e.g. \cite{Uhlhaas2009}) discussing the role of synchrony in neural function. Other work looks for example at synchronisation of groups of networks \cite{Sorrentino2007} and indeed synchrony can be measured experimentally \cite{Reyes2003} in groups of neurons using dynamic patch clamping.

We discuss some of the types of behaviour that can emerge in the collective dynamics and the response of partial synchronised states to external forcing. Clearly, any system of $N$ coupled dynamical systems can be written in the form
\begin{equation}
\frac{\d}{\d t}x_i = f_i(x_i)+\epsilon g_i(x_i;x_1,\ldots,x_N) ,
\label{eq:coupledcells}
\end{equation}
where each system is parametrised by $x_i\in\R^d$; $\epsilon=0$ corresponds to decoupling the systems and the functions $g_i$ represent drive from other systems on the $i$th system. Many approaches start with the less general case
\begin{equation}
\FD{}{t} x_i = f_i(x_i)+ \epsilon \sum_{j=0}^{N} w_{ij} G_{ij}(x_i,x_j) ,
\label{eq:pairwisecoupledcells}
\end{equation}
which can be justified for system where the collective interaction between systems can be broken into ``pairwise'' interactions $G_{ij}$ that are summed according to some linear weights $w_{ij}$ (some of which may be zero) that represent the strength of the couplings and $\epsilon$ the strength of coupling of the network. Note that there is clearly no unique way to write the system in this form; more specifically, one can without loss of generality choose $\epsilon=1$, $w_{ij}=1$ by suitable choice of $G_{ij}$. On the other hand it can be useful to be able to e.g. modulate the strength of the coupling across the whole network independently of changing individual coupling strengths. Similarly it is often useful to specialise to a case such as $G_{ij}=G$ and have similar interactions.

\subsection{Networks, motifs and coupled cell dynamics}
\label{sec:networks}

We focus now on the dynamics of pairwise coupled networks such as (\ref{eq:pairwisecoupledcells}) as this form is assumed in most cases. Under an additional assumption that the coupling between the oscillators is of the same type and either present or absent, one can consider uniformly weighted coupling of the form
\begin{equation}
w_{ij}= g A_{ij} ,
\nonumber
\end{equation}
where $g$ is fixed and $A_{ij}$ is an adjacency matrix of the graph of interactions, i.e. $A_{ij}=1$ if there is a link from $j$ to $i$ and $0$ otherwise, or more generally
\begin{equation}
w_{ij}= g_{ij} A_{ij} ,
\nonumber
\end{equation}
where $g_{ij}>0$ represents the strength of coupling and $A_{ij}$ the adjacency matrix. It is clear that the coupling structure as represented in $A_{ij}$ will influence the possible dynamics on the network and to make further progress it is useful to restrict to particular network structures. Some important literature on network structures is reviewed for example in \cite{Newman2010}, while \cite{Arenas2008} reviews work on synchrony in complex networks up to the time of their publication;  recent work includes for example \cite{DoGross2012}. For a recent review of the application of complex networks to neurological disorders in the brain, see \cite{Stam2014}.

It is interesting to try and understand the effect of network structure on synchrony, so we briefly outline some basic measures of network structure. The {\em in-degree} of the node $i$ is the number of incoming connections (i.e. $d_{\mathrm{in}}(i)=\sum_{j} A_{ji}$), while the {\em out-degree} is the number of outgoing connections (i.e. $d_{\mathrm{out}}(i)=\sum_{j} A_{ji}$) and the distribution of these degrees is often used to characterise a large graph. A {\em scale-free} network is a large network where the distribution of in (or out) degrees scales as a polynomial. This can be contrasted with highly structured homogeneous networks (for example on a lattice) where the degree may be the same at each node. Other properties commonly examined include the clustering properties and path lengths within the graph. There are also various measures of centrality that help one to determine the most important nodes in a graph - for example the {\em betweenness centrality} is a measure of centrality that is the probability that a given node is on the shortest path between two uniformly randomly chosen nodes \cite{Newman2010}. As expected, the more central nodes are typically most important if one wishes to achieve synchrony in a network.

Other basic topological properties of networks that are relevant to their dynamics include, for example, the following, most of which are mentioned in \cite{Arenas2008,Newman2010}: The network is {\em undirected} if $A_{ij}=A_{ji}$ for all $i,j$, otherwise it is {\em directed}. We say nodes $j$ and $i$ in the network $A_{ij}$ are {\em path-connected} if for some $n$ there is a path from $j$ to $i$, i.e. $(A^n)_{ij}\neq 0$ for some $n$. The network is {\em strongly connected} if for each $i,j$ is path connected in both directions while it is {\em weakly connected} if we replace $A_{ij}$ by $\max(A_{ij},A_{ji})$ (i.e. we make the network undirected) and the latter network is strongly connected. N.B. There may be some confusion in that strong and weak connectivity are properties of a directed network - while strong and weak coupling are properties of the coupling strengths for a given network!  The {\em diameter} of a network is the maximal length of a shortest path between two points on varying the endpoints. Other properties of the adjacency matrix are discussed for example in \cite{Atay2006} where spectral properties of graph Laplacians are linked to the problem of determining stability of synchronised states. Other work we mention is that of Pecora \textit{et al}. \cite{Pecoraetal2000,PecBar2005} on synchronisation in coupled oscillator arrays (and see \S~\ref{sec:MSF}), while \cite{Rothkegel2011} explores the recurrent appearance of synchrony in networks of pulse-coupled oscillators (and see \S~\ref{sec:pulse}).

Finally, we mention {\em network motifs} - these are subgraphs that are ``more prevalent'' than others within some class of graphs. More precisely, given a network one can look at the frequency with which a small subgraph appears relative to some standard class of graphs (for example random graphs) and if a certain subgraph appears more often than expected, this characterises an important property of the graph \cite{Miloetal2002}. Such analysis has been used in systems biology (such as transcription or protein interaction networks) and has been applied to study the structure in neural systems (see for example \cite{SpornsKoetter2004,Spornsetal2007}) and the implications of this for the dynamics. The have also been used to organise the analysis of the dynamics of small assemblies of coupled cells, see for example \cite{Kamei2009,Benjaminetal2012}.

\subsection{Weak and strong coupling}
\label{sec:WeakandStrong}

Continuing with systems of the form (\ref{eq:coupledcells}) or (\ref{eq:pairwisecoupledcells}), if the coupling parameter $\epsilon$ is, in some sense, small we refer to the system as ``weakly coupled''. Mathematically, the weak coupling approximation is very helpful because it allows one to use various types of perturbation theory to investigate the dynamics \cite{Hoppensteadt97}. For coupling of limit cycle oscillators it allows one to greatly reduce the dimension of phase space. Nonetheless, many dynamical effects (e.g. ``oscillator death'' where the oscillations in one or more oscillators are completely suppressed by the action of the network \cite{Ermentrout90}) cannot occur in the weak coupling limit and moreover real biological systems that often have ``strong coupling''. We will return to this topic to discuss oscillator behaviour. One can sometimes use additional structure such as weak dissipation and weak coupling of the oscillators to perform a semi-analytic reduction to phase oscillators; see for example \cite{Ashwin2000a,Sepulchre1997}.

\subsection{Clusters, exact and generalised synchrony}

If one has a notion of synchrony between the systems of (\ref{eq:coupledcells}), it is possible to discuss {\em clustering} according to mutual synchrony. Caution needs to be exercised whenever discussing synchrony - there are many distinct notions of synchrony which may be appropriate in different contexts and, in particular, synchrony is typically a property of a particular solution at a particular point in time rather than a property of the system as a whole.

An important case of synchrony is {\em exact synchrony}: we say $x_i(t)$ and $x_j(t)$ are exactly synchronised if $x_i(t)=x_j(t)$ for all $t$. {\em Generalised synchrony} is, as the name suggests, much more general and corresponds to there simply being a functional relationship of the form $x_i(t)=F(x_j(t))$. For oscillators, phases can be used to define additional notions such as {\em phase} and {\em frequency synchrony}: see \S~\ref{subsec:locking}.

Although we focus mostly on individual units with simple (periodic) dynamics, if the units have more complex dynamics (such as ``chaotic oscillators'') one can understand synchrony of the cells by analysis of the linearised differences between oscillators, and there is a sizeable literature on this; see the review \cite{Boccaletti-etal-2006}, or \cite{Nekorkin1999} for clusters in a system of globally coupled bistable oscillators. In the case of two linearly coupled identical chaotic oscillators
$$
\FD{}{t} x_1 = f(x_1)+\epsilon (x_2-x_1),~~\FD{}{t} x_2 = f(x_2)+\epsilon (x_1-x_2) ,
$$
where $(x_1,x_2)\in\R^{2d}$, if the individual oscillator $\ID{x}=f(x)$ has a chaotic attractor $A\subset \R^d$ then the coupled system will have an attracting exactly synchronised attractor $\tilde{A}=\{(x,x)~:~x\in A\}$ only if the coupling $\epsilon$ is sufficiently large in relation to the maximal Liapunov exponent of the synchronous state \cite{Boccaletti-etal-2006}.

\subsection{Synchrony, dynamics and time delay}

An area of intense interest is the role of time delay in the collective dynamics of coupled systems. In the neural context it is natural to include propagation delay between neurons explicity, for example in models such as
\begin{equation}
\frac{\d}{\d t}x_i(t) = f_i(x_i(t))+ \epsilon \sum_{j=1}^{N} w_{ij} G_{ij}(x_i(t),x_j(t-\tau)) ,
\label{eq:pairwisedelaycoupledcells}
\nonumber
\end{equation}
where the delay time $\tau>0$ represents the time of propagation of the signal from one neuron to another. This presents a number of serious mathematical challenges, not least due to the fact that delay dynamical systems are infinite dimensional: one must specify the initial condition over a time interval $t\in[t_0-\tau,t_0]$ in order to have any chance of uniquely defining the future dynamics; this means that for a state to be linearly stable, an infinite number of eigenvalues need to have real part less than zero.

Nonetheless, much can be learned about stability, control and bifurcation of dynamically synchronous states in the presence of delay; for example \cite{Campbell2007,Hoevel2008,Dhamala2004,Popovych2006,Levnajic2010}, and the volume \cite{Atay2010} includes a number of contributions by authors working in this area. There are also well-developed numerical tools such as DDE-BIFTOOL \cite{dde-biftools2002,dde-biftools2014} that allow continuation, stability and bifurcation analysis of coupled systems with delays.  For an application of these techniques to the study of a Wilson-Cowan neural population model with two delays we refer the reader to \cite{Coombes09}.

\subsection{A short introduction to symmetric dynamics}
\label{subsec:symmetry}

Although no system is ever truly symmetric, in practise many models have a high degree of symmetry.\footnote{Indeed, the human brain consists of the order of $10^{11}$ neurons, but of the order of $100-1000$ types {\tt http://neuromorpho.org/neuroMorpho/index.jsp} meaning there is a very high replication of cells that are only different by their location and exact morphology.}  Indeed many real world networks that have \textit{grown} (e.g. giving rise to tree-like structures) are expected to be well approximated by models that have large symmetry groups \cite{MacArthur2008}.

Symmetric (more precisely, equivariant) dynamics provides a number of powerful mathematical tools that one can use to understand emergent properties of systems of the form
\begin{equation}
\frac{\d}{\d t} {x}=f(x) ,
\label{eq:ode}
\end{equation}
with $x\in R^N$. We say (\ref{eq:ode}) is {\em equivariant} under the action of a group $\Gamma$ if and only if $f(gx)=gf(x)$ for any $g\in\Gamma$ and $x\in \R^N$. There is a well developed theory of dynamics with symmetry; in particular see \cite{golubitsky-schaeffer1985,golubitsky-schaeffer-stewart1988,golubitsky-stewart2001}. These give methods that help in a number of ways:
\begin{itemize}
\item {\bf Description:} one can identify symmetries of networks and dynamic states to help classify and differentiate between them
\item {\bf Bifurcation:} there is a well-developed theory of bifurcation with symmetry to help understand the emergence of dynamically interesting (symmetry broken) states from higher symmetry states
\item {\bf Stability:} bifurcation with symmetry often gives predictions about possible bifurcation scenarios that includes information about stability
\item {\bf Generic dynamics:} symmetries and invariant subspaces can provide a powerful structure with which one can understand more complex attractors such as heteroclinic cycles
\item {\bf Design:} one can use symmetries to systematically build models and test hypotheses
\end{itemize}
The types of symmetries that are often most relevant for mathematical modelling of finite networks of neurons are the permutation groups, i.e. the symmetric groups and their subgroups. Nonetheless, continuum models of neural systems may have continuous symmetries that influence the dynamics and can be used as a tool to understand the dynamics; see for example \cite{Bressloffetal2002}.

\subsection{Permutation symmetries and oscillator networks}

We review some aspects of the equivariant dynamics that have proven useful in coupled systems that are relevant to neural dynamics - see for example \cite{Ashwin92,Bressloff99}. In doing so we mostly discuss dynamics that respects some symmetry group of permutations of the systems. The full permutation symmetry group (or simply, the symmetric group) on $N$ objects, $S_N$, is defined to be the set of all possible permutations of $N$ objects. Formally it is the set of permutations $\sigma:\{1,\ldots,N\}\rightarrow \{1,\ldots,N\}$ (invertible maps of this set). To determine effects of the symmetry, not only the group must be known but also its {\em action} on phase space. If this action is linear then it is a {\em representation} of the group. The representation of the symmetry group is critical to the structure of the stability, bifurcations and generic dynamics that are equivariant with the symmetry.

For example, if each system is characterised by a single real variable, one can view the action of the permutations on $\R^N$ as a {\em permutation matrix}
$$
\left[M_{\sigma}\right]_{ij} = \left\{ \begin{array}{cl} 1 & \mbox{ if }i=\sigma(j)\\
0 & \mbox{ otherwise}\end{array}\right. ,
$$
for each $\sigma\in \Gamma$; note  that $M_{\sigma}M_{\rho}=M_{\sigma\rho}$ for any $\sigma,\rho\in \Gamma$. Table~\ref{tab:symms} lists some commonly considered examples of symmetry groups used in coupled oscillator network models.

\begin{table}
\begin{tabular}{lcp{6cm}}
Name & Symbol & Comments \\
\hline
Full permutation & $S_N$ & Global or all-to-all coupling \cite{Ashwin92,Brown03}\\
Undirected ring & $\D_N$ & Dihedral symmetry \cite{Ashwin92,Brown03}\\
Directed ring & $\Z_N$ & Cyclic symmetry \cite{Ashwin92,Brown03}\\
Polyhedral networks & $G$ & \cite{OthmerScriven1971}\\
Lattice networks & $G_1\times G_2$ & $G_1$ and $G_2$ could be $\D_k$ or $\Z_k$\\
Hierarchical networks & $G_1 \wr G_2$ & $G_1$ is the local symmetry, $G_2$ the global symmetry, and $\wr$ is the wreath product \cite{Dionneetal1996}.
\end{tabular}
\caption{Some permutation symmetry groups that have been considered as examples of symmetries of coupled oscillator networks.}
\label{tab:symms}
\end{table}

More generally, for (\ref{eq:coupledcells}) equivariance under an action of $\Gamma$ means that for all $\sigma\in\Gamma$, $x\in\R^{Nd}$ and $i=1,\ldots,N$ we have
$$
f_{\sigma(i)}(x_{\sigma(i)})+\epsilon g_{\sigma(i)}(x_{\sigma(i)};x_{1},\ldots,x_{N})=f_{i}(x_{\sigma(i)})+\epsilon g_{i}(x_{\sigma(i)};x_{\sigma(1)},\ldots,x_{\sigma(N)}).
$$
A simple consequence of this is: if $\Gamma$ acts {\em transitively} on $\{1,\ldots,N\}$ (i.e. if for any $i$ and $j$ there is a $\sigma\in\Gamma$ such that $\sigma(j)=i$) then all oscillators are identical, i.e. $f_i(x_i)=F(x_i)$ for some function $F$.

The presence of symmetries means that solutions can be grouped together into families  - given any $x$ the set $\Gamma x:= \{ gx~:~g\in\Gamma\}$ is the {\em group orbit} of $x$ and all points on this group orbit will behave in dynamically the same way.

\subsection{Invariant subspaces and symmetries}

\begin{figure}
\includegraphics[width=6cm]{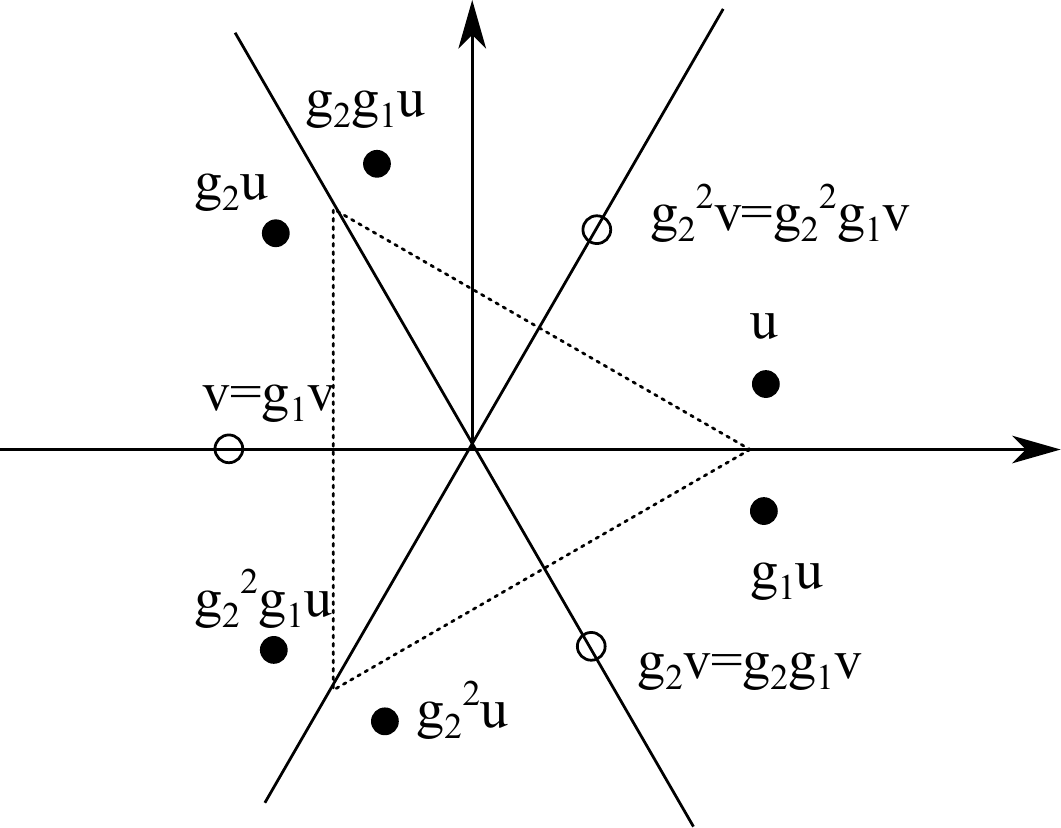}
\caption{Suppose the six-element group $\Gamma=\D_3$ of symmetries of the equilateral triangle acts on $\R^2$, generated by a rotation $g_2$ and a reflection $g_1$ in the $x$-axis. The group orbit of the point $u$ that is not fixed by any symmetries also has six elements (shown by filled circles), while any group orbit of a point $v$ that is fixed by a symmetry (e.g. $g_1$) has correspondingly fewer points (shown by open circles) in the group orbit. Bifurcation of equilibria with more symmetry typically leads to several equilibria with less (``broken'') symmetry.}
\label{fig:grouporbit}
\end{figure}

Given a point $x\in\R^{Nd}$ we define the {\em isotropy subgroup} (or simply the
symmetry) of $x$ under the action of $\Gamma$ on $\R^{Nd}$ to be
$$
\Sigma_x:=\{ g\in \Gamma~:~ gx=x\}.
$$
This is a subgroup of $\Gamma$, and the set of these groups forms a lattice (the {\em isotropy lattice}) by inclusion of subgroups. They are dynamically important in that for any trajectory $x(t)$ we have $\Sigma_{x(0)}=\Sigma_{x(t)}$ for all $t$. A converse problem is to characterise the set of all points with a certain symmetry. If $H$ is a subgroup (or more generally, a subset) of $\Gamma$ then the {\em fixed point space} of $H$ is defined to be
$$
\fix(H):=\{x\in \R^{Nd}~:~gx=x~\mbox{ for all }g\in H\}.
$$
Because all $x\in\fix(H)$ have symmetry $H$ these subspaces are dynamically invariant. See Fig.~\ref{fig:grouporbit} that illustrates this principle. Typical points $x\in\fix(H)$ have $\Sigma_x=H$, however some points may have more symmetry; more precisely, if $H\subset K$ are isotropy subgroups then $\fix(H)\supset \fix (K)$; and the partial ordering of the isotropy subgroups corresponds to a partial ordering of the fixed point subspaces.

One can identify similar types of isotropy subgroup as those that are {\em conjugate} in the group, i.e. we say $H_1$ and $H_2$ are conjugate subgroups of $\Gamma$ if there is a $g\in\Gamma$ such that $gH_1=H_2g$. If this is the case, note that
\begin{eqnarray*}
g\fix(H_1)&=& g\{x\in \R^{Nd}~:~hx=x~\mbox{ for all }h\in H_1\}\\
&=& \{x\in \R^{Nd}~:~hg^{-1}x=g^{-1}x~\mbox{ for all }h\in H_1\}\\
&=& \{x\in \R^{Nd}~:~ghg^{-1}x=x~\mbox{ for all }h\in H_1\}\\
&=& \{x\in \R^{Nd}~:~hx=x~\mbox{ for all }h\in H_2\} = \fix(H_2) ,
\end{eqnarray*}
meaning that the fixed point spaces of conjugate subgroups (and the dynamics on them) are in some sense equal.

By identifying symmetries up to conjugacy allows for a considerable reduction of the number of cases one nes to consider; note that conjugate subgroups must have fixed point subspaces of the same dimension where essentially the same dynamics will occur.

The fixed point subspaces are often used (implicitly or explicitly) to enable one to reduce the dimension of the system and to make it more analysable. As an example, to determine the existence of a fully exact synchrony solution one only need to suppose $x_i(t)=x(t)$ and determine whether there is such a solution $x(t)$ for the system (\ref{eq:coupledcells}).

For periodic orbits, the symmetry of points on the orbit to symmetries of the orbit as an invariant set are as follows. Suppose $P$ is a periodic orbit (which can be viewed as a ``loop'' in phase space $\R^N$). Let $K$ denote the symmetries that fix all points on $P$ (the ``spatial symmetries'') and $H$ denote the symmetries that fix $P$ as a set (the ``spatio-temporal symmetries''); note that $K$ will be a subgroup of $H$. Finally, let
$$
L_K := \bigcup_{g\not\in K} \fix(\{g\})\cap \fix(K),
$$
be the set of points in phase space that have strictly more symmetry than $K$.

\begin{thm}{\cite[Theorem 3.4]{golubitsky-stewart2001}} Consider ODEs on $\R^n$ with a given finite symmetry group $\Gamma$. There is a periodic orbit $P$ with spatial symmetries $K$ and spatio-temporal symmetries $H$ if and only if
\begin{itemize}
\item $H/K$ is cyclic
\item $K$ is an isotropy subgroup
\item $\dim \fix(K)\geq 2$
\item $H$ fixes a connected component of $\fix(K)/L_K$, where $L_K$ is defined as above.
\end{itemize}
\label{thm:honk}
\end{thm}

One way of saying this is that the only possible spatio-temporal symmetries of periodic orbits are cyclic extensions of isotropy subgroups. Further theory, outlined in \cite{golubitsky-stewart2001}, shows that one can characterise possible symmetries of chaotic attractors; these may include a much wider range of spatio-temporal symmetries $(H,K)$ including some that do not satisfy the hypotheses of Theorem~\ref{thm:honk}. This means that the symmetries of attractors may contain dynamical information about the attractor.

\subsection{Bifurcations with symmetry and genericity}

Bifurcations of ODEs can be classified and analysed by codimension according to methods for example in texts \cite{Kuznetsov1998,GuckenheimerHolmes1990}. This involves the following steps:
\begin{itemize}
\item[(a)] Identification of the {\em marginally unstable modes} (the directions that are losing stability).
\item[(b)] Reduction to a {\em centre manifold} parametrised by the marginally unstable modes (generically this is one or two dimensional when only one parameter is varied).
\item[(c)] Study of the dynamics of the {\em normal form} for the bifurcation under generic assumptions on the normal form coefficients.
\end{itemize}
Indeed, the only generic codimension one local bifurcations (i.e. the only one-parameter bifurcations of equilibria that  will not split into a number of simpler bifurcations for some small perturbation on the system) are the saddle-node (also called fold, or limit point) and the Hopf (also called Andronov-Hopf) bifurcation. Additional more complicated bifurcations can appear at higher codimension. This classification by codimension has enabled development of a powerful set of numerical tools to help analysis of such systems, not just for local bifurcations of equilibria but also some global bifurcations (in particular, periodic orbit and homoclinic bifurcations). Particular packages to do this include AUTO \cite{Autovolume2007}, MatCont \cite{Dhooge2003}, CONTENT \cite{Kuznetsov1996} and XPPAUT \cite{Ermentrout02} (which includes an implementation of AUTO).

If we restrict to symmetry preserving perturbations, a much wider range of bifurcations can appear at low codimension - this is because the symmetry can cause a marginal mode of instability in one direction to appear simultaneously in many other directions meaning that
\begin{itemize}
\item[(a')] Identification of the {\em marginally unstable modes} (symmetry means there can generally be several of these that will go unstable at the same time).
\item[(b')] Reduction to a {\em centre manifold} parametrised by the marginally unstable modes (these are preserved by the action of the symmetries and may be of dimension greater than two even for one parameter bifurcations).
\item[(c')] Study of the dynamics of the {\em normal form} for the symmetric bifurcation under generic assumptions on the normal form coefficients (the symmetries mean that some coefficients may be zero, some are constrained to be equal while others may be forced to satisfy nontrivial and sometimes obscure algebraic relationships).
\end{itemize}
These factors conspire to make symmetric bifurcations rich and interesting in behaviour - even in codimension one it is possible for heteroclinic cycles or chaos to bifurcate from high symmetry solutions. However, the same factors mean that many features cannot be caught by numerical path-following packages such as those listed above - the degeneracies mean that many branches may coming out of the bifurcation; it is generally a challenge to identify all of these. Essentially, bifurcation theory needs to be developed in the context of the particular group action. Examples of some consequences of this for weakly coupled oscillator networks with symmetries are considered in \S~\ref{sec:weakcoupling}.

\subsection{Robust heteroclinic attractors, cycles and networks}

The presence of symmetries in a dynamical system can cause highly nontrivial dynamics even away from bifurcation points. Of particular interest are robust invariant sets that consist of networks of equilibria (or periodic orbits, or more general invariant sets) connected via heteroclinic connections that are preserved under small enough perturbations that respect the symmetries \cite{Krupa1997}. These structures may be cycles or more generally networks. They can be robust to perturbations that preserve the symmetries and indeed they can be attracting \cite{KrupaMelbourne2004,golubitsky-stewart2001}. We are particularly interested in the attracting case in which case we call these invariant sets {\em heteroclinic attractors} and trajectories approaching such attractors show a typical intermittent behaviour - periods that are close to the dynamics of an unstable saddle-type invariant set, and switches between different behaviours.

In higher dimensional systems, heteroclinic attractors may have subtle structures such as ``depth two connections'' \cite{AshwinField1999}, ``cycling chaos'' where there are connections between chaotic saddles \cite{Dellnitzetal1995,golubitsky-stewart2001,AshwinRucklidge1998} and ``winnerless competition'' \cite{Rabinovich2006,Rabinovich2014}. Related dynamical structures are found in the literature in attractors that show ``chaotic itinerancy'' or ``slow switching''. Such complex attractors can readily appear in neural oscillator models in the presence of symmetries and have been used to model various dynamics that contribute to the function of neural systems; we consider this, along with some examples, in \S~\ref{sec:heteroclinic}.

\subsection{Groupoid and related formalisms}
\label{subsec:groupoid}

Some less restrictive structures found in some coupled dynamical networks also have many of the features of symmetric networks (including invariant subspaces, bifurcations that appear to be degenerate, and heteroclinic attractors) but without necessarily having the symmetries.

One approach \cite{Stewartetal2003} has been to use a structure of groupoids - these are mathematical structures that satisfy some, but not all, of the axioms of a group and can be useful in understanding the constraints on the dynamics of coupled cell systems of the form (\ref{eq:gen-coupledcells}). A groupoid is similar to a group except that the composition of two elements in a groupoid is not always defined, and the inverse of a groupoid element may only be locally defined. This formalism can be used to describe the permutations of inputs of cells as in \cite{Stewartetal2003,golubitsky-stewart2006}.

Suppose that we have (\ref{eq:gen-coupledcells}) with cells $\mathcal{C}=\{1,\ldots,N\}$ and suppose that there are connections $\mathcal{E}$, i.e. are pairs of cells $(i,j)$ in $\mathcal{C}$ such that cell $i$ appears in the argument of the dynamics of cell $j$. We say
$$
I(j)=\{i\in\mathcal{C}~:~(i,j)\in\mathcal{E}\} ,
$$
is the {\em input set} of cell $j$ and there is a natural equivalence relation $\sim_I$ defined by $j\sim_I k$ if there is a bijection (an input automorphism)
$$
\beta:I(j)\rightarrow I(k) ,
$$
with $\beta(j)=k$ such that for all $i\in I(j)$ we have $(i,j)\sim_E (\beta(i),k)$. The set $B(j,k)$ of input automorphisms of this type and the set of all such input automorphisms has the structure of a groupoid \cite{Stewartetal2003}.

Given a coupling structure of this type, an {\em admissible vector field} is a vector field on the product space of all cells that respects the coupling structure, and this generalises the idea of an equivariant vector field in the presence of a symmetry group acting on the set of cells.  The dynamical consequences of this have a similar flavour to the consequences one can find in symmetric systems except that fewer cases have been worked out in detail, and there are many open questions.

\begin{figure}
\includegraphics[width=9cm]{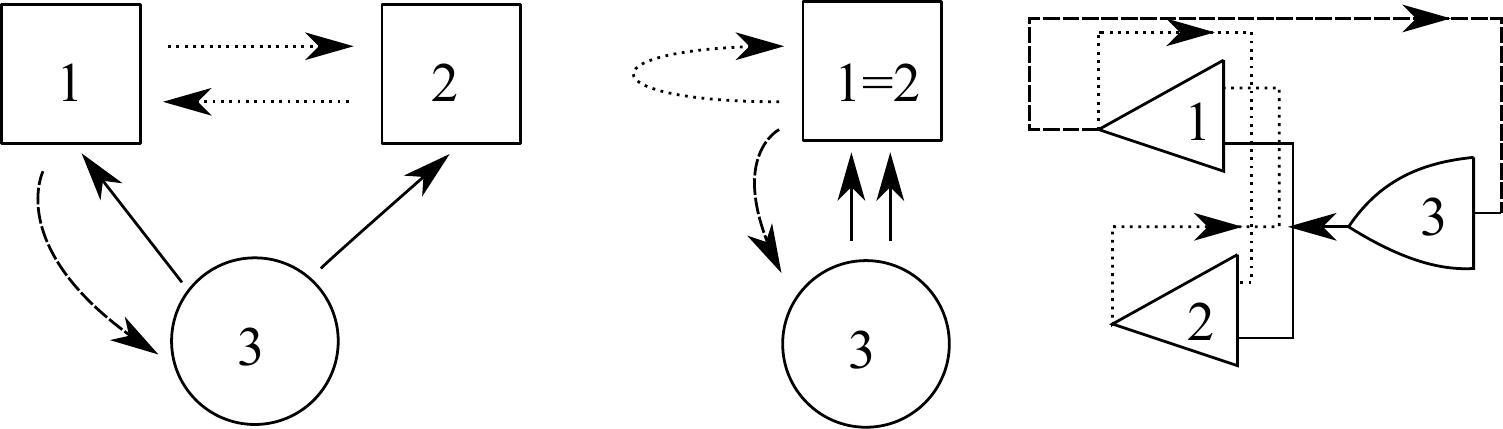}
\caption{Left: a system of three coupled cells with two cell types (indicated by the circle and square) coupled in a way that there is no permutation symmetry but there is an invariant subspace corresponding to cells 1 and 2 being synchronised. The different line styles show coupling types that can potentially be permuted {\em (after \cite[Fig. 3]{Stewartetal2003})}. Middle: the quotient two-cell network corresponding to cells $1$ and $2$ being synchronised. Right: The same network structure shown using the notation of \cite{Aguiaretal2011}.}
\label{fig:3cells}
\end{figure}

To illustrate, consider the system of three cells
\begin{equation}
\begin{split}
\FD{}{t}x_1 &= g(x_1,x_2,x_3) ,\\
\FD{}{t}x_2 &= g(x_2,x_1,x_3) ,\\
\FD{}{t}x_3 &= h(x_3,x_1) ,
\end{split}
\label{eq:threecellgroupoid}
\nonumber
\end{equation}
where $g(x,y,z)=g(x,z,y)$; this is discussed in \cite[\S 5]{golubitsky-stewart2006} in some detail. This has a coupling structure as shown in Fig.~\ref{fig:3cells}. In spite of there being no exact symmetries in the system there is a nontrivial invariant subspace where $x_1=x_2$. In the approach of \cite{golubitsky-stewart2006} the dynamically invariant subspaces that can be understood in terms of the {\em balanced colourings} of the graph where the cells are grouped in such a way that the inputs are respected - this corresponds to an admissible {\em pattern of synchrony}.

The invariant subspaces that are forced to exist by this form of coupling structure have been called {\em polydiagonals} in this formalism, which correspond to clustering of the states. For every polydiagonal one can associate a {\em quotient network} by identifying cells that are synchronised, to give a smaller network. As in the symmetric case the existence of an invariant subspace does not guarantee that it contains any attracting solutions. Some work has been done to understand generic symmetry breaking bifurcations in such networks - see for example \cite{Aguiar2009137}, or spatially periodic patterns in lattice networks \cite{Dias2009}. Variants of this formalism have been developed to enable different coupling types between the same cells to be included.

Periodic orbits in such networks can also have interesting structures associated with the presence of invariant subspaces. The so-called {\em rigid phase conjecture} \cite{StewartParker2007,golubitsky-stewart2006}, recently proved in \cite{Golubitsky2012}, states that if there is a periodic orbit in the network such that two cells have a rigid phase relation between (i.e. one that is preserved for all small enough structure-preserving perturbations) then this must be forced by either a $\Z_n$ symmetric perturbation of the cells in the network, or in some quotient network.

An alternative formalism for discussing asymmetric coupled cell networks  has been developed in \cite{Field2004,Aguiaretal2011,AgarwalField2010a,AgarwalField2010b} that also allows one to identify invariant subspaces. Each cell has one output and several inputs that may be of different types. These papers concentrate on the questions: (a) When are two cell networks formally equivalent (i.e. when can the dynamics of one cell network be found in the other, under suitable choice of cell)? (b) How can one construct larger coupled cell systems with desired properties by ``inflating'' a smaller system $S$, such that the larger system has $S$ as a quotient? (c) What robust heteroclinic attractors exist in such systems?

\section{Coupled limit cycle oscillators}
\label{sec:coupledlimitcycles}

In addition to variants on the systems in \S~\ref{sec:neuronoscillators}, we mention that some nonlinear ``conceptual'' planar limit cycle oscillators are also studied in neural models. These include:
\begin{eqnarray*}
\frac{\d^2}{\d t^2} x+ (a_0+a_1 x^2)\FD{}{t}x +\omega^2x=0 && \mbox{ van der Pol},\\
\frac{\d^2}{\d t^2} x+ (a_0+a_1 x^2)\FD{}{t}x +\omega^2x+a_2x^3=0 && \mbox{ van der Pol-Duffing},\\
\FD{}{t} z = (\lambda+i\omega) z + (a_0+ i a_1) |z|^2z && \mbox{ Stuart-Landau oscillator},
\end{eqnarray*}
where $x$ is real, $z$ is complex and the coefficients $\lambda,\omega$ and $a_j$ are real constants.

If the coupling between two or more limit cycle oscillators is relatively large, it can affect not only the phases but also the amplitudes, and a general theory of strongly interacting oscillators is likely to be no more or less complicated than a general theory of general nonlinear systems. However, the theory of \textit{weak-coupling} is relatively well developed (see \S~\ref{sec:reduced} and \S~\ref{sec:weakcoupling}) and specific progress can sometimes be made for the special choice of neuron model. Examples where one can do this include IF (see \S~\ref{sec:IFnetworks}), piece-wise linear models such as McKean \cite{Coombes2001}, caricatures of FHN, ML \cite{Coombes08} and singularly perturbed relaxation oscillators with linear \cite{Storti1986} or fast threshold modulation coupling \cite{Somers1995}.

For linear coupling of planar oscillators, much is known about the general case \cite{Aronson1990,Kurrer1997}.  If the linear coupling is proportional to the difference between two state variables this is referred to as ``diffusive", and otherwise it is called ``direct".  The difference between the two cases is most strongly manifest when considering the mechanism of \textit{oscillator death} (see \S~\ref{sec:WeakandStrong}).
The diffusive case is more natural in a neuroscience context as it can be used to model electrical gap junction coupling (which depends on voltage-differences).  The existence of synchronous states in networks of identical units is inherited from the properties of the underlying single neuron model since in this case coupling vanishes, though the stability of this solution will depend upon the pattern of gap-junction connectivity.

Gap junctions are primarily believed to promote synchrony, though this is not always the case and they can also lead to robust stable asynchronous states \cite{Sherman1992}, as well as ``bursting'' generated by cyclic transitions between coherent and incoherent network states \cite{Han95}. For work on gap junctions and their role in determining network dynamics see for example \cite{Postnov1999,Pfeuty03,Ermentrout06,Daido2006,Mancilla2007,Coombes08}.

\subsection{Stability of the synchronised state for complex networks of identical systems}
\label{sec:MSF}

There is one technique specific to the analysis of the synchronous state in a quite large class of network models that is valid for strongly coupled identical systems, namely the \textit{master stability function} (MSF) approach. For networks of coupled systems or oscillators with identical components the MSF approach of Pecora and Carroll  \cite{Pecora1998} can be used to determine the stability of the synchronous state in terms of the eigenstructure of the network connectivity matrix.
To introduce the MSF formalism it is convenient to consider $N$ nodes (oscillators)\footnote{In this section we assume little about the dynamics of the nodes - they may be ``chaotic oscillators''.} and let $x_i \in \RSet^d$ be the $d$ dimensional vector of dynamical variables of the $i$th node with isolated (uncoupled) dynamics $\ID{x_i} = {F}({x}_i)$, with $i=1,\ldots, N$.  The output for each node is described by a vector function ${H} \in \RSet^d$ (which is not necessarily linear).  For example for a three dimensional system with ${x}=(x^{(1)},x^{(2)},x^{(3)})$ and linear coupling only occurring through the $x^{(3)}$-component then we would set ${H}({x}) = (0,0,x^{(3)})$. For a given adjacency matrix $A_{ij}$ and associated set of connection strengths $g_{ij} \geq 0$ and a global coupling strength $\sigma$ the network dynamics of coupled identical systems, to which the MSF formalism applies, is specified by
\begin{equation}
\frac{\d}{\d t} {{x}}_i = {F}({x}_i) + \sigma \sum_{j=1}^N A_{ij} g_{ij} \left [ {H}({x}_j)-{H}({x}_i) \right ] \equiv {F}({x}_i) - \sigma \sum_{j=1}^N \mathcal{G}_{ij} {H}({x}_j) .
\nonumber
\end{equation}
Here the matrix $\mathcal{G}$ with blocks $\mathcal{G}_{ij}$ has the graph-Laplacian structure $\mathcal{G}_{ij} = -A_{ij}g_{ij} + \delta_{ij} \sum_{k}A_{ik} g_{ik}$.
The $N-1$ constraints ${x}_1(t)={x}_2(t)=\ldots ={x}_N(t) = s(t)$ define the (invariant) synchronisation manifold, with ${s}(t)$ a solution in $\R^d$ of the uncoupled system, namely $\ID{s}={F}({s})$.  Although we will not discuss in detail here, we assume that the asymptotic behaviour $s(t)$ is such that averages along trajectories converge, i.e. the behaviour of $s(t)$ for the uncoupled system is governed by a natural ergodic (SRB) measure for the dynamics.

To assess the stability of this state we perform a linear stability analysis expanding a solution as ${x}_i(t) = {s}(t) +\delta {x}_i(t)$ to obtain the variational equation
\begin{equation}
\delta \frac{\d}{\d t} {{x}_i} = D {F}({s}) \delta {x}_i - \sigma D {H}({s}) \sum_{j=1}^N \mathcal{G}_{ij} \delta {x}_j .
\nonumber
\end{equation}
Here $D {F}({s})$ and $D {H}({s})$ denote the Jacobian of ${F}({s})$ and ${H}({s})$ around the synchronous solution respectively.
The variational equation has a block form that can be simplified by projecting $\delta {x}$ into the eigenspace spanned by the (right) eigenvectors of the matrix $\mathcal{G}$.  This yields a set of $N$ decoupled equations in the block form
\begin{equation}
\frac{\d}{\d t} {{\xi}_l} = \left [ D {F}({s}) - \sigma \lambda_l D {H}({s}) \right ] {\xi}_l, \qquad l=1,\ldots,N,
\nonumber
\end{equation}
where ${\xi}_l$ is the $l$th (right) eigenmode associated with the eigenvalue $\lambda_l$ of $G$ (and $D {F}({s})$ and $D {H}({s})$ are independent of the block label $l$).  Since $\sum_i \mathcal{G}_{ii}=0$ there is always a zero eigenvalue, say $\lambda_1=0$, with corresponding eigenvector $(1,1,\ldots,1)$, describing a perturbation parallel to the synchronisation manifold.  The other $N-1$ transverse eigenmodes must damp out for synchrony to be stable.  For a general matrix $\mathcal{G}$ the eigenvalues $\lambda_l$ may be complex, which brings us to consideration of the system
\begin{equation}
\frac{\d}{\d t} {{\xi}} = \left [ D {F}({s}) - \alpha D {H}({s}) \right ] {\xi}, \qquad \alpha = \sigma \lambda_l \in \CSet .
\label{variational}
\end{equation}
For given $s(t)$, the MSF is defined as the function which maps the complex number $\alpha$  to the greatest Lyapunov exponent of (\ref{variational}).
The synchronous state of the system of coupled oscillators is stable if the MSF is negative at  $\alpha=\sigma \lambda_l$  where  $\lambda_l$  ranges over the eigenvalues of the matrix $\mathcal{G}$ (excluding $\lambda_1=0$).

For a ring of identical (or near identical) coupled periodic oscillators in which the connections have randomly heterogeneous strength  Restrepo \textit{et al}. \cite{Restrepo2004} have used the MSF method to determine the possible patterns at the desynchronisation transition that occurs as the coupling strengths are increased. Interestingly they demonstrate Anderson localisation of the modes of instability, and show that this could organise waves of desynchronisation that would spread to the whole network.  For a further discussion about the use of the MSF formalism in the analysis of synchronisation of oscillators on complex networks we refer the reader to \cite{Arenas2008,Porter2014}, and for the use of this formalism in a non-smooth setting see \cite{Thul2010}. This approach has recently been extended to cover the case of cluster states by making extensive use of tools from computational group theory to determine admissible patterns of synchrony \cite{Pecora2014} (and see also \S~\ref{subsec:groupoid}) in unweighted networks.

\subsection{Pulse-coupled oscillators}
\label{sec:pulse}

Another example of a situation in which analysis of network dynamics can be carried out without the need for any reduction or assumption is that of \textit{pulse coupled oscillators}, in which interactions between neurons are mediated by instantaneous ``kicks" of the voltage variable.

Networks of $N$ identical oscillators with global (all-to-all) strong pulse coupling were first studied by Mirollo and Strogatz \cite{Mirollo90}. They assumed that each oscillator is defined by a state variable $v$ and is of integrate--and--fire type with threshold $v_{th}=1$ and reset value $v_R=0$. When oscillator $i$ in the network fires the instantaneous pulsatile coupling pulls all other oscillators $j \neq i$ up by a fixed amount $\epsilon$ or to firing, whichever is less, i.e.
$$
\text{If}~ v_i(t)=1\quad \mbox{ then } \quad v_j(t^+)= \min(1, v_j(t) +\epsilon) \quad \text{for all $j \neq i$} .
$$
Mirollo and Strogatz assume that the coupling is excitatory ($\epsilon>0$). If $m$ oscillators fire simultaneously then the remaining $N-m$ oscillators are pulled up by $m\epsilon$, or to firing threshold.

In the absence of coupling each oscillator has period $\Delta$ and there is a natural phase variable $\phi(t) =t/\Delta \mod 1$ such that $\phi=0$ when $v=0$ and $\phi=1$ when $v=1$. Mirollo and Strogatz further assume that the dynamics of each (uncoupled) oscillator is governed by $v(t)=f(\phi)$ where $f$ is a smooth function satisfying $f(0)=0$, $f(1)=1$, $f^\prime(\phi)>0$ and $f^{\prime\prime}(\phi)<0$ for all $\phi \in [0,1]$. Because of these hypotheses on $f$, it is invertible with inverse $\phi=g(v)$.

Note that for the leaky (linear) integrate--and--fire model (LIF) where
\[ \tau \FD{}{t} v = -v +I , \] we have $f(\phi) = I(1-\e^{-\Delta \phi/\tau})$ where $\Delta = \tau \ln(I/(I-1))$, for $I>1$, which satisfies the above conditions. However, quadratic IF models do not satisfy the concavity assumption.

If an oscillator is pulled up to firing threshold due to the coupling and firing of a group of $m$ oscillators which have already synchronised then the oscillator is `absorbed' into the group and remains synchronised with the group for all time. (Here synchrony means firing at the same time). Since there are now more oscillators in the synchronised group, the effect of the coupling on the remaining oscillators is increased and this acts to rapidly pull more oscillators into synchronisation. Mirollo and Strogatz \cite{Mirollo90} proved that for pulsatile coupling and $f$ satisfying the conditions above, the set of initial conditions for which the oscillators do not all become synchronised has zero measure.
Here we briefly outline the proof for two pulse coupled oscillators. See Mirollo and Strogatz \cite{Mirollo90} for the generalisation of this proof to populations of size $N$.

Consider two oscillators labelled $A$ and $B$ with $\phi_{A}$ and $v_A$ denoting respectively the phase and state of oscillator $A$ and similarly for oscillator $B$. Suppose that oscillator $A$ has just fired so that $\phi_A=0$ and $\phi_B=\phi$ as in Fig.~\ref{fig:Pulse} (a). The return map $R(\phi)$ is the phase of $B$ immediately after $A$ next fires. It can be shown that the return map has a unique, repelling fixed-point:

\begin{figure}[h]
\includegraphics[width=12.5cm]{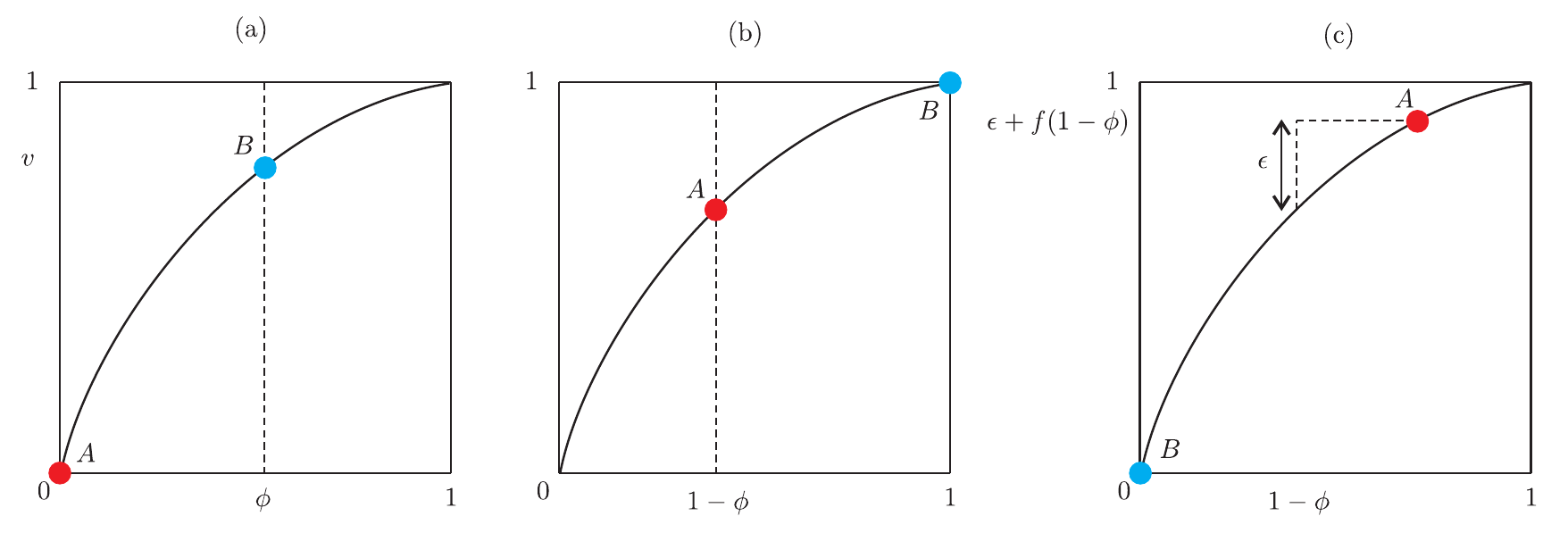}\\
  \caption{A system of two oscillators governed by $v = f(\phi)$, and interacting by pulse coupling.
(a) The state of the system immediately after oscillator $A$ has fired. (b) The state of the
system just before oscillator $B$ reaches the firing threshold. (c) The state of the system just after $B$ has fired. $B$ has
jumped back to zero, and the state of $A$ is now $\min(1, \epsilon + f(1 - \phi))$.
}\label{fig:Pulse}
\end{figure}

Oscillator $B$ reaches threshold when $\phi_A=1-\phi$ and an instant later $B$ fires and the pulsatile coupling makes $v_A=  \min(1, \epsilon + f(1-\phi))$. If $v_A=1$ then the oscillators have synchronised. Assuming that $v_A= \epsilon + f(1-\phi)<1$ then $\phi_A = g(\epsilon + f(1-\phi))$ and after one firing the system has moved from $(\phi_A, \phi_B)= (0, \phi)$ to $(h(\phi), 0)$ where $h(\phi) = g(\epsilon + f(1-\phi))$ is the firing map (see Fig.~\ref{fig:Pulse} (c)). The return map is given by a further application of the firing map so that $R(\phi)=h(h(\phi))$. The assumption that
$\epsilon + f(1-\phi)<1$ is satisfied when $\epsilon \in [0,1)$ and $\phi \in (\delta, 1)$ where $\delta=1-g(1-\epsilon)$. Thus the domain of $h$ is $(\delta,1)$ and the domain of $R$ is $(\delta, h^{-1}(\delta))$. Since $h$ is monotonically decreasing, $\delta<h^{-1}(\delta)$ for $\epsilon<1$ and the interval is nonempty. Thus on the whole of $[0,1)$ the return map is defined as
\[ R(\phi) = \left\{ \begin{array}{ll} 1 & \phi>h^{-1}(\delta),\\ h(h(\phi)) &  \phi \in (\delta, h^{-1}(\delta)), \\ 0 & \phi<\delta. \end{array} \right.  \] Since the points $0$ and $1$ are identified, if $\phi <\delta$ or $\phi>h^{-1}(\delta)$ then the two oscillators will become synchronised.

It can be shown that almost all initial conditions eventually become synchronised since (i) $R$ has a unique fixed point $\overline{\phi} \in (\delta, h^{-1}(\delta))$ and (ii) this fixed point is unstable (i.e. $|R^\prime(\overline{\phi})|>1$).
To see that $R$ has a unique fixed point, observe that fixed points $\overline{\phi}$ are roots of $F(\phi)\equiv \phi-h(\phi)$. Now $F(\delta)=\delta-1<0$ and $F(h^{-1}(\delta))= h^{-1}(\delta)-\delta>0$ so $F$ has a root in $(\delta, h^{-1}(\delta))$ and this root is unique since $F^\prime(\phi)=1-h^\prime(\phi)>2$.

Extensions to the framework of Mirollo and Strogatz include the introduction of a time delay in the transmission of pulses and the consideration of inhibitory coupling. It has been observed that delays have a dramatic effect on the dynamics in the case of excitatory coupling. Considering first a pair of oscillators, Ernst \textit{et al}. \cite{Ernst1995} demonstrate analytically that inhibitory coupling with delays gives stable in-phase synchronisation while for excitatory coupling, synchronisation with phase lag occurs. As the number of globally coupled oscillators increases, so does the number of attractors which can exist for both excitatory and inhibitory coupling.

In the presence of delays many different cluster state attractors can coexist. The dynamics settle down onto a periodic orbit with clusters reaching threshold and sending pulses alternately \cite{Ernst1995, Ernst1998, Timme2003}. Under the addition of weak noise when the coupling is inhibitory, the dynamics stay near this periodic orbit indicating that all cluster state attractors are stable \cite{Timme2003}. However, the collective behaviour shows a marked difference when the coupling is excitatory. In this case, weak noise is sufficient to drive the system away from the periodic orbit and results in persistent switching between unstable (Milnor) attractors.

These dynamics are somewhat akin to heteroclinic switching and the relationship between networks of unstable attractors and robust heteroclinic cycles has been addressed by a number of authors \cite{AshTim2005, Broer2008, Kirst2008}. In particular, Broer \textit{et al}. \cite{Kirst2008} highlight a situation in which there is a bifurcation from a network of unstable attractors to a heteroclinic cycle within a network of pulse coupled oscillators with delays and inhibitory coupling. They note that the model used in previous work \cite{Ernst1995, Ernst1998, Timme2003} is locally noninvertible since the original phase of an oscillator cannot be recovered once it has received an input which takes it over threshold causing the phase to be reset. Kirst and Timme \cite{Kirst2008} employ a framework in which a reset function $R(\zeta)=c\zeta$, $c \in [0,1]$ is introduced which ensures that the flow becomes locally time invertible when $c>0$. They demonstrate that for $c=0$ (where the locally noninvertible dynamics are recovered), the system has a pair of periodic orbits $A_1$ and $A_2$ which are unstable attractors enclosed by the basin of each other. When $c>0$, $A_1$ and $A_2$ are non-attracting saddles with a heteroclinic connection from $A_1$ to $A_2$. Furthermore, there is a continuous bifurcation from the network of two unstable attractors when $c=0$ to a heteroclinic two cycle when $c>0$.

For an interesting dynamical systems perspective on the differences between ``kick" synchronisation (in pulsatile coupled systems) and ``diffusive" synchronisation \cite{Steur2009} and the lack of mathematical work on the former problem see \cite{Mauroy2012a}. For example, restrictions on the dynamics of symmetrically coupled systems of oscillators when the coupling is time-continuous can be circumvented for pulsatile coupling leading to more complex network dynamics \cite{Kielblock2011}.

In the real world of synaptic interactions, however, pulsatile kicks are probably the exception rather than the rule, and the biology of neurotransmitter release and uptake is better modelled with a distributed delay process, giving rise to a post synaptic potential with a finite rise and fall time.  For spike-time event driven synaptic models, described in \S~\ref{sec:coupling}, analysis at the network level is hard for a general conductance based model (given the usual expectation the single neuron model will be high dimensional and nonlinear), though far more tractable for LIF networks, especially when the focus is on phase-locked states \cite{Vreeswijk1996,VanVreeswijkC2000,VanVreeswijk2001}.  Indeed in this instance many results can be obtained in the strongly coupled regime \cite{Bressloff2000}, without recourse to any approximation or reduction.

\subsection{Synaptic coupling in networks of IF neurons}
\label{sec:IFnetworks}

The instantaneous reaction of one neuron to the firing of another, as in the pulse-coupled neurons above, does not account for the role of synapses in the transmission of currents. Bressloff and Coombes \cite{Bressloff2000} consider a network of $N$ identical, LIF neurons that interact via synapses by transmitting spike trains to one another. Let $v_i(t)$, the state of neuron $i$ at time $t$, evolve according to
\begin{equation}
\FD{}{t} v_i = -v_i +I_i + X_i(t),
\label{IFN}
\end{equation}
 where $I_i$ is a constant external bias and $X_i(t)$ is the total synaptic current into the cell. As before, we supplement this with the reset condition that whenever $v_i=1$ neuron $i$ fires and is reset to $v_i=0$. The synaptic current $X_i(t)$ is generated by the arrival of spikes from other neurons $j$ and can be taken to have the form
\[ X_i(t) = \epsilon \sum_{j=1}^N w_{ij} \sum_{m \in \ZSet} J(t-T_j^m),\] where $\epsilon w_{ij}$ represents the weight of the connection from the $j$th neuron to the $i$th neuron with $\epsilon$ characterising the overall strength of synaptic interactions, $T_j^m$ denotes the sequence of firing times of the $j$th neuron and $J(t)$ determines the course of postsynaptic response to a single spike. A biologically motivated choice for $J(t)$ is
\[ J(t) = \eta(t)\Theta(t), \quad \eta(t)=\alpha^2 t{\rm e}^{-\alpha t},\] where $\Theta(t)=1$ if $t>0$ and zero otherwise. Here $\eta$ is an alpha function (see also \S~\ref{sec:coupling}) and the maximum synaptic response occurs at a nonzero delay $t=\alpha^{-1}$. Note that in the limit of large inverse rise time $\alpha$, $J(t)$ approximates a delta function (more like pulse-coupling).

Bressloff and Coombes \cite{Bressloff2000} show that the behaviour of the network of oscillators differs depending on the strength of the coupling. This is another instance in which information is lost through making weak-coupling assumptions.
To see this one may integrate (\ref{IFN}) between $T_i^{n}$ and $T_i^{n+1}$, exploiting the linearity of the equations and solving with variation of parameters, to obtain a map of the firing times.  Since the drive $X_i(t)$ depends upon all previous firing events of the other neuron this is an implicit map that relates all the network firing events to one another.  It is convenient to introduce the set of inter-spike intervals (ISIs) $\Delta_{ij}^{n,m}= T_i^n-T_j^m$, so that we may write the firing map in the convenient form
\begin{equation}
I_i (1-\e^{-\Delta_{ii}^{n+1,n}})-1 + \sum_{j=1}^N \sum_{m \in \ZSet} F_{ij} (\Delta_{ii}^{n+1,n},\Delta_{ij}^{n,m}) = 0 ,
\label{implicitmap}
\end{equation}
where $T_i^{n+1}> T_j^m$ for all $j,m$, and
\begin{equation}
F_{ij}(x,y) = \epsilon w_{ij} \e^{-x} \int_0^x \e^{s} J(s+y) \d s.
\nonumber
\end{equation}
Phase-locked solutions may be found, for an arbitrary coupling strength $\epsilon$, using the ansatz $T_j^m=(m-\phi_j) \Delta$ for some self-consistent ISI $\Delta$ and constant phases $\phi_j$.  Substitution into (\ref{implicitmap}) yields
\begin{equation}
1= I_i (1-\e^{-\Delta}) + \epsilon \sum_{j=1}^N w_{ij} K(\phi_j-\phi_i),
\label{IFphases}
\end{equation}
where
\begin{equation}
K(\phi) = \e^{-\Delta} \int_0^\Delta \e^{s} P(s+\phi\Delta) \d s, \quad P(t) = \sum_{m \in \ZSet} J(t-m \Delta) ,
\nonumber
\end{equation}
and $P(t)=P(t+\Delta)$.
Choosing one of the phases, say $\phi_1$, as a reference then (\ref{IFphases}) provides a set of $N$ equations for the unknown period $\Delta$ and the remaining $N-1$ relative phases $\phi_j-\phi_1$.

In order to investigate the linear stability of phase-locked solutions of equation (\ref{IFphases}), we consider perturbations of the firing times
which we write in the form $T_j^m =(m-\phi_j) \Delta + \delta_j^m$.  Linearisation of the firing map (\ref{implicitmap}) gives an explicit map for these perturbations that can be written as
\begin{equation}
I_i\e^{-\Delta}(\delta_i^{n+1}-\delta_i^n) +  \sum_{j=1}^N \sum_{m \in \ZSet} \PD{F_{ij}}{x}(\delta_i^{n+1}-\delta_i^n) + \PD{F_{ij}}{y} (\delta_i^{n}-\delta_j^m) = 0,
\nonumber
\end{equation}
where the partial derivatives of $F_{ij}$ are evaluated at the phase-locked state $(\Delta_{ii}^{n+1,n},\Delta_{ij}^{n,m})=(\Delta, (n-m)\Delta +(\phi_j-\phi_i)\Delta)$.  For solutions of the form $\delta_j^m = \lambda^{m}\delta_j$ this reduces to
\begin{equation}
(\lambda-1) P_i \delta_i =
\epsilon \sum_{j=1}^N H_{ij}(\lambda) \delta_j ,
\label{explicitmap}
\end{equation}
where $P_i=I_i-1+\epsilon \sum_j W_{ij} P((\phi_j-\phi_i)\Delta)$, $H_{ij}(\lambda) = W_{ij} G((\phi_j-\phi_i)\Delta,\lambda) - \delta_{ij} \sum_{k} W_{ik}G((\phi_k-\phi_i)\Delta,1)$,\\
and
\begin{equation}
G(t,\lambda) = \sum_{m \in \ZSet} \lambda^{-m} \e^{-\Delta} \int_0^\Delta \e^s J'(s+t+m\Delta) \d s .
\nonumber
\end{equation}
One solution to equation (\ref{explicitmap}) is $\lambda=1$ with $\delta_i = \delta$ for all $i=1,\ldots,N$. This reflects the invariance of the dynamics with respect to uniform phase shifts in the firing times. Thus the condition for linear stability of a phase-locked state is that all remaining solutions $\lambda$ of equation (\ref{explicitmap}) are within the unit disc. For $\lambda -1 \sim O(\epsilon)$, and $\epsilon$ small, we may expand (\ref{explicitmap}) as
\begin{equation}
(\lambda-1) (I_i-1)\delta_i = \epsilon \sum_{j=1}^N \mathcal{H}_{ij} (\Phi) \delta_j + O(\epsilon^2) ,
\nonumber
\end{equation}
where $\mathcal{H}_{ij} (\Phi) = H_{ij}(1) = [W_{ij} K'(\phi_j-\phi_i) - \delta_{ij} \sum_{k} W_{ik}K'(\phi_k-\phi_i)]/\Delta$ and we have used the result that $G(\phi,1)=K'(\phi)/\Delta$.
Suppose for simplicity that $I_i=I>1$ for all $i$, so that $\Delta= \ln[I/(I-1)] + O(\epsilon)$.  Then the weak-coupling stability condition is that $\text{Re}\, \lambda_p <0$, where $\lambda_p$, $p=1,\ldots,N$, are the eigenvalues of the $N \times N$ matrix with elements $\epsilon \mathcal{H}_{ij}(\Phi)$.

\begin{figure}[h!]
\begin{center}
\includegraphics[height=1.65in]{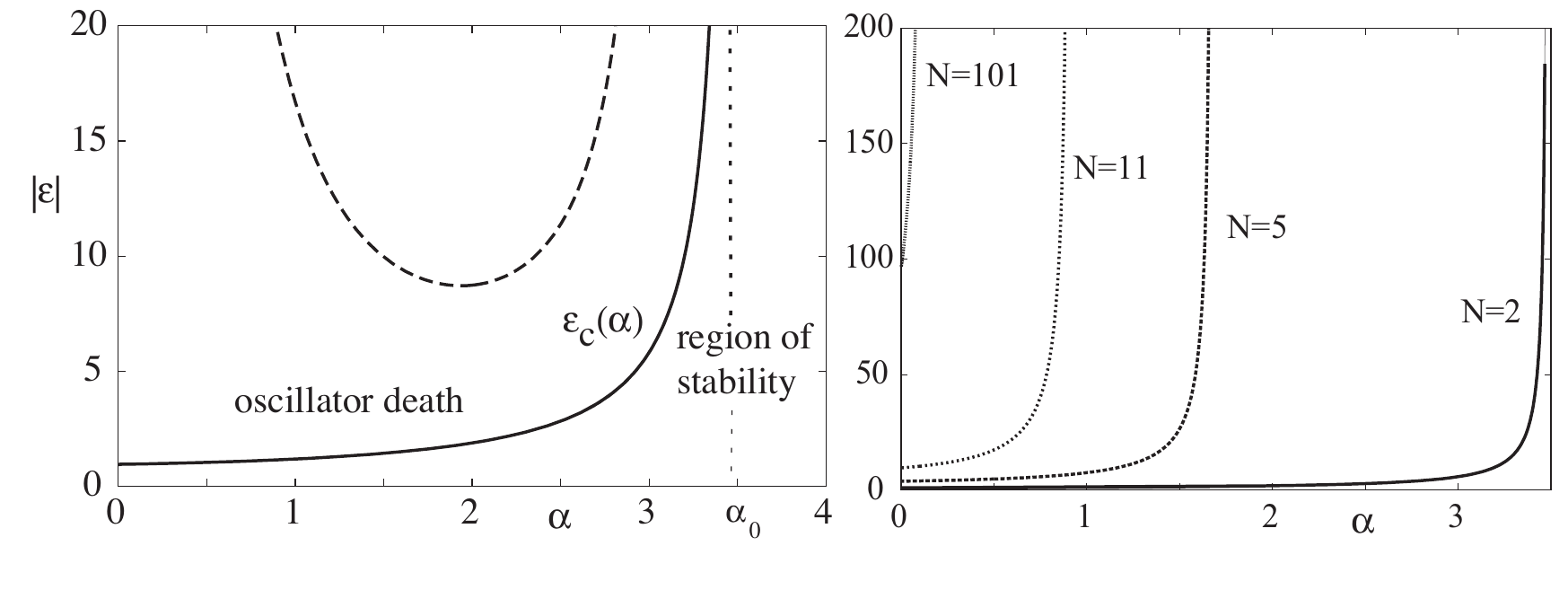}
\caption{
Left: Region of stability for a synchronised pair of identical IF neurons with inhibitory coupling and collective period $T=\ln 2$.  The solid curve $|\epsilon| = \epsilon_c(\alpha)$ denotes the boundary of the stability region. Crossing the boundary from below signals excitation of the linear mode
$(1,-1)$ leading to a stable state in which one neuron becomes quiescent (oscillator death). For $\alpha >
\alpha_0$ the synchronous state is stable for all $\epsilon$. The dashed curve denotes corresponds to the eigenvalue with $\nu=1$.
Right: Plot of critical coupling $\epsilon_c$ as a function of $\alpha$ for various network sizes $N$. The critical inverse rise-time
$\alpha_0(N)$ is seen to be a decreasing function of $N$ with $\alpha_0(N)\rightarrow 0$ as $N\rightarrow \infty$.
\label{Fig:StrongIF}
}
\end{center}
\end{figure}

As an explicit example let us consider the synchronous state ($\phi_i=\phi$ for all $i$). From (\ref{IFphases}) we see that this is guaranteed to exist for the choice $\sum_j W_{ij}=\gamma$ and $I_i=I[1-\epsilon K(0) \gamma]$ for some constant $I>1$ which sets the period as $\Delta = \ln[I/(I-1)]$.  Using the result that $K'(\phi)=-\Delta K(\phi) + \Delta P(\phi \Delta)/I$ the spectral problem for the synchronous state then takes the form
\begin{equation}
\left [
(\lambda-1) (I-1 + \epsilon I  \gamma K'(0)/\Delta ) + \epsilon  \gamma K'(0)/\Delta
\right ] \delta_i=
\epsilon G(0,\lambda) \sum_{j=1}^N W_{ij} \delta_j .
\nonumber
\end{equation}
We may diagonalise this equation in terms of the eigenvalues of the weight matrix, denoted by $\nu_p$ with $p=1,\ldots, N$ (and we note that $\gamma$ is the eigenvalue with eigenvector $(1,1,\ldots,1)$ corresponding to a uniform phase-shift).
Looking for non-trivial solutions then gives us the set of spectral equations $\mathcal{E}_p(\lambda)=0$, where
\begin{equation}
\mathcal{E}_p(\lambda) = (\lambda-1)(I-1+\epsilon I \gamma K'(0)/\Delta) + \epsilon \gamma K'(0)/\Delta - \epsilon \nu_p G(0,\lambda).
\label{spectral}
\end{equation}
We may use (\ref{spectral}) to determine bifurcation points defined by $|\lambda|=1$.
For sufficiently small $\epsilon$, solutions to (\ref{spectral})  will either be in a neighbourhood of the real solution $\lambda=1$ or in a neighbourhood of one of the poles of $G(0,\lambda)$. Since the latter lie in the 
 unit disc, the stability of the synchronous solution (for weak coupling) is determined by $\epsilon K'(0)[\text{Re}\, \nu_p-\gamma] <0$.  For strong coupling the characteristic equation has been solved numerically in \cite{Bressloff2000} to illustrate the possibility of Hopf bifurcations ($\lambda=\e^{i \theta}$, $\theta \neq 0$, $\theta \neq \pi$) with increasing $|\epsilon|$, leading to oscillator death in a globally coupled inhibitory network for a sufficiently slow synapse and bursting behaviour in asymmetrically coupled networks.
Bifurcation diagrams illustrating these phenomenon are shown in Fig.~\ref{Fig:StrongIF}
To see how these phenomena can occur from a more analytical perspective it is useful to consider the Fourier representation $J(t) = (2 \pi)^{-1} \int_{-\infty}^\infty \d \omega \widetilde{J}(\omega) \e^{i \omega t}$, where $\widetilde{J}(\omega)=\alpha^2/(\alpha +i \omega)^2$, so that $G$ may be easily evaluated with $\lambda=\e^z$ as
\begin{equation}
G(0,\e^z) =
\frac{1}{\Delta} \sum_{n \in \ZSet} \frac{\widetilde{J}(\omega_n-iz/\Delta) (i \omega_n + z/\Delta) (\e^z-\e^{-\Delta}), }{1+i \omega_n + z/\Delta}, \quad \omega_n = 2 \pi n/\Delta , \label{G}
\end{equation}
from which it is easy to see a pole at $z=-\alpha \Delta$.
This suggests writing $z$ in the form
$z=-\alpha(1+\kappa_p) \Delta$ and expanding the spectral equation in powers of $\alpha$ to find a solution.  For small $\alpha$ we find from (\ref{G}) that $G(0,\e^z) =  -\alpha (1+\kappa_p) (1-\e^{-\Delta})/(\kappa_p^2 \Delta)$.  Balancing terms of order $\alpha$ then gives $\kappa^2_p=\epsilon \nu_p (1-\e^{-\Delta})/(\Delta^2(I-1))$, where we use the result that $G(0,\e^0) = O(\alpha^2)$.  Thus for small $\alpha$, namely slow synaptic currents, we have that $K'(0)=0$, so that a weak-coupling analysis would predict neutral stability (consistent with the notion that a set of IF neurons with common constant input would frequency lock with an arbitrary set of phases).  However, our strong coupling analysis predicts that the synchronous solution will only be stable if $\text{Re}\, z^\pm_p<0$ with $z^\pm_p=[-1\pm \kappa_p] \alpha \Delta$.
Introducing the firing rate function $f=1/\Delta$ then $z^\pm_p$ can be written succinctly as
\begin{equation}
z^\pm_p=\left [-1 \pm \sqrt{\epsilon \nu_p f'(I)} \right] \alpha \Delta.
\nonumber
\end{equation}
Thus for an asymmetric network (with at least one complex conjugate pair of eigenvalues) it may occur that as $|\epsilon|$ is increased a pair of eigenvalues determined by $z^\pm_p$ may cross the imaginary axis to the right hand complex plane signalling a discrete Hopf bifurcation in the firing times.  For a symmetric network with real eigenvalues an instability may occur as some $\kappa_p \in \RSet$ increases through $1$, signalling a breakdown of frequency locking.  The above results (valid for slow synapses) can also be obtained using a firing rate approach, as described in \cite{Bressloff2000}.

The results above, albeit valid for strong coupling, are only valid for LIF networks.  To obtain more general results for networks of limit cycle oscillators it is useful to consider a reduction to phase models.

\section{Reduction of limit cycle oscillators to phase-amplitude and phase models}
\label{sec:reduced}

Consider a system of the form
\begin{equation}
	\frac{\d}{\d t} {x} = f(x) + \epsilon g(x,t) , \quad x\in\R^n ,
	\label{eq:driven}
\end{equation}
such that for $\epsilon=0$ the system possesses a periodic orbit
$$
\Gamma=\{ u(t)~:~t\in\R\} ,
$$
with minimal period $T>0$ (such that $u(t)=u(t+T)$ for all $t\in\R$ but $u(t)\neq u(s)$ for $0<s<T$). We will assume that $\Gamma$ is attracting and hyperbolic, i.e. linearly stable so there one zero Floquet exponent and the others with negative real part. Then we say that (\ref{eq:driven}) is a {\em limit cycle oscillator}. We will reduce this to a description that involves a {\em phase} that lives on a topological circle that can be thought of an interval $[0,T)$ with the topology that comes from identifying the ends of the interval: $\epsilon$ and $T-\epsilon$ are close for $\epsilon$ small. This circle is sometimes called the one-torus $\T$.

There are a number of conventions used in the literature to represent this phase. We discuss these conventions before looking at general reduction methods. One convention is to define a phase $\vartheta$  modulo $T$ such that
$$
\frac{\d}{\d t} {\vartheta}=1 ,
$$
where $\vartheta(u(t))=t$ interpreted modulo $T$ and such that $\vartheta(u_0)=0$ for some chosen point $u_0\in\Gamma$. Another convention is to define a phase $\theta$ modulo $2\pi$ such that
$$
\FD{}{t} \theta = \omega ,
$$
where $\omega=2\pi/T$ is the {\em angular frequency} of the oscillator. Again we pick a point $u_0\in\Gamma$ and require that $\theta=0$ at $u=u_0$.

In the remainder of the article we will use $\vartheta$ and $\theta$ as an indication of the convention we are using for phase, in the special case $T=2\pi$ both conventions coincide and we use $\theta$. Typically one of these descriptions is more convenient than the others but all can in principle be adapted to any phase reduction. Table~\ref{tab:phases} expresses some features of the conventions which commonly used for phase variables.

\begin{table}
\begin{tabular}{|l|p{10mm}|l|p{15mm}|p{25mm}|p{25mm}|}
\hline
Symbol & Phase space & Period & Uncoupled equation & Advantages & Disadvantages \\
\hline
$\vartheta$ & $\R/T\Z$, $[0,T)$ & $T$ & $\FD{}{t}\vartheta = 1$ & Simplicity of uncoupled equation, interpretation of $\vartheta$ as ``time'' & Phase space depends on parameters and initial conditions.\\
\hline
$\theta$ & $\R/2\pi\Z$, $[0,2\pi)$ & $2\pi$  & $\FD{}{t}\theta = \omega$ & Phase space fixed, good for heterogeneous oscillators. & Equations need scaling.\\
\hline
$\phi$ & $\R/\Z$, $[0,1)$ & $1$  & $\FD{}{t}\phi = \frac{1}{\Delta} $ & Phase space fixed, good for heterogeneous oscillators. & Equations need scaling.\\
\hline
\end{tabular}

\caption{Comparison of the three conventions for a phase variable that we use in this review. The $\phi$ is used for IF models.}
\label{tab:phases}
\end{table}

We now review some techniques of reduction which can be employed to study the dynamics of \eqref{eq:driven} when $\epsilon \neq 0$ so that the perturbations may take the dynamics away from the limit cycle.
 In doing so we will reduce for example to an ODE for $\vartheta(t)$ taken modulo $T$. Clearly any solution of an ODE must be continuous in $t$ and typically $\vartheta(t)$ will be unbounded in $t$ growing at a rate that corresponds to the frequency of the oscillator. Strictly speaking, the coordinate we are referring to in this case is on the {\em lift} of the circle $\T$ to a covering space $\R$, and for any phase $\vartheta\in[0,T)$ there are infinitely many lifts to $\R$ given by $\vartheta+ kT$ for $k\in \Z$. However, in common with most literature in this area we will not make a notational difference between whether the phase is understood on the unit cell e.g. $\theta\in[0,2\pi)$ or on the lift, e.g. $\theta\in\R$ modulo $2\pi$.

\subsection{Isochronal coordinates}
\label{subsec:isochrons}

Consider \eqref{eq:driven} with $\epsilon=0$. The \textit{asymptotic} (or \textit{latent}) \textit{phase} of a point $x_0$ in the basin of attraction of the limit cycle $\Gamma$ of period $T$ is the value of $\vartheta(x_0)$ such that
\begin{equation}
\label{eq:asymphase}
\lim_{t\to \infty} |x(t)- u(t+\vartheta(x_0))|=0 ,
\nonumber
\end{equation} 
where $x(t)$ is a trajectory starting at $x_0$. Thus if $u(t)$ and $x(t)$ are trajectories on and off the limit cycle respectively, they have the same asymptotic phase if the distance between $u(t)$ and $x(t)$ vanishes as $t\to \infty$.
The locus of all points with the same asymptotic phase is called an \textit{isochron}.  Thus an isochron extends the notion of phase
off the cycle (within its basin of attraction).  Isochrons can also be interpreted as the leaves of the stable manifold of a hyperbolic limit cycle. They fully specify the dynamics in the absence of perturbations \cite{Guckenheimer1975}.

There are very few instances where the isochrons can be computed in closed form (though see the examples in \cite{Winfree01} for plane-polar models where the radial variable decouples from the angular one). Computing the isochron foliation of the basin of attraction of a limit cycle is a major challenge since it requires knowledge of the limit cycle and therefore can only be computed in special cases or numerically.

One computationally efficient method for numerically determining the isochrons is backward integration, however it is unstable and in particular for strongly attracting limit cycles the trajectories determined by backwards integration may quickly diverge to infinity. See Izhikevich \cite{Izhikevich05} for a MATLAB code which determines smooth curves approximating isochrons. Other methods include the continuation based algorithm introduced by Osinga and Moehlis \cite{Osinga10}, the geometric approach of Guillamon and Huguet to find high order approximations to isochrons in planar systems \cite{Guillamon2009}, quadratic and higher order approximations \cite{Demir2010,Takeshita2010},
and the forward integration method using the Koopman operator and Fourier averages as introduced by Mauroy and Mezi\'c \cite{Mauroy2012}.  This latter method is particularly appealing and given its novelty we describe the technique below

\begin{figure}[h!]
\begin{center}
\includegraphics[width=3.5in]{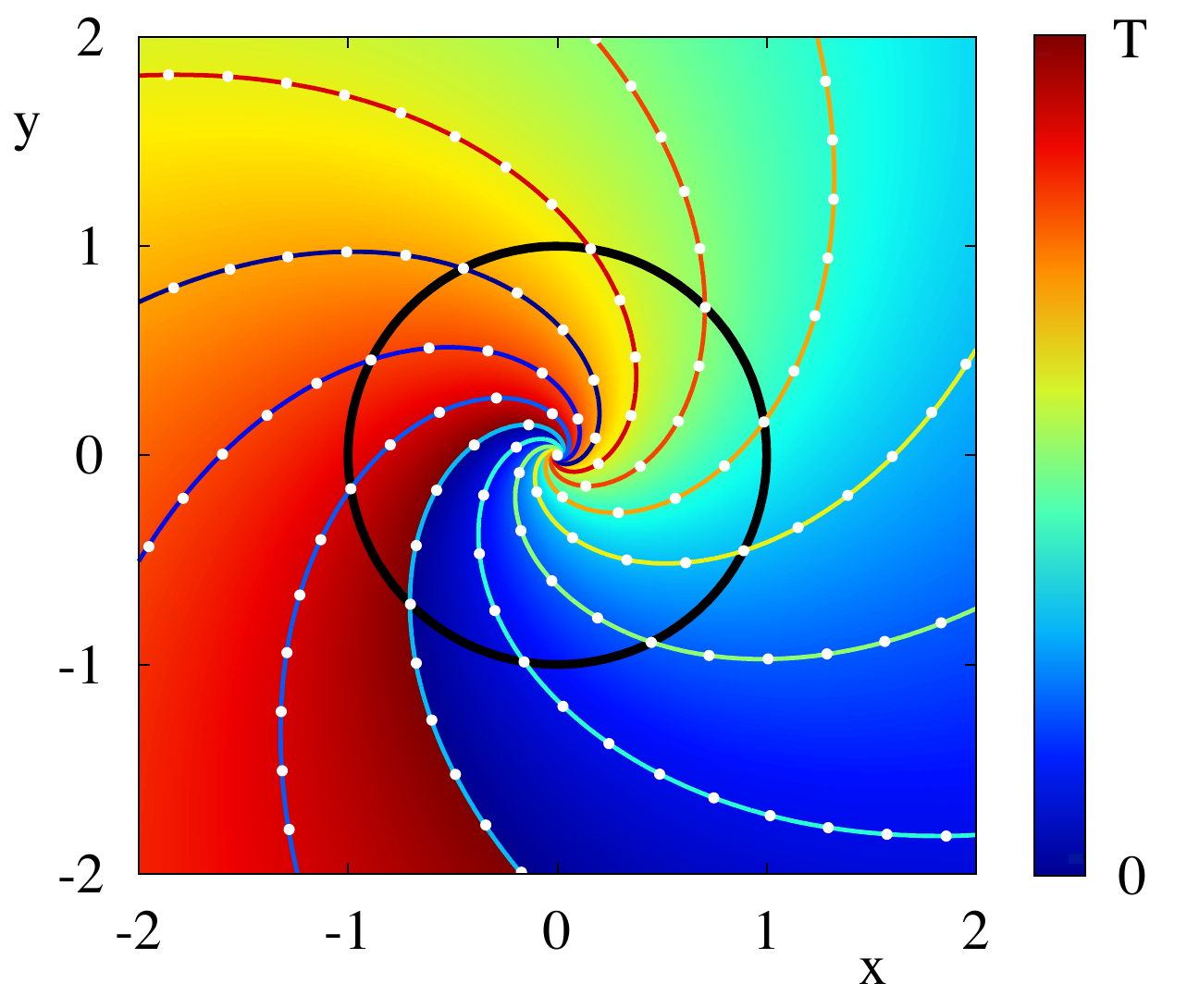}
\caption{Isochrons found using the method of Fourier averages for the Stuart-Landau oscillator:  $\frac{\d}{\d t} {z}=z[\lambda(1+ic)/2+i \omega] - z |z|^2 (1+i c)/2$, $z=x+iy$, with $\lambda=2$, $c=1$ and $\omega=1$ so that $T=2\pi$.  The black curve represents the periodic orbit of the system; in this case, we have $\theta=\vartheta$.
The background colour represents the Fourier average, whilst the coloured lines are the isochrons, given as level sets of the Fourier average. The white dots are the actual isochrons, computed analytically as $(x,y)=((1+r) \cos(\theta + c \ln(1+r)),(1+r) \sin (\theta + c \ln (1+r)))$, $r \in (-1,\infty)$.
\label{Fig:isochron}
}
\end{center}
\end{figure}

The Koopman operator approach for constructing isochrons for a $T$-periodic orbit focuses on tracking observables (or measures on a state space) rather than the identification of invariant sets.  The Koopman operator, $K$, is defined by $K = z \circ \Phi_t(x)$, where $z: \RSet^n \rightarrow \RSet$ is some observable of the state space and $\Phi_t(x)$ denotes the flow evolved for a time $t$, staring at a point $x$.  The Fourier average of an observable $z$ is defined as
\begin{equation}
\widehat{z}(x;\omega) = \lim_{T \rightarrow \infty} \frac{1}{T}\int_0^T (z \circ \Phi_t)(x) \e^{-i \omega t} \d t .
\label{Fourier}
\end{equation}
For a fixed $x$,  (\ref{Fourier}) is equivalent to a Fourier transform of the (time-varying) observable computed along a trajectory. Hence, for a dynamics with a stable limit cycle (of frequency $2\pi/T$),
it is clear that the Fourier average can be nonzero only for the frequencies $\omega_n= 2 \pi n/ T$, $n \in \ZSet$. The Fourier averages are the eigenfunctions of $K$, so that
\begin{equation}
K \, \widehat{z}(x;\omega_n) = \e^{i \omega_n t}\, \widehat{z}(x;\omega_n), \qquad n \in \ZSet .
\nonumber
\end{equation}
Perhaps rather remarkably the isochrons are level sets of $\widehat{z}(x;\omega_n)$ for almost all observables. The only restriction being that the first Fourier coefficient of the Fourier observable evaluated along the limit cycle is nonzero over one period.  An example of the use of this approach is shown in Fig.~\ref{Fig:isochron}, where we plot the isochrons of a Stuart-Landau oscillator.

\subsection{Phase-amplitude models}
\label{subsec:phase-amplitude}

An alternative (non isochronal) framework for studying oscillators with an attracting limit cycle is to make a transformation to a moving orthonormal coordinate system around the limit cycle where one coordinate gives the phase on the limit cycle while the other coordinates give a notion of distance from the limit cycle. It has long been known in the dynamical systems community how to construct such a coordinate transformation, see \cite{Hale1969} for a discussion.  The importance of considering the effects of both the phase and amplitude interactions of neural oscillators has been highlighted by several authors including Ermentrout and Kopell \cite{Ermentrout91} and Medvedev \cite{Medvedev2011}, and that this is especially pertinent when considering phenomenon such as oscillator death (and see \S~\ref{sec:WeakandStrong}).  Phase-amplitude descriptions have already successfully been used to find equations for the evolution of the energies (amplitudes) and phases of  weakly coupled weakly dissipative networks of nonlinear planar oscillators (modelled by small dissipative perturbations of a Hamiltonian oscillator) \cite{Ashwin1989,Ashwin2000a,Ashwin2005}. Lee \textit{et al}. \cite{Lee2013} use the notion of phase and amplitudes of large networks of globally coupled Stuart-Landau oscillators to investigate the effects of a spread in amplitude growth parameter (units oscillating with different amplitudes and some not oscillating at all) and the effect of a homogeneous shift in the nonlinear frequency parameter.

We now discuss the phase-amplitude coordinate transformation detailed by Hale \cite{Hale1969}, some of the situations where it has been employed and other areas in which a phase-amplitude description of oscillators is necessary to reveal the dynamics of the system.
Consider again the system \eqref{eq:driven} with $\epsilon=0$ which has an attracting hyperbolic periodic orbit. Make a transformation to a moving orthonormal coordinate system as follows. Choose one axis of the coordinate system to be in the direction of unit tangent vector along the periodic orbit, $\xi$, given by
\begin{equation}
\xi(\vartheta) = \FD{u}{\vartheta}\Big / \left|\FD{u}{\vartheta}\right|.
\label{eq:xi}
\nonumber
\end{equation}
The remaining coordinate axes can be expressed as the columns of an $n\times (n-1)$ matrix $\zeta$.  We can then write an arbitrary point $x$ in terms of its phase $\vartheta\in[0,T)$ and its amplitude $\rho$:
\begin{equation}
	x(\vartheta,\rho)=u(\vartheta)+\zeta(\vartheta)\rho.
\label{eq:trans}
\nonumber
\end{equation}
Here, $|\rho|$ represents the Euclidean distance from the limit cycle. The vector of amplitudes $\rho \in \R^{n-1}$ allows us to consider points away from the limit cycle.

Upon projecting the dynamics onto the moving orthonormal system, we obtain the dynamics of the transformed system for $\vartheta\in[0,T)$ and $\rho\in \R^{n-1}$
\begin{align}
	\frac{\d}{\d t} {\vartheta}=1+f_1(\vartheta,\rho) , \quad  \frac{\d}{\d t} {\rho}=A(\vartheta)\rho+f_2(\vartheta,\rho) .
	\nonumber
\end{align}
where
\begin{align}
	f_1(\vartheta,\rho)&=-h^T(\vartheta,\rho)\FD{\zeta}{\vartheta}\rho+h^T(\vartheta,\rho)\left[f(u+\zeta\rho)-f(u)\right] , \nonumber\\
	f_2(\vartheta,\rho)&=-\zeta^T\FD{\zeta}{\vartheta}\rho f_1+\zeta^T\left[f(u+\zeta\rho)-f(u)-\Df\zeta\rho\right] ,  \nonumber
\end{align}
\begin{align}
	h(\vartheta,\rho) = \left[\left| \FD{u}{\vartheta}\right| + \xi^T \FD{\zeta}{\vartheta}\rho \right]^{-1} \xi , \quad A(\vartheta)=\zeta^T\left[-\FD{\zeta}{\vartheta}+\Df\zeta\right] \label{eq:h} ,
\end{align}
and $\Df$ is the Jacobian of the vector field $f$, evaluated along the periodic orbit $u$.
The technical details to specify the orthonormal coordinates forming $\zeta$ can be found in the appendix of \cite{Wedgwood2013}.
It is straight-forward to show that $f_1(\vartheta,\rho) \rightarrow 0 \mbox{ as } |\rho |\rightarrow 0$, $f_2(\vartheta,0) = 0$ and that $\partial f_2 (\vartheta,0)/\partial\rho = 0$.
In the above, $A(\vartheta)$ describes the $\vartheta$-dependent rate of attraction or repulsion from cycle and $f_1$ captures the {\it shear} present in the system, that is, whether the speed of $\vartheta$ increases or decreases dependent on the distance from cycle. A precise definition for shear is given in \cite{Ott2010} and will be discussed further in \S~\ref{sec:shear}.

Some caution must be exercised when applying this transformation as it will break down when the determinant of the Jacobian of the transformation vanishes. This never occurs on cycle (where $\rho=0$) but it may do so for some $|\rho|=k>0$, setting an upper bound on how far from the limit cycle these phase-amplitude coordinates can be used to describe the system. In \cite{Wedgwood2013} it is noted that for the planar ML model the value of $k$ can be relatively small for some values of $\vartheta$, but that breakdown occurs where the orbit has high curvature. In higher dimensional systems this issue would be less problematic.

Similarly, the coordinate transformation can be applied to driven systems (i.e. \eqref{eq:driven} with $\epsilon \neq 0$) where $\epsilon$ is not necessarily small. In this case the dynamics in $(\vartheta, \rho)$ coordinates, where $\vartheta\in[0,T)$ and $\rho\in\R^{n-1}$, are
\begin{align}
	\frac{\d}{\d t} {\vartheta}&=1+f_1(\vartheta,\rho)+\epsilon h^T(\vartheta,\rho)g(u(\vartheta)+\zeta(\vartheta)\rho,t),  \label{drivetheta}\\
	\frac{\d}{\d t} {\rho}&=A(\vartheta)\rho+f_2(\vartheta,\rho)+\epsilon\zeta^T B(\vartheta,\rho) g(u(\vartheta)+\zeta(\vartheta)\rho,t), \label{driverho}
\end{align}
with
\begin{equation}
	B(\vartheta,\rho) = \mathrm{I}_n-\FD{\zeta}{\vartheta}\rho h^T(\vartheta,\rho) \label{eq:B},
\end{equation}
and $\mathrm{I}_n$ is the $n\times n$ identity matrix. Here, $h$ and $B$ describe the effect in terms of $\vartheta$ and $\rho$ that the perturbations have. For planar models, $B=\mathrm{I}_2$.

Medvedev \cite{Medvedev2011} has employed this phase-amplitude description to determine conditions for stability of the synchronised state in a network of identical oscillators with separable linear coupling. Medvedev \cite{Medvedev2010} has also used the framework to consider the effects of white noise on the synchronous state, identifying the types of linear coupling operators lead to synchrony in a network of oscillators provided that the strength of the interactions is sufficiently strong.

\subsection{Dynamics of forced oscillators:  shear-induced chaos}
\label{sec:shear}

Since phase-amplitude coordinates can capture dynamics a finite distance away from the limit cycle, (and additionally have the advantage over isochronal coordinates of being defined outside of the basin of attraction of the limit cycle) they can be used to model dynamical phenomena in driven systems where the perturbations necessarily push the dynamics away from the limit cycle. There is no need to make any assumptions about the strength of the forcing $\epsilon$.

The phase-amplitude description of a forced oscillator is able to detect the presence of other structures in the phase space. For example if the system were multistable, phase-amplitude coordinates would track trajectories near these other structures and back again, should another perturbation return the dynamics to the basin of attraction of the limit cycle. These coordinates would also detect the presence of other non-attracting invariant structures such as saddles in the unperturbed flow. Orbits passing near the saddle will remain there for some time and forcing may act to move trajectories near to this saddle before returning to the limit cycle. It may also be the case that the forcing acts to create trapping regions if the forcing is strong compared to the attraction to the limit cycle.

\begin{figure}[htbp]
\begin{center}
\includegraphics[width=4.7in]{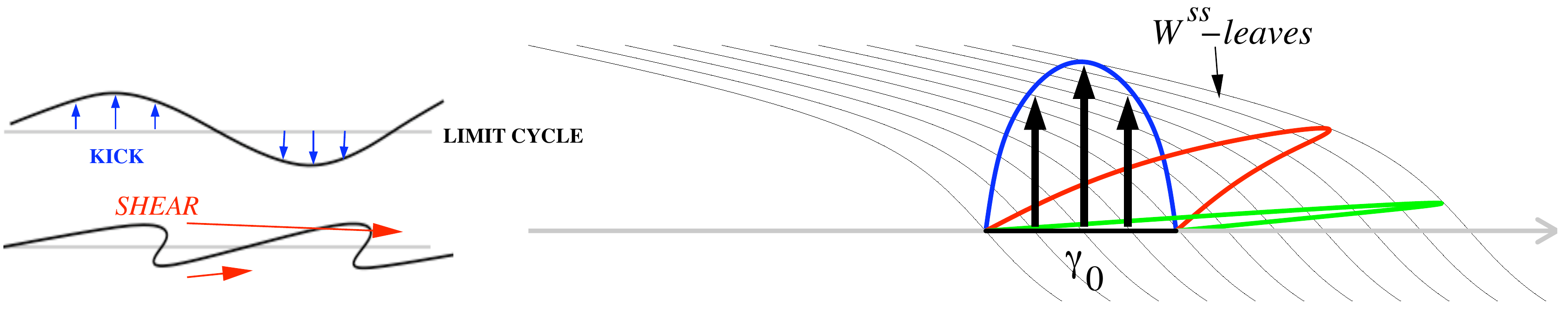}
\caption{The stretch-and-fold action of a kick followed by relaxation in the presence of shear.
Left:  A non-constant kick moves data away from the limit cycle (horizontal grey line).  As the image of the cycle under the kick relaxes back to the cycle (under the unperturbed flow) the action of shear causes folds to appear.
Right:  The geometry of folding in relation to the isochron foliation (black lines).  An initial segment $\gamma_0$ is kicked to the blue curve and is then allowed to relax back to cycle, passing through the red and green curves, which are images of the blue curve under the unperturbed flow.
\label{Fig:shear}
}
\end{center}
\end{figure}

Another dynamical phenomenon which phase-amplitude coordinates are able to capture is the occurrence of shear-induced chaos. {\em Shear} in a system near a limit cycle $\Gamma$ is the differential in components of the velocity tangent to the limit cycle as one moves further from $\Gamma$ in phase space. Shear forces within the system act to speed up (or slow down) trajectories further away from the limit cycle compared to those closer to it. This phenomenon is illustrated in Fig.~\ref{Fig:shear}.

As we show below, an impulsive force is applied (the system is kicked) a `bump' in $\Gamma$ is produced. 
If there is sufficient shear in the system then the bump is folded and stretched as it is attracted back to the limit cycle. Such folding can potentially lead to the formation of horseshoes and strange attractors. However, if the attraction to the limit cycle is large compared to the shear strength or size of the kick then the bumps will dissipate before any significant stretching occurs.

Shear-induced chaos is most commonly discussed in the context of discrete time kicking of limit cycles. Wang and Young \cite{Wang2001,Wang2002,Wang2003} prove rigorous results in the case of periodically kicked limit cycles with long relaxation times. Their results provide details of the geometric mechanism for producing chaos. Here we briefly review some of these results. More detailed summaries can be found in \cite{Lin2008} and \cite{Lin2010}.

Let $\Phi_t$ be the flow of the unperturbed system, which has an attracting hyperbolic limit cycle $\Gamma$. We can think of a kick as a mapping $\kappa$. The dynamics of the system with periodic kicking every $\tau$ units of time can be obtained by iterating the map $F_\tau=\Phi_\tau \circ \kappa$. Provided that there is a neighbourhood $\mathcal{U}$ of $\Gamma$ such that points in $\mathcal{U}$ remain in the basin of attraction of $\Gamma$ under kicks and $\tau$ is long enough that the kicked points can return to $\mathcal{U}$, then $\Gamma_\tau= \cap_{n\geq 0} F_\tau^n(\mathcal{U})$ is an attractor for the periodically kicked system. If the kicks are weak then $\Gamma_\tau$ is generally expected to be a slightly perturbed version of $\Gamma$ and we should expect fairly regular behaviour. In this case $\Gamma_\tau$ is an invariant circle. To obtain more complicated dynamics it is necessary to break the invariant circle. This idea is illustrated by the following linear shear model considered in \cite{Wang2002}:

\begin{align}
\frac{\d}{\d t} {\vartheta} & = 1+\sigma y ,\nonumber\\
\frac{\d}{\d t} {y}& =-\lambda y+A  H(\vartheta)  \sum_{n=0}^\infty \delta(t+n\tau),\nonumber
\end{align}
where $(\vartheta, y) \in [0,T)\times \mathbb{R} = S^1 \times \mathbb{R}$ are coordinates in the phase space, $\lambda, \sigma, A>0$ are constants and $H:S^1 \to \mathbb{R}$ is a nonconstant smooth function. The unforced system ($A=0$) has a limit cycle $\Gamma=S^1 \times \{0\}$. If the quantity
\[
\frac{\sigma}{\lambda} A \equiv \frac{\mbox{shear}}{\mbox{contraction rate}}\times (\mbox{kick `amplitude'}) ,
\]
is sufficiently large, then it is possible to show that there is a positive measure set $\mathcal{T}\subset \mathbb{R}^+$ such that for all $\tau \in \mathcal{T}$, $\Gamma_\tau$ is a strange attractor of $F_\tau$ \cite{Wang2003}. How large the quantity must be depends on the function $H(\vartheta)$. Since $H(\vartheta)$ is nonconstant (to create the bumps), the larger shear $\sigma$ and the kick amplitude $A$, the more folding will occur as the bump is attracted back to the limit cycle. Also note that weaker limit cycles (smaller $\lambda$) create more favourable conditions for shear induced chaos as the shear has longer to act before the bump is attracted back to the limit cycle.

For a more general system with periodic forcing the shear may not appear explicitly as a parameter. To elucidate what kind of kicks may cause shear induced chaos in this case we appeal to the isochrons of the system. We think of the isochrons of $\Gamma$ as the strong stable manifold of $\Gamma$. Suppose, as illustrated in Fig.~\ref{Fig:shear}, that a section $\Gamma_0$ of the limit cycle is kicked upwards with the end points held fixed and assume $\tau = np$ for some $n,p \in \mathbb{Z}^+$. Since the isochrons are flow invariant, during relaxation the flow moves each point of the kicked curve $\kappa(\Gamma_0)$ back towards $\Gamma$ along the isochrons. In Fig.~\ref{Fig:shear} we can clearly see the effect of the shear with a fold forming.

From Fig.~\ref{Fig:shear} one can see that kicks along isochrons or in directions roughly parallel to the isochrons will not produce strange attractors, nor will kicks that carry one isochron's manifold to another. The cause of the stretching and folding is the variation in how far points $x \in \Gamma$ are moved by $\kappa$ in the direction transverse to the isochrons (i.e. the ordering of points in terms of asymptotic phase is altered by the action of the kick). Lin and Young \cite{Lin2008} emphasise that the occurrence of shear induced chaos depends on the interplay between the geometries of the kicks and the dynamical structures of the unforced system.

In the case of the linear shear model above, the isochrons of the unforced system are simply the lines with slope $-\lambda/\sigma$ in $(\vartheta, y)$ coordinates. Variation in kick distances in directions transverse to these isochrons is guaranteed with any nonconstant function $H$, with greater variation given by larger values of $\sigma/\lambda$ and $A$.

Beyond the rigorous results proved by Wang and Young \cite{Wang2001,Wang2002,Wang2003,Lin2008} concerning periodically kicked limit cycles of the linear shear model and for supercritical Hopf bifurcations, Ott and Stenlund \cite{Ott2010} prove that shear-induced chaos may exist near general limit cycles. In addition, Lin and Young \cite{Lin2008} have carried out numerical studies including random kicks at times given by a Poisson process and systems driven by white noise. They also consider forcing of a pair of coupled oscillators. In all cases, shear-induced chaos occurs when the shearing and amplitude of the forcing are large enough to overcome the effects of damping.

Lin \textit{et al}. \cite{Lin2013} demonstrate that the ML model can exhibit shear-induced chaos near the homoclinic bifurcation when periodically forced, by plotting images of the periodic orbit under successive applications of the kick map and calculating the maximum Lyapunov exponent. They also emphasise that the phenomenon of shear-induced chaos cannot be detected by the perturbative techniques such as those outlined in \S~\ref{subsec:phase} and \S~\ref{subsec:phaseresponse} below.

Wedgwood \textit{et al}. \cite{Wedgwood2013} go on to show that the phase-amplitude description is well-suited to understanding the behaviour of neural oscillators in response to external stimuli. They consider a number of neural oscillator models in addition to a generic planar model and examine the response of the system to periodic pulsatile forcing by taking $x\in\R^2$,
\[
g(x, t) = \sum_{n\in \mathbb{Z}} (\delta(t-n\tau),0)^T ,
\]
and a variety of choices for $f(x)$ in \eqref{eq:driven}. Their numerical simulations indicate that for both the FHN and ML models the behaviour remains qualitatively similar when the relevant functions $f_1(\vartheta, \rho)$ are replaced by $\sigma \rho$, $f_2(\vartheta, \rho)$ is dropped and $A(\vartheta)$ is replaced with $-\lambda$ (for a wide range of $\sigma, \lambda>0$). They then focus on the effect of the form of the forcing function in the phase-amplitude coordinates. Evaluating the functions $h(\vartheta, \rho)$ and $B(\vartheta, \rho)$ of equations \eqref{eq:h} and \eqref{eq:B} respectively for the particular model and denoting the first component of $h$ as $P_1$ and the first component of $\xi$ as $P_2$ they find that forcing each system at the same ratio of the natural frequency of the underlying periodic orbit and implementing the choices above gives the following ODE on $\vartheta\in[0,T)$, $\rho\in\R^{n-1}$
\begin{equation}
\frac{\d}{\d t} {\vartheta} = 1 +\sigma \rho + \epsilon P_1(\vartheta,\rho) \sum_{n\in \mathbb{Z}} \delta(t-n\tau), \qquad \frac{\d}{\d t} {\rho} = -\lambda \rho +\epsilon P_2(\vartheta)\sum_{n\in \mathbb{Z}} \delta(t-n\tau).
\nonumber
\end{equation}
By developing a stroboscopic map for this system, Wedgwood \textit{et al}. \cite{Wedgwood2013} numerically evaluate the largest Lyapunov exponent which is found to be positive for the Morris-Lecar model but negative for the FHN model.

The analysis of the behaviour of generic networks of oscillators within a phase-amplitude framework is a challenging open problem but such a description would allow for greater accuracy (compared to the phase only methods traditionally used and described below) in elucidating a richer variety of the complex dynamics of oscillator networks.

\subsection{Phase oscillator models}
\label{subsec:phase}

To obtain a phase description, one can consider the limit of strong attraction in equations (\ref{drivetheta})-(\ref{driverho}).  However, it is more appealing to consider a phase variable with a uniform rotation rate that assigns a phase coordinate $\vartheta\in[0,T)$ to each point $x \in \Gamma$ according to
\begin{equation}
\FD{}{t}\vartheta(x(t)) = 1, \qquad x \in \Gamma .
\label{phasedot}
\end{equation}
This reduction to a phase description gives a simple dynamical system, albeit one that cannot describe evolution of trajectories in phase-space that are away from the limit cycle.  However, the phase reduction formalism is useful in quantifying how a system (on or close to a cycle) responds to weak forcing, via the construction of the infinitesimal phase response curve (iPRC).  This can be written, for a given ODE model, as the solution to an \textit{adjoint equation}.  For a given high dimensional conductance based model this can be solved for numerically, though for some normal form descriptions closed form solutions are also known \cite{Brown04}.

The iPRC at a point on cycle is equal to the gradient of the (isochronal) phase at that point.  Writing this vector quantity as $Q$ means that weak forcing $\epsilon g(t)$ in the original high dimensional models transforms as $\ID{\vartheta}=1+\epsilon \langle Q, g \rangle$ where $\langle\cdot,\cdot \rangle$ defines the standard inner product and $\epsilon$ is a small parameter.  For periodic forcing such equations can be readily analysed, and questions relating to synchronisation, mode-locking and Arnol'd tongues can be thoroughly explored \cite{Pikovsky01}.  Moreover, this approach forms the basis for constructing models of weakly interacting oscillators, where the external forcing is pictured as a function of the phase of a firing neuron.  This has led to a great deal of work on phase-locking and central pattern generation in neural circuitry and see for example \cite{Hoppensteadt97}.

However, the assumption that phase alone is enough to capture the essentials of neural response is one made more for mathematical convenience than being physiologically motivated.  Indeed for the popular type I ML firing model with standard parameters, direct numerical simulations with pulsatile forcing show responses that cannot be explained solely with a phase model \cite{Lin2013}, as just highlighted in \S~\ref{sec:shear} (since strong interactions will necessarily take one away from the neighbourhood of a cycle where a phase description is expected to hold).

\subsection{Phase response}
\label{subsec:phaseresponse}

It is common practice in neuroscience to characterise a neuronal oscillator in terms of its phase response to a
perturbation \cite{Rinzel1989}.  This gives rise to the notion of a so-called phase response curve (PRC), which for a real neuron can be determined experimentally \cite{Galan2005,Tateno2007}, and can also be related to the poststimulus time histogram \cite{Gutkin2005}.  Consider a dynamical system $\ID{x}=f(x)$, $x\in\R^N$ with a $T$-periodic solution $u(t)=u(t+T)$ and introduce an infinitesimal perturbation $\Delta x_0$ to the trajectory $u(t)$ at time $t=0$.
This perturbation evolves according to the linearised equation of
motion:\begin{equation}
\FD{}{t} \Delta x = Df(u(t)) \Delta x , \qquad
\Delta x(0) = \Delta x_0.
\label{linearized}
\nonumber
\end{equation}
Here $Df(u)$ denotes the Jacobian of $f$ evaluated along $u$.
Introducing a time-independent \textit{isochronal} phase shift $\Delta \vartheta$ as $\vartheta(u(t)+\Delta u(t))-\vartheta(u(t))$, we have to first order in $\Delta x$
that\begin{equation}
\Delta \vartheta = \langle Q(t), \Delta x(t) \rangle ,
\label{PRC1}
\end{equation}
where $\langle\cdot,\cdot \rangle$ defines the standard inner product, and $Q=\nabla_{u} \vartheta$ is the gradient of $\vartheta$ evaluated at $u(t)$.
Taking the time-derivative of (\ref{PRC1})
gives
\begin{equation}
\left \langle \FD{}{t} Q, \Delta x \right \rangle  = -\left \langle Q, \FD{}{t} \Delta x \right \rangle
= -\left \langle Q, Df(u) \Delta x \right \rangle = -\left \langle Df^T(u)Q,  \Delta x \right \rangle .
\nonumber
\end{equation}
Since the above equation must hold for arbitrary perturbations, we see that the gradient $Q=\nabla_{u} \vartheta$ satisfies the linear
equation\begin{equation}
\FD{}{t} Q = D(t) Q , \qquad
D(t) = - Df^T(u(t)) ,
\label{PRC2}
\end{equation}
subject to the boundary conditions
$$
\langle \nabla_{u(0)} \vartheta , f(u(0)) \rangle =1~~\mbox{and}~~Q(t)=Q(t+T).
$$
The first condition simply guarantees that $\ID{\vartheta}=1$ (at any point on the periodic orbit), and the second enforces periodicity.  Note that the notion of phase response can also be extended to time-delayed systems \cite{Kotani2012,Novicenko2012a}.

\begin{figure}[htbp]
\begin{center}
\includegraphics[width=4.5in]{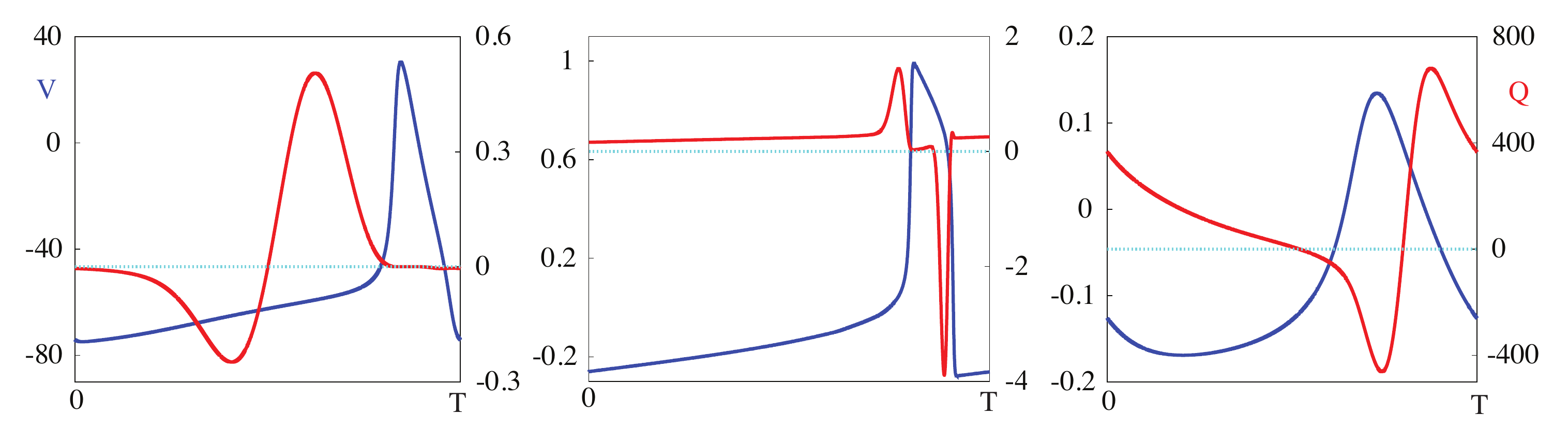}
\caption{From left to right: Phase response curves for the Hodgkin-Huxley model, the Wilson model and the Morris-Lecar model corresponding to the discussion in \S~\ref{sec:neuronoscillators}.  Orbits are in blue and iPRCs in red.
\label{Fig:PRC}
}
\end{center}
\end{figure}

In general equation (\ref{PRC2}) must be solved numerically to obtain the iPRC, say, using the \textit{adjoint} routine in XPPAUT \cite{Ermentrout02} or MatCont \cite{Govaerts2006}.  However, for the case of a nonlinear IF model, defined by (\ref{eq:one}), the PRC is given explicitly by
\begin{equation}
Q(\vartheta) = \frac{1}{I+f \circ \Psi^{-1}(\vartheta)}, \qquad \Psi(v) =
\tau \int_{v_\text{R}}^{v(t)} \frac{\d v'}{I+f(v')} ,
\label{RIF}
\end{equation}
where the period of oscillation is found by solving $\Psi(v_\text{th})=T$ (and the response function is valid for finite perturbations).
For example, for the quadratic IF model, obtained from (\ref{eq:one}) with the choice with $f(v)=v^2$, and taking the limit $v_\text{th} \rightarrow \infty$ and $v_\text{R} \rightarrow -\infty$ we find $Q(\vartheta) = \sin^2(\vartheta \pi/T)/I$ with $T=\pi \tau /\sqrt{I}$, recovering the iPRC expected of an oscillator close to a SNIC bifurcation \cite{Ermentrout1996,Ermentrout2012}.
For a comprehensive discussion of iPRCs for various common oscillator models see the excellent survey article by
Brown \textit{et al}. \cite{Brown04}.
The iPRC for planar piecewise linear IF models can also be computed explicitly \cite{Coombes12}, although we do not discuss this further here.
In Fig.~\ref{Fig:PRC} we show the numerically computed iPRCs for the Hodgkin-Huxley model, the Wilson cortical model and the Morris-Lecar model previously discussed in \S~\ref{sec:neuronoscillators}.
For completeness it is well to note the iPRC for a planar oscillator close to a supercritical Hopf bifurcation has an adjoint with a shape given by $(-\sin(2 \pi t/T),\cos(2 \pi t/T))$ for an orbit shape $(\cos(2 \pi t/T), \sin(2 \pi t/T))$.

Having defined the phase in some neighbourhood of the limit cycle, we can write (\ref{phasedot}) as
$$
\FD{}{t} \vartheta(x) = \left \langle \nabla_x \vartheta  , \frac{\d}{\d t} {x} \right \rangle .
$$
For $\ID{x}=f(x)$ this gives $\langle \nabla_x \vartheta , f(x) \rangle = 1$.
We now consider the effect of a small external periodic force on the self sustained oscillations as in (\ref{eq:driven}), with $g(x,t)=g(x,t+\Delta)$,
where in general $\Delta$ is different to $T$.
For weak perturbations ($\epsilon \ll 1$) the state point will leave $\Gamma$, though will stay in some neighbourhood, which we denote by $\mathcal{U}$.  We extend the phase off cycle using isochronal coordinates so that $\langle \nabla_x \vartheta , f(x) \rangle = 1$ holds for any point $x \in \mathcal{U}$.
For a perturbed oscillator, in these coordinates we have
\begin{equation}
\frac{\d}{\d t} {\vartheta} = \left \langle \nabla_x \vartheta , \frac{\d}{\d t} {x} \right \rangle = \left \langle \nabla_x \vartheta , f(x) + \epsilon g(x,t) \right \rangle
= 1 + \epsilon \left \langle \nabla_x \vartheta , g(x,t) \right \rangle .
\nonumber
\end{equation}
As a first approximation we evaluate the right hand side of the above on the limit-cycle to get an equation for the phase dynamics:
\begin{equation}
\frac{\d}{\d t} {\vartheta} = 1 + \epsilon I(\vartheta,t), \qquad I(\vartheta,t) = \langle Q(u(\vartheta)) , g(u(\vartheta),t) \rangle .
\label{phase dynamics}
\end{equation}
The phase dynamics (\ref{phase dynamics}) is still very hard to analyse, and as such it is common
to resort to a further simplification, namely \textit{averaging}.
First let us introduce a rotating phase $\psi=\vartheta- Tt/\Delta$,
in which case (\ref{phase dynamics}) becomes
\begin{equation}
\frac{\d}{\d t} {\psi} = -\delta  + \epsilon I(\psi+T t/\Delta,t), \qquad \delta  = T/\Delta-1.
\nonumber
\end{equation}
If both $\epsilon$ and $\delta$ are small then $\ID{\psi} \simeq 0$ and $\psi$ evolves \textit{slowly}.  Thus we may set $\psi(s) \simeq \psi(t)$ for $s \in [t, t+T]$. Averaging the above over one period gives
\begin{equation}
\frac{\d}{\d t} {\psi} \simeq -\delta  + \epsilon H(\psi),\qquad H(\psi) = \frac{1}{T} \int_0^T \left \langle Q (u(\psi+ s)), g (u(\psi+ s),s) \right \rangle \d s ,
\label{averaged}
\end{equation}
where we have used the result that $I(\psi +s,t)=I(\psi +s+T,t+T)$.  The function $H(\psi)$ is $T$-periodic and can be written as a Fourier series $H(\psi) = \sum_n H_n \e^{2 \pi i n \psi/T}$, with the simplest example of an averaged phase-dynamics being
\begin{equation}
\frac{\d}{\d t} {\psi} = -\delta  + \epsilon \sin \psi
\nonumber
\end{equation}
which is called the Adler equation \cite{Adler1946}.
If we denote the maximum and minimum of $H$ by $H_\text{min}$ and $H_\text{max}$ respectively then for a \textit{phase-locked} $1$:$1$ state defined by $\frac{\d}{\d t} {\psi}=0$ we require $\epsilon H_\text{min} < \delta  < \epsilon H_\text{max}$.  In this case there are two fixed points $\psi_\pm$ defined by $H(\psi_\pm) = \delta$.
One of these is unstable (say $\psi_-$, so  that $\epsilon H'(\psi_-)>0$) and the other stable ($\psi_+$, with $\epsilon H'(\psi_+)<0$).  This gives rise to a rotating solution with constant rotation frequency so that $\vartheta=(1+\delta)t+ \psi_+$. The two solutions coalesce in a saddle-node bifurcation when $\delta = \epsilon H_\text{min}$ and $\delta = \epsilon H_\text{max}$ (or equivalently when $H'(\psi_\pm)=0$).
In the case of the Adler model the parameter region for phase-locking is given explicitly by a triangular wedge defined by $\epsilon = |\delta |$ -- a so-called Arnol'd tongue.  Outside of this tongue solutions \textit{drift} (they cannot lock to the forcing period) according to (\ref{averaged}), and the system evolves quasi-periodically. We treat weakly coupled phase oscillators in \S~\ref{sec:weakcoupling}.


\section{Weakly coupled phase oscillator networks}
\label{sec:weakcoupling}

The theory of weakly coupled oscillators \cite{Kuramoto84,Ermentrout84} is now a standard tool of dynamical systems  theory and has been used by many authors to study oscillatory neural networks, see for example \cite{Ermentrout84,vanVreeswijk94,Chow98,Lewis03}.  The book by Hoppensteadt and Izhikevich provides a very comprehensive review of this framework \cite{Hoppensteadt97}, which can also be adapted to study networks of relaxation oscillators (in some singular limit) \cite{Izhikevich00,Coombes2001}.

Consider, for illustration, a system of interacting limit-cycle oscillators (\ref{eq:pairwisecoupledcells}). Following the method in \S~\ref{subsec:phaseresponse}, similar to (\ref{averaged}) we obtain the network's phase dynamics in the form
\begin{equation}
\frac{\d}{\d t} {\theta_i} = \omega_i + \epsilon \sum_{j=1}^N w_{ij} \langle Q_i(u_i(\theta_i)) , G(u_j(\theta_j)) \rangle
\label{phase dynamicsnetwork},
\end{equation}
where the frequency $\omega_i$ allows for the fact that oscillators are not identical and, for this reason, we will assume that $\theta_i\in[0,2\pi)$. Precisely this form of network model was originally suggested by Winfree to describe populations of coupled oscillators.
The Winfree model \cite{Winfree67} assumes a separation of time-scales so that an oscillator can be solely characterised by its phase on cycle (fast attraction to cycle) and is described by the network equations
\begin{equation}
\frac{\d}{\d t} {\theta}_i = \omega_i + \epsilon R(\theta_i) \frac{1}{N} \sum_{j=1}^N P(\theta_j) ,
\label{Winfree}
\end{equation}
describing a globally coupled network with a biologically realistic PRC $R$ and pulsatile interaction function $P$.  Using a mixture of analysis and numerics Winfree found that with large $N$ there was a transition to macroscopic synchrony at a critical value of the \textit{homogeneity} of the population.  Following this Kuramoto \cite{Kuramoto84} introduced a simpler model with interactions mediated by phase differences, and showed how the transition to collective synchronisation could be understood from a more mathematical perspective.  For an excellent review of the Kuramoto model see \cite{Strogatz2000} and \cite{Acerbron2005}.

The natural way to obtain a phase-difference model from (\ref{phase dynamicsnetwork}) is, as in \S~\ref{subsec:phaseresponse}, to average over one period of oscillation.  For simplicity let us assume that all the oscillators are identical, and $\omega_i=\omega$ $\forall i$, in which case we find that
\begin{equation}
\frac{\d}{\d t} {\theta}_i = \omega + \epsilon \sum_{j=1}^N w_{ij} H(\theta_j-\theta_i),
\label{phasenetwork}
\end{equation}
where
\begin{equation}
H(\psi) = \frac{1}{2\pi}\int_0^{2\pi} \left \langle Q (u(s)), G (u(\psi+ s)) \right \rangle \d s .
\label{H}
\end{equation}
The $2\pi$-periodic function $H$ is referred to as the \textit{phase interaction function}.
If we write complex Fourier series for $Q$ and $G$ as
$$
Q(t)=\sum_{n\in\Z} Q_n \e^{i n t}~~\mbox{ and }~~G(t)=\sum_{n\in\Z} G_n \e^{ i n t}
$$
respectively then
\begin{equation}
H(\psi) = \sum_{n\in\Z} H_n \e^{i n \psi}
\label{eq:Hfourier}
\end{equation}
with $H_n = \langle Q_{-n}, G_n \rangle$. Note that certain caution has to be exercised in applying averaging theory. In general, one can only establish that a solution of the unaveraged equations
is $\epsilon$-close to a corresponding solution of the averaged system for times of
$O(\epsilon^{-1})$. No such problem arises in the case of hyperbolic fixed points corresponding to phase-locked states.

When describing a piece of cortex or a central pattern generator circuit with a set of
oscillators, the biological realism of the model typically resides in the phase interaction function.
The simplest example is $H(\psi)=\sin(\psi)$, which when combined with a choice of global coupling defines the well known Kuramoto model \cite{Kuramoto84}.  However, to model realistic neural networks one should calculate (\ref{H}) directly, using knowledge of the single neuron iPRC and the form of interaction.  As an example consider synaptic coupling, described in \S~\ref{sec:coupling}, that can be written in the form $G(u(\psi)) = \sum_m \eta(\psi+m 2\pi)$,
and a single neuron model for which the iPRC in the voltage variable $v$ is given by $R$ (say experimentally or from numerical investigation).  In this case
\begin{equation}
H(\psi) = \int_0^\infty R (2 \pi s-\psi) \eta(2 \pi s) \d s .
\label{Hsynapse}
\end{equation}
If instead we were interested in diffusive (gap-junction) coupling then we would have
\begin{equation}
H(\psi) = \frac{1}{2 \pi}\int_0^{2 \pi} R(s) [v(s+\psi)-v(s)] \d s .
\label{Hgap}
\nonumber
\end{equation}
For the HH model $R(t)$ is known to have a shape like $-\sin(t)$ for a spike centred on the origin (see Fig.~ \ref{Fig:PRC}).  Making the further choice that $\eta(t)=\alpha^2 t \e^{-\alpha t}$ then (\ref{Hsynapse}) can be evaluated as
\begin{equation}
H(\psi) = \frac{[1-(1/\alpha)^2] \sin (\psi)-2/\alpha  \cos(\psi)}{2 \pi [1+(1/\alpha)^2]^2}.
\label{HalphaHH}
\end{equation}
In the particular case of two oscillators with reciprocal coupling and $\omega=1$ then
\begin{equation}
\begin{split}
\FD{}{t} {\theta}_1 &= 1 + \epsilon H(\theta_2-\theta_1) ,\\
\FD{}{t} {\theta}_2 &= 1 + \epsilon H(\theta_1-\theta_2) ,
\end{split}
\nonumber
\end{equation}
and we define $\varphi:=\theta_2(t)-\theta_1(t)$. A phase-locked solution satisfies
constant phase difference $\phi$ that is a zero of the (odd) function
$$
K(\varphi)=\epsilon[H(-\varphi)-H(\varphi)].
$$
A given phase-locked state is then stable provided that $K'(\varphi)<0$.
Note that by symmetry both the in-phase ($\varphi=0$) and anti-phase ($\varphi
=\pi$) states are guaranteed to exist.  For the form of phase interaction function given by $(\ref{HalphaHH})$, The stability of the synchronous solution is governed by the sign of $K'(0)$:
\begin{equation}
\text{sgn}~ K'(0) = \text{sgn} ~\{-\epsilon H'(0)\} = \text{sgn}~ \left \{  -\epsilon [(1-(1/\alpha))^2] \right \} .
\nonumber
\end{equation}
Thus for inhibitory coupling ($\epsilon < 0$) synchronisation will occur if  $1/\alpha > 1$, namely
when the synapse is \textit{slow} ($\alpha \rightarrow 0$).  It is also a simple matter to show that the anti-synchronous solution ($\varphi=\pi$) is stable for a sufficiently \textit{fast} synapse  ($\alpha \rightarrow \infty$).
It is also possible to develop a general theory for the existence and stability of phase-locked states in larger networks than just a pair.

\subsection{Phase, frequency and mode locking}
\label{subsec:locking}

Now suppose we have a general population of $N\geq 2$ coupled phase oscillators
\begin{equation}
\frac{\d}{\d t} {\theta}_j=f_j(\theta_1,\ldots,\theta_N),
\label{eq:phasepop}
\nonumber
\end{equation}
described by phases $\theta_j\in \R/2\pi\Z$. For a particular continuous choice of phases $\theta(t)$ for the trajectory one can define the {\em frequency} of the $j$th oscillator as
$$
\Omega_j=\lim_{T\rightarrow\infty} \frac{1}{T} [\theta_j(T)-\theta_j(0)].
$$
This limit will converge under fairly weak assumptions on the dynamics \cite{Karabacak2009}, though it may vary for different attractors in the same system, for different oscillators $j$ and in some cases it may vary even for different trajectories within the same attractor.

We say two oscillators $j$ and $k$ are {\em phase locked} with ratio $(n:m)$ for $(n,m)\in \Z^2\setminus (0,0)$ with no common factors of $n$ and $m$, if there is an $M$ such that
$$
|n\theta_j(t)-m\theta_k(t)|<M ,
$$
for all $t>0$. The oscillators are {\em frequency locked} with ratio $(n:m)$ if
$$
n\Omega_j-m\Omega_k=0.
$$
If we say they are simply phase (or frequency locked)  without explicit mention of the $(n:m)$ ratio, we are using the convention that they are $(1:1)$ phase (or frequency) locked. The definition of $\Omega_j$ means that if two oscillators are phase locked then they are frequency locked. The converse is not necessarily the case: two oscillators may be frequency locked but not phase locked if the phase difference $n\theta_j(t)-m\theta_k(t)$ grows sublinearly with $t$.

For the special case globally coupled averaged networks ($w_{ij}=1/N$ for the system (\ref{phasenetwork})) is $S_N \times \T$ equivariant.  By topological arguments, maximally symmetric solutions describing synchronous, splay, and a variety of cluster states are expected to exist generically~\cite{Ashwin92}. The system (\ref{phasenetwork}) with global coupling is in itself an interesting subject of study in that it is of arbitrarily high dimension $N$ but is effectively determined by the single function $H(\varphi)$ that is computable from a single pair of oscillators. The system (and variants thereof) have been productively studied by thousands of papers since the seminal work of Kuramoto \cite{Kuramoto84}.

\subsection{Dynamics of general networks of identical phase oscillators}
\label{subsec:phase-locked}

The collective dynamics of phase oscillators have been investigated for a range of regular network structures including linear arrays and rings with uni- or bi-directional coupling e.g. \cite{Ermentrout84,Ermentrout85,Ashwin92,Brown03}, and hierarchical networks \cite{Skardal2012}. In some cases the systems can be usefully investigated in terms of permutation symmetries of (\ref{phasenetwork}) with global coupling, for example $\Z_N$ or $\D_N$ for uni- or bi-directionally coupled rings. In other cases a variety of approaches have been developed adapted to particular structures though these have not in all cases been specifically applied to oscillators; some of these approaches are discussed in \S~\ref{sec:networks}

We recall that the form of the coupling in (\ref{phasenetwork}) is special in the sense that it assumes the interactions between two oscillators are independent of any third - pairwise coupling \cite{Ashwin92,Brown03}. If there are degeneracies such as
\begin{equation}
\label{eq:degenerate}
\sum_{k=0}^{m-1} H\left(\varphi+\frac{2\pi k}{m}\right)=0 ,
\end{equation}
which can appear when some of the Fourier components of $H$ are zero, this can lead to degeneracies in the dynamics. For example \cite{Watanabe94}, while \cite[Theorem 7.1]{Ashwin92} shows that if $H$ satisfies (\ref{eq:degenerate}) for some $m\geq 2$ and $N$ is a multiple of $m$ then (\ref{phasenetwork}), with global coupling, will have $m$-dimensional invariant tori in phase space that are foliated by neutrally stable periodic orbits. This degeneracy will disappear on including either non-pairwise coupling or introducing small but non-zero Fourier components in the expansion of $H$ but as noted in \cite{Kori2014} this will typically be the case for the interaction of oscillators even if they are near a Hopf bifurcation.

We examine in more detail some of the phase locked states that can arise in weakly coupled networks of identical phase oscillators described by (\ref{phasenetwork}).
We define a $1$:$1$ phase-locked solution to be of the form
$\theta_i(t)=\varphi_i+ \Omega t$, where
$\varphi_i$ is a constant phase and $\Omega$ is
the collective frequency of the coupled oscillators. Substitution into the averaged
system (\ref{phasenetwork}) gives
\begin{equation}
\Omega = \omega +{\epsilon}\sum_{j=1}^N w_{ij} H(\varphi_{j} - \varphi_i) .
\label{Omega}
\end{equation}
After choosing some reference oscillator, these $N$ equations
determine the collective frequency $\Omega$ and $N-1$
relative phases with the latter independent of $\epsilon$.

It is interesting to compare the weak coupling theory for phase-locked states with the analysis of LIF networks from \S~\ref{sec:IFnetworks}.  Equation (\ref{IFphases}) has an identical structure to that of equation (\ref{Omega}) (for
$I_i = I$ for all $i$), so that the classification of solutions using group theoretic methods is the same in both situations.  There are, however, a number of significant differences between phase-locking equations (\ref{Omega}) and (\ref{IFphases}). First, equation (\ref{IFphases}) is exact, whereas equation (\ref{Omega}) is valid only to $O(\epsilon)$ since it is derived under the assumption of weak coupling.
Second, the collective period of oscillations $\Delta$ must be determined self-consistently in equation (\ref{IFphases}).

In order to
analyse the local stability of a phase-locked
solution $\Phi=(\phi_1,\ldots,\phi_N)$, we linearise the system by setting
$\theta_i(t)={\varphi}_i+\Omega t+ \widetilde{\theta}_i(t)$
and expand to first-order in $\widetilde{\theta}_i$:
\begin{equation}
\FD{}{t} \widetilde{\theta}_i= {\epsilon} \sum_{j=1}^N\widehat{\mathcal
H}_{ij}(\Phi) \widetilde{\theta}_{j} ,
\nonumber
\end{equation}
where
\begin{equation}
\widehat{\mathcal{H}}_{ij}(\Phi)=w_{ij}H'(\varphi_j-\varphi_i)
-\delta_{i,j}\sum_{k=1}^N w_{ik}H'(\varphi_k-\varphi_i) ,
\nonumber
\end{equation}
and $H'(\varphi) = \d H(\varphi)/\d\varphi$.
One of the eigenvalues of the Jacobian $\widehat{\mathcal H}$ is always zero, and the corresponding eigenvector points in the direction of the flow, that is $(1,1,\ldots,1)$. The phase-locked solution will be stable provided that
all other eigenvalues have a negative real part.  We note that the Jacobian has the form of a graph-Laplacian mixing both anatomy and dynamics, namely it is the graph-Laplacian of the matrix with components $-w_{ij}H'(\varphi_j-\varphi_i)$.

\subsubsection{Synchrony}
\label{sec:synchrony}

Synchrony (more precisely, exact phase synchrony) is where $\theta_1=\theta_2= \ldots = \theta_{N-1}=\theta_N = \Omega t+ t_0$ for some fixed frequency $\Omega$ is a classic example of a phase-locked state.  Substitution into (\ref{phasenetwork}), describing a network of identical oscillators, shows that $\Omega$ must satisfy the condition
\begin{equation}
\Omega = \omega + \epsilon H(0)\sum_{j=1}^N w_{ij} , \qquad \forall i .
\nonumber
\end{equation}
One way for this to be true for all $i$ is if $H(0)=0$, which is the case say for $H(\theta)=\sin (\theta)$ or for diffusive coupling, which is linear in the difference between two state variables so that $H(0)=0$. The existence of synchronous solutions is also guaranteed if $\sum_{j=1}^N w_{ij}$ is independent of $i$.  This would be the case for global coupling where $w_{ij}=1/N$, so that the system has permutation symmetry.

If the synchronous solution exists then the Jacobian is given by $\epsilon H'(0)\mathcal{L}$ where $\mathcal{L}$ is the graph-Laplacian with components $\mathcal{L}_{ij}=w_{ij}-\delta_{ij}\sum_{k} w_{ik}$.
\begin{figure}[htbp]
\begin{center}
\includegraphics[width=6cm]{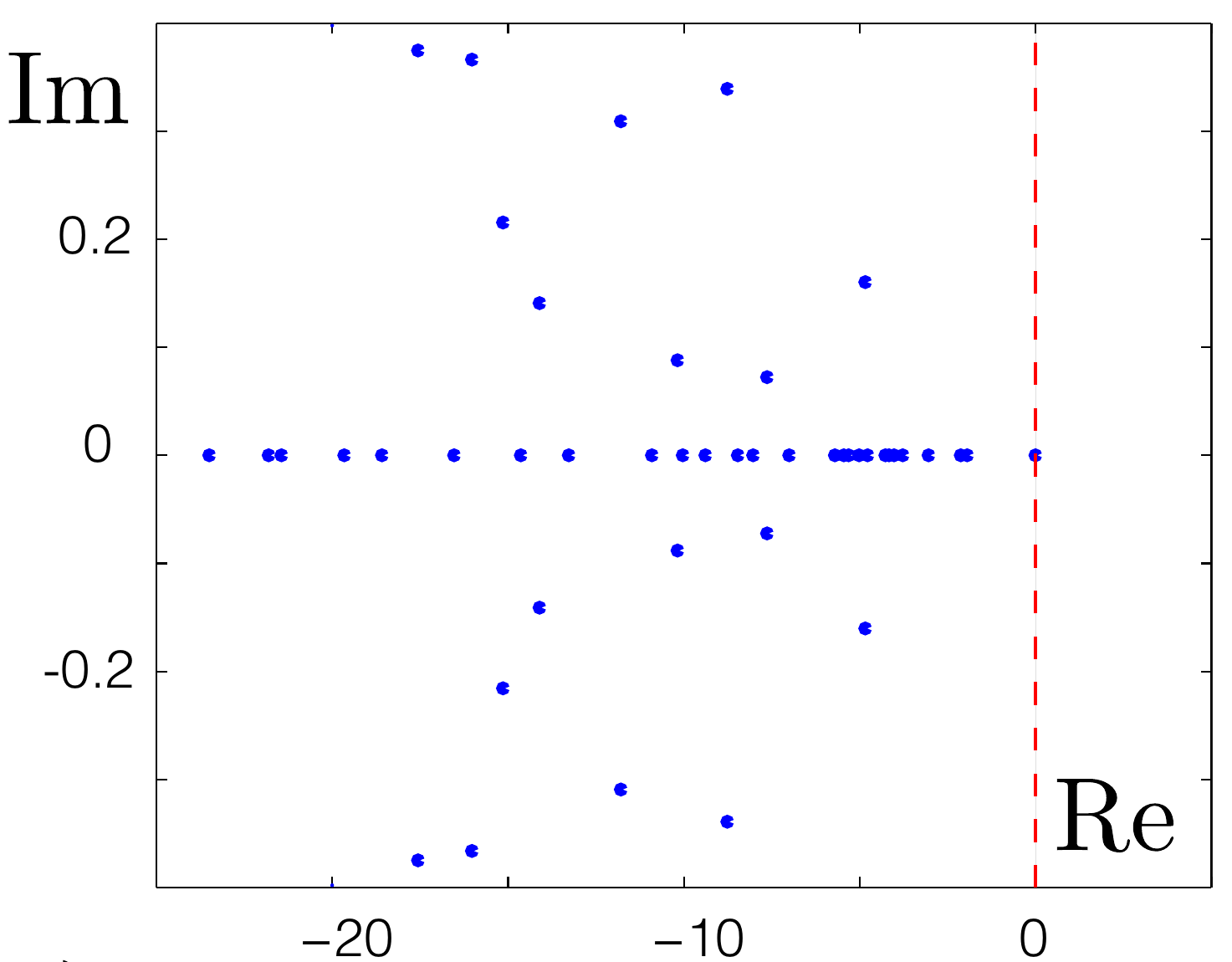}
\caption{Eigenvalues of the graph-Laplacian for the anatomical network structure of the Macaque monkey brain --- generated from the \href{http://cocomac.g-node.org}{CoCoMac database}.
\label{CoCoMac}}
\end{center}
\end{figure}
We note that $\mathcal{L}$ has one zero eigenvalue, with eigenvector $(1,1,\ldots,1,1)$.
Hence if all the other eigenvalues of $\mathcal{L}$ lie on one side of the imaginary axis then stability is solely determined by the sign of $\epsilon H'(0)$.  In Fig.~\ref{CoCoMac} we show the eigenvalues of the graph-Laplacian of the anatomical network structure of the Macaque monkey brain, as determined from the CoCoMac database \cite{Stephan13}.  Here all the eigenvalues lie to the left of the imaginary axis so that the stability of the synchronous solution (should it exist) is solely determined by the sign of $\epsilon H'(0)$.
For global coupling we have that $\mathcal{L}_{ij} = N^{-1}- \delta_{ij}$, and the ($N-1$ degenerate) eigenvalue is $-1$.  Hence the synchronous solution will be stable provided $\lambda = -\epsilon H'(0)<0$.

\subsubsection{Asynchrony}
\label{sec:asynchrony}

Another example of a phase-locked state is the purely asynchronous solution whereby all phases are uniformly distributed around the unit circle. This is sometimes referred to as a {\em splay state} or {\em splay-phase state} and can be written $\ID{\theta_i}=\Omega$ with $\theta_{i+1}-\theta_i=2\pi/N ~\forall i$.  Like the synchronous solution it will be present but not necessarily stable in networks with global coupling, with an emergent frequency that depends on $H$:
\begin{equation}
\Omega = \omega + \epsilon \frac{1}{N}\sum_{j=1}^N H\left(\frac{2\pi j}{N}\right).
\nonumber
\end{equation}
In this case the Jacobian takes the form
\begin{equation}
\widehat{\mathcal{H}}_{nm}(\Phi)=  \frac{\epsilon}{N} \left [A_{n-m}
-\delta_{nm} \Gamma \right ] ,
\nonumber
\end{equation}
where $\Gamma=\sum_{k} H'(2\pi k/N)$ and $A_{n}=H'(2\pi n/N)$.  Hence the eigenvalues are given by $\lambda_p = \epsilon [\nu_p -\Gamma]/N$, $p=0,\ldots,N-1$ where $\nu_p$ are the eigenvalues of $A_{n-m}$:
$\sum_m A_{n-m} a _m^p = \nu_p a_n^p$,
where $a_n^p$ denote the components of the $p$th eigenvector.  This has solutions of the form $a_n^p = \e^{-2 \pi i np/N}$ so that $\nu_p = \sum_m A_m \e^{2 \pi i m p/N}$, giving
\begin{equation}
\lambda_p = \frac{\epsilon}{N} \sum_{m=1}^N H'\left(\frac{2\pi m}{N}\right) [\e^{2 \pi i m p/N}-1] ,
\nonumber
\end{equation}
and the splay state will be stable if $\text{Re} \, (\lambda_p) < 0~ \forall p \neq 0$.

In the large $N$ limit $N \rightarrow \infty$ we have the useful result that (for global coupling) network averages may be replaced by time averages:
\begin{equation}
\lim_{N \rightarrow \infty} \frac{1}{N} \sum_{j=1}^N F\left(\frac{2\pi j}{N}\right) = \frac{1}{2\pi} \int_0^{2\pi} F(t) \d t = F_0 ,
\nonumber
\end{equation}
for some $2\pi$-periodic function $F(t)=F(t+2\pi)$ (which can be established using a simple Riemann sum argument), with a Fourier series $F(t) = \sum_n F_n \e^{i n t}$.
Hence in the large $N$ limit the collective frequency of a splay state (global coupling) is given by
$\Omega = \omega + \epsilon H_0$, with
eigenvalues\begin{equation}
\label{lambdan}
\lambda_p = \frac{\epsilon}{2\pi} \int_0^{2\pi} H'(t) \e^{ i p t} \d t = -2 \pi i p \epsilon H_{-p} .
\nonumber
\end{equation}
Hence a splay state is stable if $-p \epsilon \text{Im} \, H_p < 0$, where we have used the fact that since $H(\theta)$ is real, then $\text{Im} \, H_{-p}=- \text{Im} \, H_p$.
As an example consider the case $H(\theta) = \theta(\pi-\theta)(\theta-2 \pi)$ for $\theta\in[0,2\pi)$ and $\epsilon>0$ (where $H$ is periodically extended outside $[0,2\pi)$).  It is straight forward to calculate the Fourier coefficients (\ref{eq:Hfourier}) as $H_n=6i/n^3$, so that $-p \epsilon \text{Im} \, H_p = - 6 \epsilon/p^2 < 0 ~\forall p \neq 0$.  Hence the asynchronous state is stable.  If we flip any one of the coefficients $H_m \rightarrow - H_m$ then $\text{Re} \, \lambda_m >0$ and the splay state will develop an instability to an eigenmode that will initially destabilise the system in favour of an $m$-cluster.
\begin{figure}[htbp]
\begin{center}
\includegraphics[width=10cm]{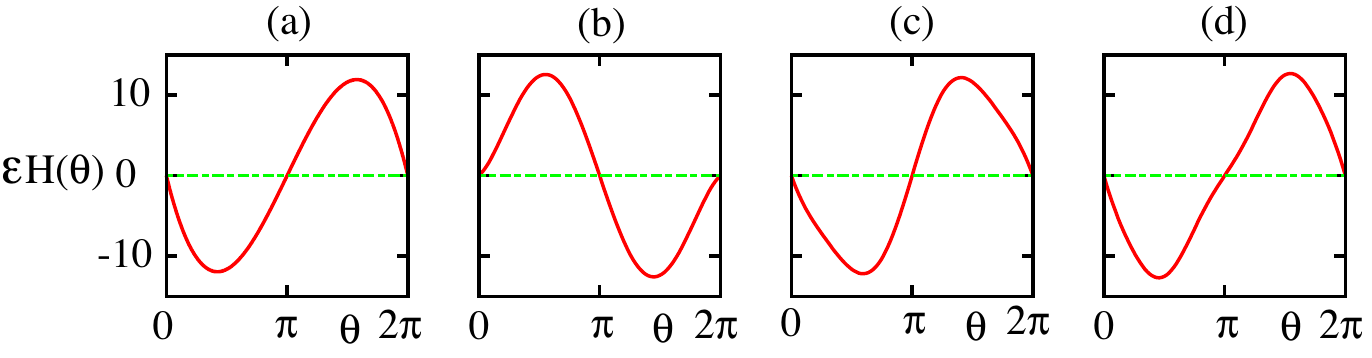}
\caption{(a)  A phase interaction given by $\epsilon H(\theta) = \theta(\pi-\theta)(\theta-2 \pi)$ for $\theta\in[0,2\pi)$ with complex Fourier series coefficients given by $H_n=6i/n^3$.  The remaining panels show the effect of flipping the sign of just one of the co-efficients, namely $H_m \rightarrow -H_m$.  (b)  $m=1$, and the asynchronous solution will destabilise in favour of the synchronous solution since $H'(0)>0$.  (c) $m=2$, and the asynchronous solution will destabilise in favour of a $2$-cluster.  (d) $m=3$, and the asynchronous solution will destabilise in favour of a $3$-cluster.  Note that only small changes in the shape of $H$, as seen in panels (c)-(d), can lead to a large change in the emergent network dynamics.
\label{OddH}}
\end{center}
\end{figure}

\subsubsection{Clusters for globally coupled phase oscillators}

For reviews of the stability of cluster states (in which subsets of the oscillator population synchronise, with oscillators belonging to different clusters behaving differently) we refer the reader to \cite{Ashwin92,Golomb94,Brown03}; here we use the notation of \cite{Oroszetal2009}. If a group of $N$ oscillators is neither fully synchronised nor desynchronised it may be clustered. We say $\cA=\{A_1,\ldots,A_M\}$ is an $M$-cluster partition of
$\{1,\ldots,N\}$ if
\begin{equation}
\label{eq:partition}
\{1,\ldots,N\}=\bigcup_{p=1}^M A_p\,,
\nonumber
\end{equation}
where $A_p$ are pairwise disjoint sets ($A_p\cap A_q=\emptyset$ if $p \neq q$).
NB if $a_p=|A_p|$ then
\begin{equation}
\sum_{p=1}^M a_p = N\,.
\nonumber
\end{equation}
One can refer to this as a cluster of type $(a_1,\ldots,a_M)$. It is possible to show that any clustering can be realised as a stable periodic orbit of the globally coupled phase oscillator system \cite{Oroszetal2009} for suitable choice of phase interaction function; more precisely, there is a coupling function $H$ for the system (\ref{phasenetwork}), with global coupling, such that for any $N$ and any given $M$-cluster partition $\cA$ of $\{1,\ldots, N\}$ there is a linearly stable periodic orbit realising that partition (and all permutations of it). See also \cite{Skardal2011} where the authors consider clustering in this system where $H(\varphi)=\sin M\varphi$. More generally, there are very many invariant subspaces corresponding to spatio-temporal clustering that we can characterise in the following Theorem.

\begin{thm}{\bf \cite[Theorem 3.1]{Ashwin92}}
\label{thm:ASiso}
The subsets of $\T^N$ that are invariant for (\ref{phasenetwork}), with global coupling, because of symmetries of $S_N\times \T$ correspond to isotropy subgroups in the conjugacy class of
$$
\Sigma_{k,m}:=(S_{k_1}\times\cdots\times S_{k_{\ell}})^m \times_s \Z_m
$$
where $N=mk$, $k=k_1+\cdots+k_{\ell}$ and $\times_s$ denotes the semidirect product. The points with this isotropy have $\ell m$ clusters that are split into $\ell$ groups of $m$ clusters of the size $k_i$. The clusters within these groups are cyclically permuted by a phase shift of $2\pi/m$. The number of isotropy subgroups in this conjugacy class is $N!/[m(k_1!\ldots k_\ell!)]$.
\end{thm}

It is a nontrivial problem to discover which of these subspaces contain periodic solutions. Note that in-phase corresponds to $\ell=m=1$, $k_1=N$ while splay phase corresponds to $\ell=k_1=1$, $m=N$. The stability of several classes of these solutions can be computed in terms of properties of $H(\varphi)$; see for example \S~\ref{sec:synchrony} and \S~\ref{sec:asynchrony} and for other classes of solution \cite{Ashwin92,Brown03,Oroszetal2009}.

\subsubsection{Generic loss of synchrony in globally coupled identical phase oscillators}

Bifurcation properties of the globally coupled oscillator system (\ref{phasenetwork}) on a varying parameter that affects the coupling $H(\varphi)$ are surprisingly complicated  because of the symmetries present in the system; see \S~\ref{subsec:symmetry}. In particular, the high multiplicity of the eigenvalues for loss of stability of the synchronous solution means:
\begin{itemize}
\item Path following numerical bifurcation programmes such as AUTO or XPPAUT need to be done with great care when applying to problems with $N\geq 3$ identical oscillators - these typically will not be able to find all solutions branching from one that loses stability.
\item A large number of branches with a range of symmetries may generically be involved in the bifurcation; indeed, all 2-cluster states $S_k \times S_{N-k}$.
\item Local bifurcations may have global bifurcation consequences owing to the presence of connections that are facilitated by the nontrivial topology of the torus \cite{AshKinSwi1990,Ashwin92}.
\item Branches of degenerate attractors such as heteroclinic attractors may appear at such bifurcations for $N\geq 4$ oscillators.
\end{itemize}

\begin{figure}[htbp]
\begin{center}
\includegraphics[width=4in]{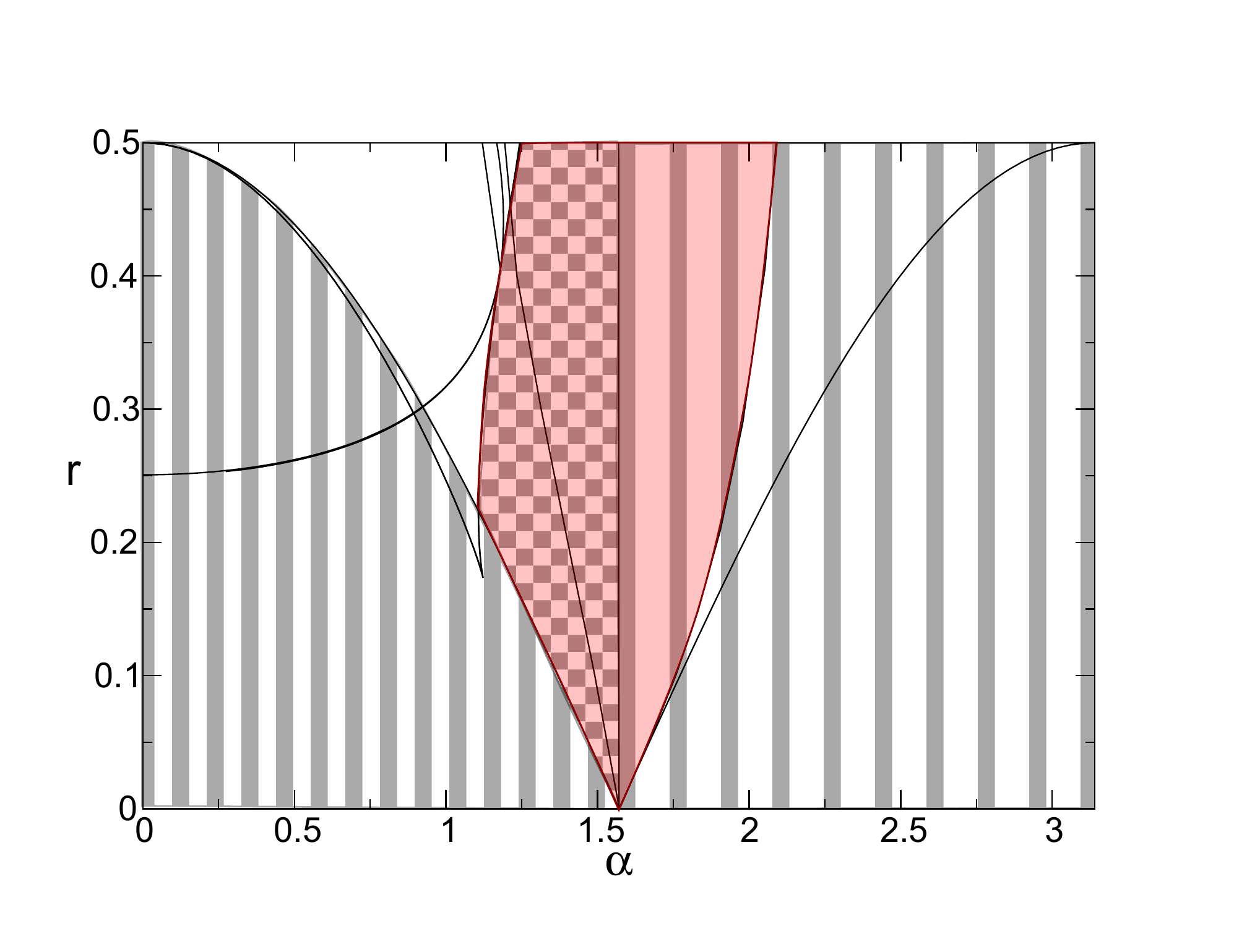}
\caption{
Regions of stability for the globally coupled $N=4$-oscillator system (\ref{phasenetwork}) with phase interaction function (\ref{eq:HMMcoupling}) and parameters in the region $r>0$, $0<\alpha<\pi$ \cite{Ashwin2008}. The narrow stripes show the region of stability of synchrony, while the wide stripes show the region of stability of the splay phase solution. The pink shaded area shows a region of existence of a robust heteroclinic network that is an attractor with in the checkerboard region; the boundaries are described in \cite{Ashwin2008}.
\label{Fig:fouroscbifs}
}
\end{center}
\end{figure}

Hansel \textit{et al}. \cite{Hanseletal1993} consider the system (\ref{phasenetwork}) with global coupling and phase interaction function of the form
\begin{equation}
H(\varphi)=-\sin(\varphi+\alpha)+r\sin (2\varphi) ,
\label{eq:HMMcoupling}
\end{equation}
for $(r,\alpha)$ fixed parameters; detailed bifurcation scenarios in the cases $N=3,4$ are shown in \cite{Ashwin2008}. As an example, Figure~\ref{Fig:fouroscbifs} shows regions on stability of synchrony, splay phase solutions and robust heteroclinic attractors as discussed later in \S~\ref{sec:heteroclinic}.

\subsection{Phase waves}
\label{sec:waves}

The phase reduction method has been applied to a number of important biological systems, including the study of travelling waves in chains of weakly coupled oscillators that model processes such as the generation and control of rhythmic activity in central pattern generators (CPGs) underlying locomotion \cite{Cohen1982,Kopell1995} and peristalsis in vascular and intestinal smooth muscle \cite{Ermentrout84}. Related phase models have been motivated by the observation that synchronisation and waves of excitation can occur during sensory processing in the cortex \cite{Ermentrout2001a}.  In the former case the focus has been on dynamics on a lattice and in the latter continuum models have been preferred.  We now present examples of both these types of model, focusing on \textit{phase wave} solutions \cite{Ermentrout1992}.

\subsubsection*{Phase waves: a lattice model}

The lamprey is an eel-like vertebrate which swims by generating travelling waves of neural activity that pass down its spinal cord.  The spinal cord contains about $100$ segments, each of which is a simple half-center neural circuit capable of generating alternating contraction and relaxation of the body muscles on either side of body during swimming.  In a seminal series of papers, Ermentrout and Kopell carried out a detailed study of the dynamics of a chain of weakly coupled limit cycle oscillators \cite{Ermentrout84,Kopell1990,Kopell86}, motivated by the known physiology of the lamprey spinal cord. They considered $N$ phase-oscillators arranged on a chain with nearest-neighbour anisotropic interactions  and identified a travelling wave as a phase-locked state with a constant phase difference between adjacent segments.  The intersegmental phase differences are defined as
$\varphi_i = \theta_{i+1} - \theta_i$.  If $\varphi_i < 0$ then the wave travels from head to tail whilst for $\varphi_i>0$ the wave travels from the tail to the head.
\begin{figure}[htbp]
\begin{center}
\includegraphics[width=4.5in]{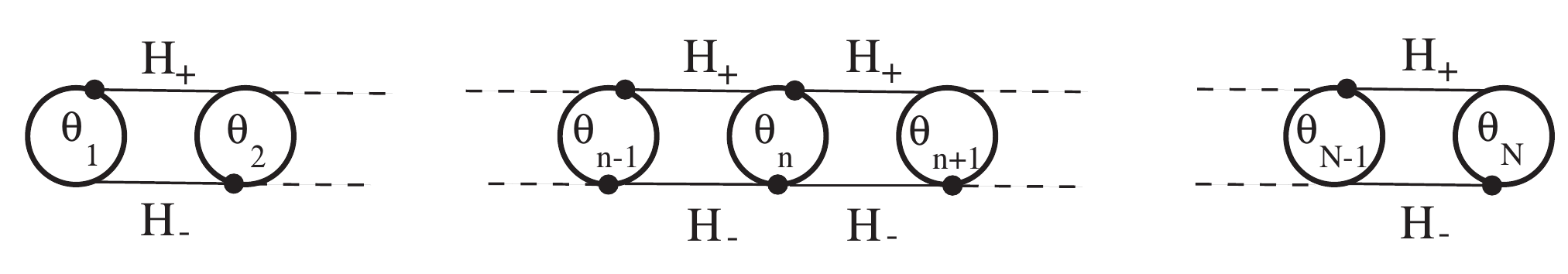}
\caption{
A chain of $N$ phase oscillators  $\varphi_i=\theta_{i+1}-\theta_i$ with $H_\pm(\varphi) =W_{\pm}H(\varphi)$.
}
\end{center}
\end{figure}
For a chain we set $W_{ij} = \delta_{i-1,j}W_- + \delta_{i+1,j}W_+$  to obtain
\begin{align}
\frac{\d}{\d t} {\theta}_1&= \omega_1 + W_+ H(\theta_2 -\theta_1) , \nonumber \\
\frac{\d}{\d t} {\theta}_i &= \omega_i + W_+ H(\theta_{i+1} -\theta_i) + W_- H (\theta_{i-1}- \theta_i), \qquad i=2,\ldots,N-1 , \nonumber \\
\frac{\d}{\d t} {\theta}_N &= \omega_N + W_- H(\theta_{N-1} -\theta_N) , \nonumber
\end{align}
where $\theta_i\in \R/2\pi \Z$. Pairwise subtraction and substitution of $\varphi_i = \theta_{i+1} - \theta_i$ leads to an $N-1$ dimensional system for the phase differences
\begin{equation}
\frac{\d}{\d t} {\varphi}_i = \Delta  \omega_i + W_+ [H(\varphi_{i+1})- H(\varphi_i)] +
W_- [H (-\varphi_i)-H (-\varphi_{i-1})],
\nonumber
\end{equation}
for $i=1 \ldots N-1$,
with boundary conditions $H(-\varphi_0) = 0 = H(\varphi_{N+1})$, where $\Delta \omega_i = \omega_{i+1} - \omega_i$.
There are at least two different mechanisms that can generate
travelling wave solutions.

The first is based on the presence of a gradient of
frequencies along the chain, that is, $\Delta \omega_i$ has the same sign for all $i$, with the
wave propagating from the high frequency region to the low frequency region.
This can be established explicitly in the case of an isotropic, odd interaction
function, $W_\pm =1$ and $H(\varphi) = -H(-\varphi)$, where we have
\begin{equation}
\frac{\d}{\d t} {\varphi}_i = \Delta  \omega_i + H(\varphi_{i+1}) + H (\varphi_{i-1}) - 2 H(\varphi_i) .
\nonumber
\end{equation}
The fixed points $\Phi=(\varphi_1,\ldots,\varphi_N)$ satisfy the matrix equation ${ H} ( \Phi ) = -{ A}^{-1} { D}$,
where ${ H} (\Phi) = (H(\varphi_1), \ldots, H(\varphi_N))^\top$, ${ D} = (\Delta \omega_1, \ldots, \Delta \omega_N)^\top$, and
${ A}$ is a tridiagonal matrix with elements $A_{ii}=-2$, $A_{i,i+1}=A_{1+1,i}=1$.
For the sake of
illustration suppose that $H(\varphi) = \sin(\varphi+\sigma)$.
Then a solution $\Phi$ will exist if every
component of ${ A}^{-1} { D}$ lies between $\pm1$.
Let $a_0 = \max \{ | ({ A}^{-1} { D})_i | \}$.
If $a_0 < 1$ then for each $i=1,\ldots,N$ there are two distinct solutions $\varphi_i^\pm$ in the interval
$[0, 2\pi]$ with $H'(\varphi_i^- ) > 0$ and $H'(\varphi_i^+ ) < 0$. In other words, there are $2^N$ phase-locked
solutions. Linearising about a phase-locked solution
and exploiting the structure of the matrix ${ A}$, it can be proven that only the
solution $\Phi^-=(\varphi_1^-, \ldots,\varphi_N^-)$ is stable. Assuming that the frequency gradient is
monotonic, this solution corresponds to a stable travelling wave. When the
gradient becomes too steep to allow phase-locking (i.e. $a_0 > 1$), two or more
clusters of oscillators (frequency plateaus) tend to form that oscillate at different
frequencies.
Waves produced
by a frequency gradient do not have a constant speed or, equivalently, constant
phase lags along the chain.

Constant speed waves in a chain of identical
oscillators can be generated by considering phase-locked solutions defined by $\varphi_i = \varphi$ for all $i$, with a collective period of oscillation $\Omega$ determined
using $\ID{\theta_1} = \Omega$ to give $\Omega =\omega_1 + W_+ H(\varphi_1)$.
The steady state equations are then $\Delta \omega_1 +W_+ H(-\varphi) =0$, $\Delta \omega_{N-1} -W_-H(\varphi) =0
$ and $\Delta \omega_i =0$, for  $i=2,\ldots, N-2$.  Thus, a travelling wave solution requires that all frequencies must be the same except at the ends of the chain.  One travelling solution is given by $\Delta \omega_{N-1}=0$ with
$\Delta \omega_1 = -W_- H(-\varphi)$ and  $H(\varphi) = 0$.
For the choice $H(\varphi)=\sin (\varphi+\sigma)$ we have that $\varphi=-\sigma$ and $\Delta \omega_1   = -W_- \sin(2 \sigma)$.  If $2 \sigma < \pi$ then $\Delta \omega_1 =\omega_2 - \omega_1<0$ and $\omega_1$ must be larger than $\omega_2$ and hence all the remaining $\omega_i$ for a forward travelling wave to exist.  Backward swimming can be generated by setting $\omega_1=0$ and solving in a similar fashion.

\subsubsection*{Phase waves: a continuum model}
\label{sec:phasewave}

There is solid experimental evidence for electrical waves in awake and aroused vertebrate preparations, as well as semi-intact and active invertebrate preparations, as nicely described by Ermentrout and Kleinfeld \cite{Ermentrout2001a}.  Moreover, these authors argue convincingly for the use of phase models in understanding waves seen in cortex and speculate that they may serve to label simultaneously perceived features in the stimulus stream with a unique phase.  More recently it has been found that cortical oscillations can propagate as travelling waves across the surface of the motor cortex of monkeys (Macaca mulatta) \cite{Rubino2006}.
Given that to a first approximation the cortex is often viewed as being built from a dense reciprocally interconnected network of corticocortical axonal pathways, of which there are typically $10^{10}$ in a human brain it is natural to develop a \textit{continuum} phase model, along the lines described by Crook \textit{et al}. \cite{Crook1997}.  These authors further incorporated axonal delays into their model to explain the seemingly contradictory result that synchrony is stable for short range excitatory coupling, but unstable for long range.  To see how a delay induced instability may arise we consider a continuum model of identical phase oscillators with space-dependent delays
\begin{equation}
\PD{}{t} \theta(x,t) = \omega + \epsilon \int_D W(x,y) H(\theta(y,t)-\theta(x,t) - \tau(x,y)) \d y ,
\label{phasecontinuum}
\end{equation}
where $x \in D \subseteq \RSet$ and $\theta\in \R/2\pi\Z$.
This model is naturally obtained as the continuum limit of (\ref{phasenetwork}) for a network arranged along a line with communication delays $\tau(x,y)$ set by the speed of an action potential $v$ that mediates the long range interaction over a distance between points in the tissue at $x$ and $y$.  Here  we have used the result that for weak coupling then delays manifest themselves as phase-shifts.  The function $W$ sets the anatomical connectivity pattern.  It is convenient to assume that interactions are homogeneous and translation invariant, so that $W(x,y) = W(|x-y|)$ with $\tau(x,y)=|x-y|/v$, and either assume periodic boundary conditions or take $D=\RSet$.

\begin{figure}[htbp]
\begin{center}
\includegraphics[width=2.5in]{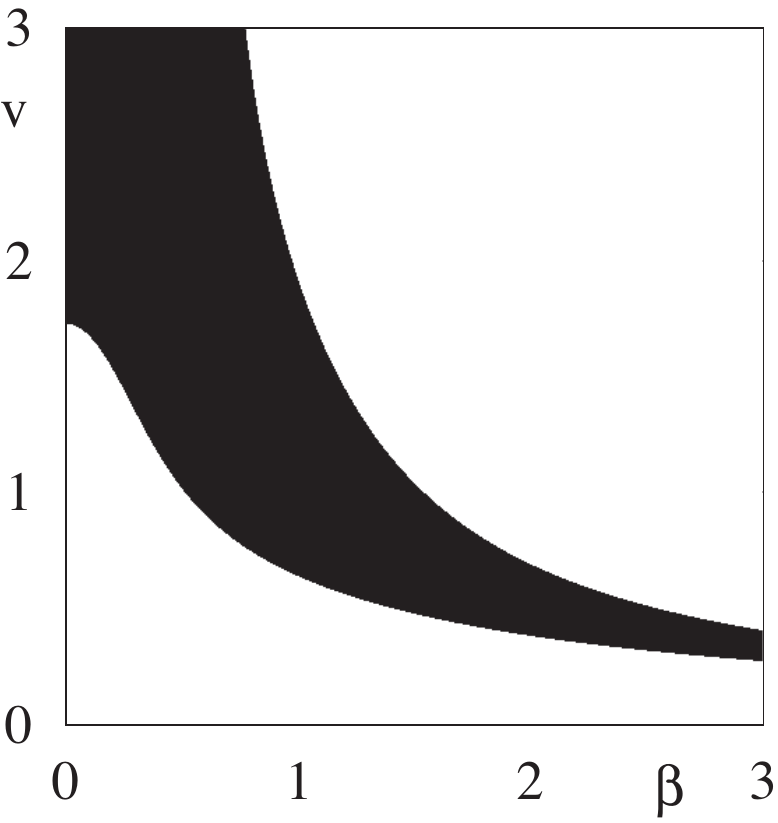}
\caption{
Stability region (black) for a phase wave $\theta(x,t) = \Omega t + \beta x$ in the $(\beta,v)$ plane for the choice $H(\theta) = \sin \theta$, $W(y)=\exp(-|y|)/2$ and $\epsilon>0$.
\label{fig:Phasewave}
}
\end{center}
\end{figure}

Following \cite{Crook1997} we construct travelling wave solutions of equation (\ref{phasecontinuum}) for $D=\RSet$ of the form
$\theta(x,t) = \Omega t + \beta x$, with the frequency $\Omega$ satisfying the dispersion relation
\begin{equation}
\Omega = \omega+\epsilon \int_{-\infty}^\infty \d y W(|y|) H (\beta y - |y|/v) .
\nonumber
\end{equation}
When $\beta=0$, the solution is synchronous. To explore the stability of the travelling wave we linearise (\ref{phasecontinuum}) about $\Omega t + \beta x$ and consider perturbations of the form $\e^{\lambda t} \e^{ipx}$, to find that travelling phase waves solutions are stable if $\text{Re} \, \lambda(p) <0$ for all $p \neq 0$, where
\begin{equation}
\lambda(p)=  \epsilon \int_{-\infty}^\infty W(|y|) H'(\beta y - |y|/v) [\e^{i p y} -1] \d y .
\nonumber
\end{equation}
Note that the neutrally stable mode $\lambda(0) =0$ represents constant phase-shifts.
For example, for the case that $H(\theta)=\sin \theta$ then we have that
\begin{equation}
\text{Re} \, \lambda(p)=\pi \epsilon [\Lambda(p, \beta_+) + \Lambda(-p, \beta_-) ] ,
\nonumber
\end{equation}
where
\begin{equation}
\Lambda (p,\beta) = W_c(p+\beta) +W_c(p-\beta)-2 W_c(\beta), \qquad W_c(p) = \int_0^\infty W(y) \cos(p y) \d y ,
\nonumber
\end{equation}
and $\beta_\pm = \pm \beta - 1/(v)$.  A plot of the region of wave stability for the choice $W(y)=\exp(-|y|)/2$ and $\epsilon>0$ in the $(\beta,v)$ plane is shown in Fig.~\ref{fig:Phasewave}.  Note that the synchronous solution $\beta=0$ is unstable for small values of $v$.
For a discussion of more realistic forms of phase interaction, describing synaptic interactions, see \cite{Bressloff97}.

\subsubsection{Phase turbulence}
\label{sec:turbulence}

\begin{figure}[htbp]
\begin{center}
\includegraphics[width=2.5in]{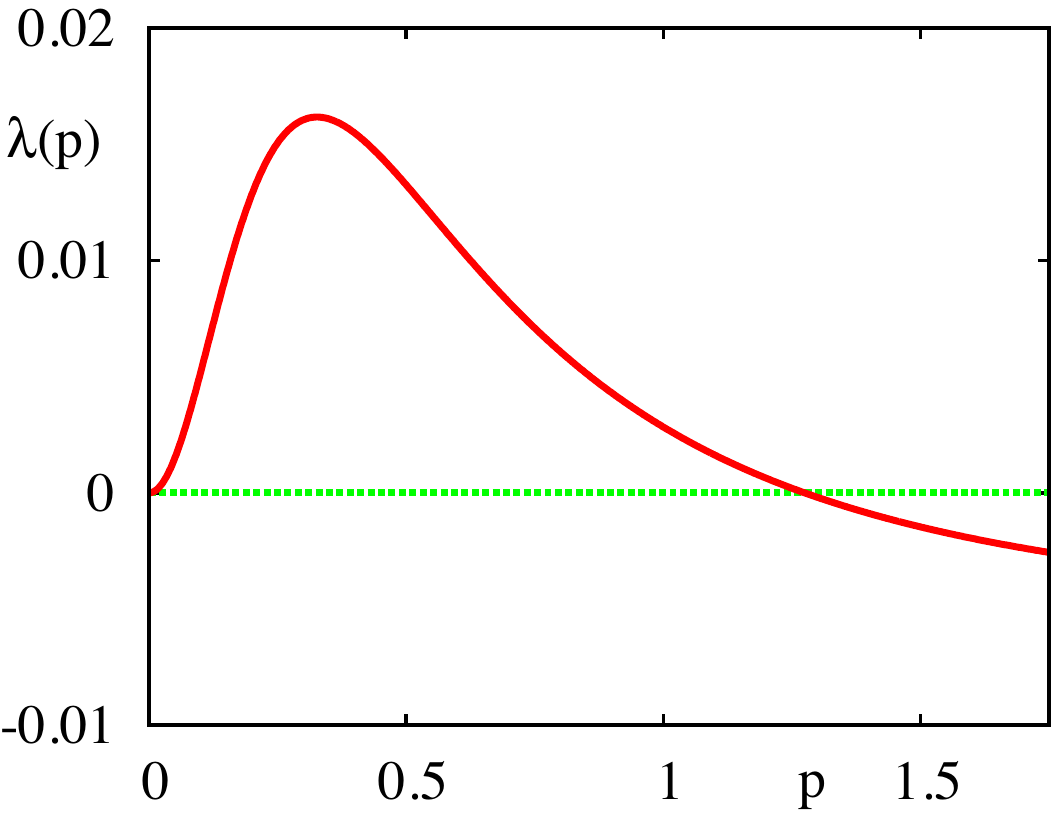}
\caption{Spectrum for the phase oscillator continuum model (\ref{sumcontinuum}) with a mixture of spatial scales and nonlinearities.  Here $H_1(\theta)=H_2(\theta)=\sin(\theta + \alpha)$, $H_3=\sin(2\theta + \alpha)$ and $W_\mu(x)=\gamma_\mu \exp(-\gamma_\mu|x|)/2$ with $\gamma_2=\gamma_3$.  Parameters are $\epsilon_1=0.5$, $\epsilon_2=0.15$, $\epsilon_3=-0.3$, $\gamma_1=1/2$, $\gamma_2=1/4$, and $\alpha=-1.45$.  There is a band of unstable wave-numbers with $p\in (0,p_c)$, with $p_c \simeq 1.25$.
\label{fig:BattogtokhSpectrum}
}
\end{center}
\end{figure}

\begin{figure}[htbp]
\begin{center}
\includegraphics[width=4in]{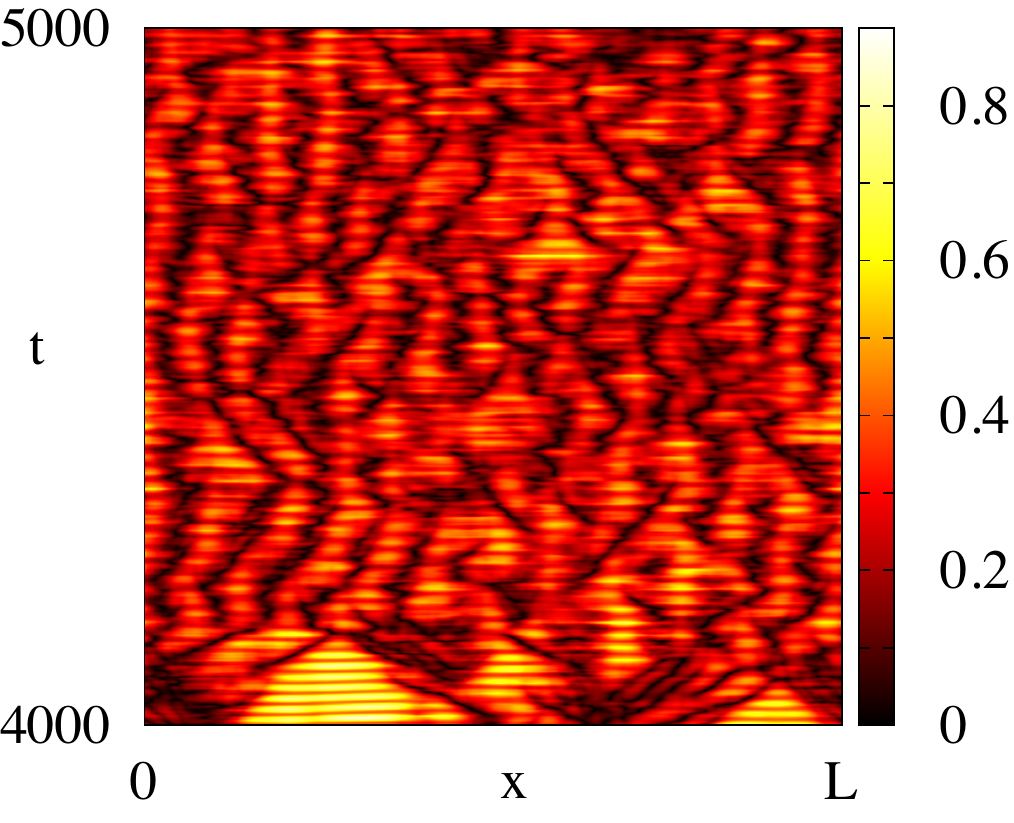}
\caption{
The emergence of a turbulent phase state in a phase oscillator continuum model.  The parameters are those as in Fig.~\ref{fig:BattogtokhSpectrum} with $\omega=0$ for which $\alpha=0.63$, $\beta=5.16$ and $\gamma=0.096$.
The physical domain size is $2^7$ and we have used a numerical mesh of $2^{12}$ points with Matlab {\tt ode45} to evolve equation (\ref{zt}) with convolutions computed using fast Fourier transforms.  As an order parameter describing the system we have chosen
$|\Psi|$, where $\Psi = (\epsilon_1 W_1+\epsilon_2 W_2 )\otimes z + \epsilon_3 W_3 \otimes z^2$.
\label{fig:Battogtokh}
}
\end{center}
\end{figure}

For appropriate choice of the anatomical kernel and the phase interaction function, continuum models of the form (\ref{phasecontinuum}) can also support weak turbulent solutions reminiscent of those seen in the Kuramoto-Sivashkinsky (KS) equation.  The KS equation generically describes the dynamics near long wavelength primary instabilities in the presence of appropriate (translational, parity and Galilean) symmetries, and is of the form
\begin{equation}
\theta_t = -\alpha \theta_{xx}+\beta (\theta_x)^2 -\gamma\theta_{xxxx} ,
\label{KS}
\end{equation}
where $\alpha,\beta,\gamma>0$.
For a further discussion of this model see \cite{Pikovsky01}.
For the model (\ref{phasecontinuum}) with decaying excitatory coupling excitation and purely sinusoidal phase coupling, simulations on a large domain show a marked tendency to generate phase-slips and spatiotemporal pattern shedding, resulting in a loss of spatial continuity of $\theta(x,t)$.  However, Battogtokh \cite{Battogtokh2002} has shown that a mixture of excitation and inhibition with higher harmonics in the phase interaction can counteract this tendency and allow the formation of turbulent states.  To see how this can arise consider an extension of (\ref{phasecontinuum}) to allow for a mixing of spatial scales and nonlinearities in the form
\begin{equation}
\PD{}{t} \theta(x,t) = \omega + \sum_{\mu=1}^M \epsilon_\mu \int_{\RSet} W_\mu(x-y) H_\mu(\theta(y,t)-\theta(x,t)) \d y ,
\label{sumcontinuum}
\end{equation}
where we drop the consideration of axonal delays.  Using the analysis of \S~\ref{sec:phasewave} the synchronous wave solution will be stable provided $\lambda(p) < 0$ for all $p \neq 0$ where
\begin{equation}
\lambda(p) = \sum_{\mu=1}^M \epsilon_\mu H_\mu'(0) \left [ \widehat{W}_\mu(p)-\widehat{W}_\mu(0) \right ], \qquad \widehat{W}_\mu(p) = \int_{\RSet} W_\mu(|y|) \e^{i p y} \d y .
\nonumber
\end{equation}
After introducing the complex function $z(x,t) = \exp(i \theta(x,t))$ and writing the phase interaction functions as Fourier series of the form $H_\mu(\theta) = \sum_n H_n^\mu \e^{i n \theta}$ then we can re-write (\ref{sumcontinuum}) as
\begin{equation}
z_t = i z \left \{ \omega + \sum_{\mu=1}^M \sum_{n \in \ZSet} \epsilon_\mu H_n^\mu z^{-n} \psi_n^\mu 
\right \}, \label{zt} 
\end{equation}
where
\begin{equation}
\psi_n^\mu(x,t) = \int_{\RSet} W_\mu(x-y) z^n (y,t) \d y \equiv W_\mu \otimes z^n . \nonumber
\end{equation}
The form above is useful for computational purposes, since $\psi_n^\mu$ can easily be computed using a fast Fourier transform (exploiting its convolution structure).
Battogtokh  \cite{Battogtokh2002} considered the choice $M=3$ with $H_1(\theta)=H_2(\theta)=\sin(\theta + \alpha)$, $H_3=\sin(2\theta + \alpha)$ and $W_\mu(x)=\gamma_\mu \exp(-\gamma_\mu|x|)/2$ with $\gamma_2=\gamma_3$.  In this case $\widehat{W}_\mu(p) = \gamma_\mu^2/(\gamma_\mu^2+p^2)$, so that
\begin{equation}
\lambda(p) =  -p^2 \cos(\alpha) \left (\frac{\epsilon_1}{\gamma_1^2+p^2}+\frac{\epsilon_2+2 \epsilon_3}{\gamma_2^2+p^2} \right ).
\nonumber
\end{equation}
By choosing a mixture of positive and negative coupling strengths the spectrum can easily show a band of unstable wave-numbers from zero up to some maximum as shown in Fig.~\ref{fig:BattogtokhSpectrum}.
Indeed this shape of spectrum is guaranteed when $\sum_\mu \epsilon_\mu H_\mu'(0)>0$ and $\sum_\mu \epsilon_\mu H_\mu'(0)/\gamma_\mu^2<0$.
Similarly the KS equation (\ref{KS}) has a band of unstable wave-numbers between zero and one (with the most unstable wave-number at $1/\sqrt{2}$).  For the case that all the spatial scales $\gamma_\mu^{-1}$ are small compared to the system size then we may develop a \textit{long wavelength} argument to develop local models for $\psi_n^\mu$.  To explain this we first construct the Fourier transform $\widehat{\psi}_n^\mu(p,t) = \widehat{W}_\mu(p) f_n(p,t)$, where $f_n$ is the Fourier transform of $z^n$  and use the expansion $\widehat{W}_\mu(p) \simeq 1-\gamma_\mu^{-2} p^2 +\gamma_\mu^{-4} p^4 + \ldots$.  After inverse Fourier transforming we find
\begin{equation}
\psi_n^\mu \simeq \left [1 + \gamma_\mu^{-2}\partial_{xx} - \gamma_\mu^{-4}\partial_{xxxx} + \ldots \right ] z^n .
\nonumber
\end{equation}
Noting that $H^1_1=H_2^1=H_3^2=\exp(i \alpha)/(2i) \equiv \Gamma$ with all other Fourier coefficients zero then (\ref{zt}) yields
\begin{align}
\theta_t &= \Omega + 2 \text{Re} \, \Gamma \sum_{\mu=1,2} \epsilon_\mu \e^{-i \theta}
\left (\gamma_\mu^{-2}\partial_{xx} - \gamma_\mu^{-4}\partial_{xxxx} + \ldots \right ) 	\e^{i \theta} \nonumber \\
&+ \epsilon_3 \e^{-2 i \theta} \left (\gamma_3^{-2}\partial_{xx} - \gamma_3^{-4}\partial_{xxxx} + \ldots \right ) \e^{2 i \theta} ,
\label{thetat}
\end{align}
where $\Omega=\omega + \sum_{\mu} \epsilon_\mu H_\mu(0)$.
Expanding (\ref{thetat}) to second order gives $\theta_t = \Omega -\alpha \theta_{xx} +\beta (\theta_x)^2$, where
$\alpha=-\sum_\mu \epsilon_\mu H_\mu'(0) \gamma_\mu^{-2}$ and $\beta=-\sum_\mu \epsilon_\mu H_\mu''(0) \gamma_\mu^{-2}$.  Going to higher order yields fourth order terms and we recover an equation of KS type with the coefficient of $-\theta_{xxxx}$ given by $\gamma=\sum_\mu \epsilon_\mu H_\mu'(0) \gamma_\mu^{-4}$.  To generate phase turbulence we thus require $\alpha>0$, which is also a condition required to generate a band of unstable wave-numbers, and $\beta,\gamma>0$.  A direct simulation of the model with the parameters for Fig.~\ref{fig:BattogtokhSpectrum}, for which  $\alpha,\beta,\gamma>0$, shows the development of a phase turbulent state.  This is represented in Fig.~\ref{fig:Battogtokh} where we plot the absolute value of the complex function $\Psi = (\epsilon_1 W_1+\epsilon_2 W_2 )\otimes z + \epsilon_3 W_3 \otimes z^2$.

\section{Heteroclinic attractors}
\label{sec:heteroclinic}

In addition to dynamically simple periodic attractors with varying degrees of clustering, the emergent dynamics of coupled systems such as (\ref{phasenetwork}) can be remarkably complex even in the case of global coupling. In this case, the dynamical complexity depends only on the phase interaction function $H$ and the number of oscillators $N$. Chaotic dynamics \cite{Bick2011} can appear in four or more globally coupled phase oscillators for phase interaction functions of sufficient complexity. We focus now on attractors that are robust and apparently prevalent in many such systems:  robust heteroclinic attractors.

In a neuroscience context such attractors have been investigated under several related names, including {\em slow switching} \cite{Hanseletal1993,KoriKuramoto2001,KoriKuramoto2003,Kiss2007} where the system evolves towards an attractor that displays slow switching between cluster states  where the switching is on a timescale determined by the noise, heteroclinic networks \cite{Aguiaretal2011,Ashwin2008,AshPos2013} or {\em winnerless competition} \cite{Rabinovich2006a,Rabinovich2010} where there are a number of possible patterns of activity that compete with each other but such that each pattern is unstable to some perturbations that take it to another pattern - this can be contrasted to {\em winner-takes-all competition} where there is attraction to a asymptotically stable pattern.

\begin{figure}[htbp]
\begin{center}
\includegraphics[width=3.5in]{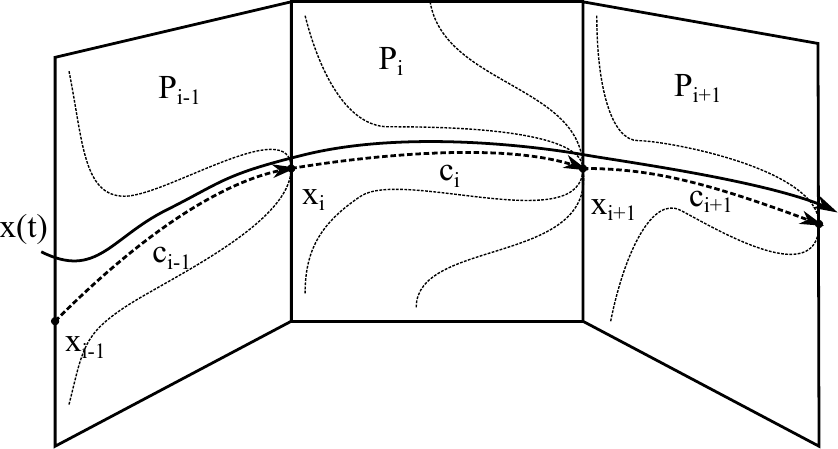}
\caption{
Schematic diagram showing a trajectory $x(t)$ (solid line) approaching part of a robust heteroclinic network (bold dashed lines). The nodes $x_i$ represent equilibria or periodic orbits of saddle type and the invariant subspaces $P_i$ are forced to exist by model assumptions and there are connecting (heteroclinic) orbits $c_i$ that connect the nodes within the $P_i$ in a robust manner. A neighbourhood of the connecting orbits $c_i$ is an absorbing {\em stable heteroclinic channel} that can be used to describe various aspects of neural system function in systems with this dynamics; see for example \cite{Rabinovich2012}.
\label{Fig:SHC}
}
\end{center}
\end{figure}

These attractors obtain their dynamical structure from the presence of invariant subspaces for the dynamics that allow the possibility of robust connections between saddles. These subspaces may be symmetry-induced fixed point subspaces, spaces with a given clustering, subspaces forced by multiplicative coupling or subspaces forced by assumptions on the nature of the coupling. In all cases, there will be a number of dynamically simple nodes, usually equilibria or periodic orbits, say $x_i$, $i=1,\ldots,k$ each of which has saddle type. These nodes have unstable manifolds that, within some invariant subspace, limit to other nodes within the network - usually because there is a robust (transverse) saddle-to-sink connection between the nodes within some invariant subspace; see \cite{AshKar2011}. It can be verified that such heteroclinic networks can be attracting if, in some sense, the rate of expansion away from the nodes is not as strong as the rate of contraction towards the nodes - see \cite{KrupaMelbourne2004} for some precise results in this direction. Figure~\ref{Fig:SHC} illustrates some of the ingredients for a robust heteroclinic attractor.

\subsection{Robust heteroclinic attractors for phase oscillator networks}

Hansel \textit{et al}. \cite{Hanseletal1993} considered the dynamics of (\ref{phasenetwork}) with global coupling and phase interaction function of the form (\ref{eq:HMMcoupling}) for $(r,\alpha)$ fixed parameters. For large $N$, they find an open region in parameter space where typical attractors are heteroclinic cycles that show slow switching between states where the clustering is into two clusters of macroscopic size. This dynamics is examined in more depth in \cite{KoriKuramoto2001} where the simulations for typical initial conditions show a long and intermittent transient to a two-cluster state that, surprisingly, is unstable. This is a paradox because only low dimensional subset of initial conditions (the stable manifold) should converge to a saddle. The resolution of this paradox is a numerical effect: as the dynamics approaches the heteroclinic cycle where the connection is in a clustered subspace, there can be numerical rounding into the subspace. For weak perturbation of the system by additive noise, they find that the heteroclinic cycle is approximated by a roughly periodic transition around the cycle whose approximate period scales as the logarithm of the noise amplitude.

\begin{figure}[htbp]
\begin{center}
\includegraphics[width=4in,clip=]{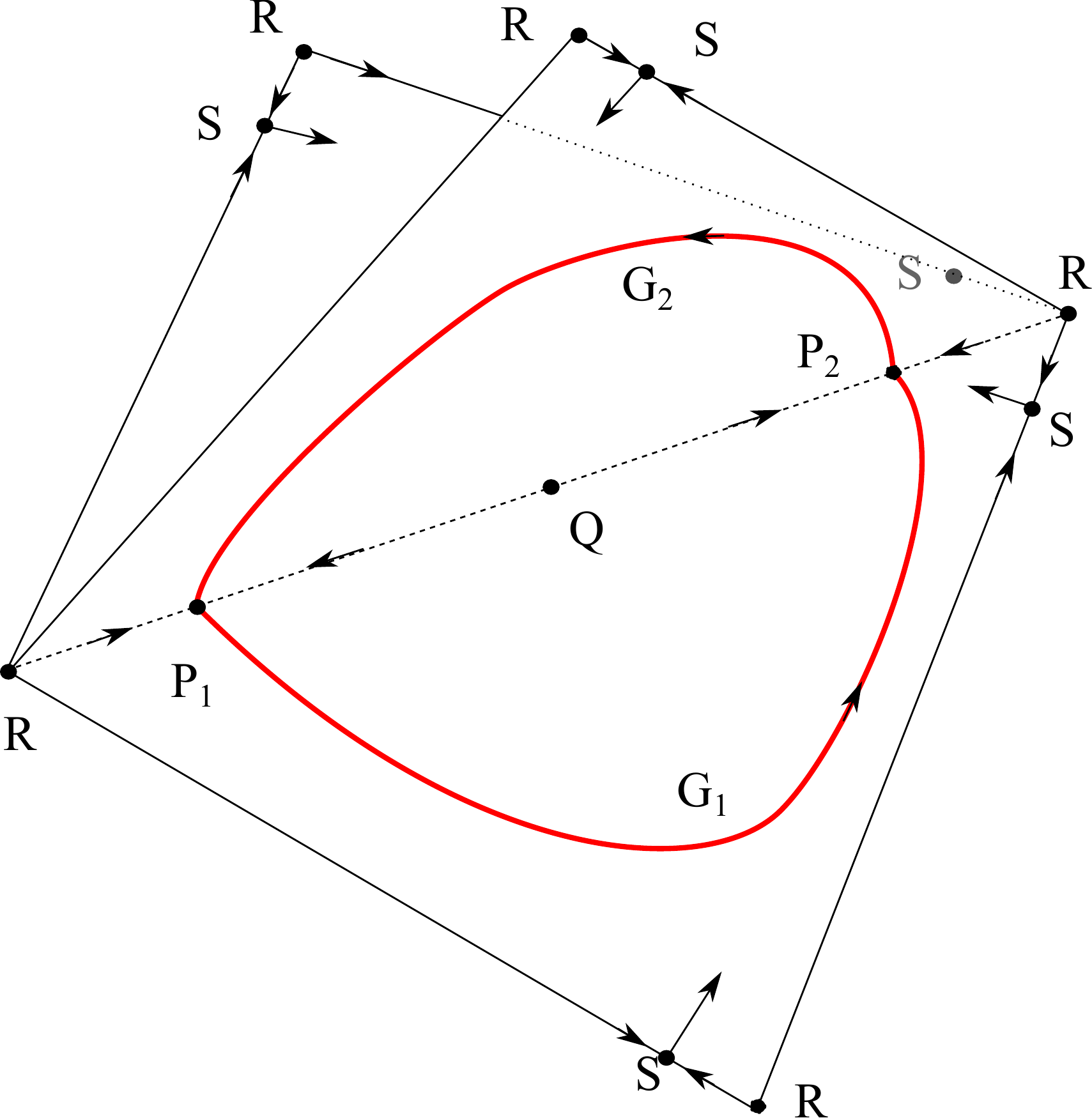}
\caption{
A robust heteroclinic cycle attractor for the all-to-all coupled $4$-oscillator system (\ref{phasenetwork}) with phase interaction function (\ref{eq:HMMcoupling}) and an open region of parameter space, as in \cite{Ashwin2008}. The figure shows the cycle in the 3 dimensional space of phase differences; the vertices $R$ all represent the fully symmetric (in-phase) oscillations $(\varphi,\varphi,\varphi,\varphi)$, varying by $2\pi$ in one of the arguments. The point $Q$ represents solutions $(\varphi,\varphi,\varphi+\pi,\varphi+\pi)$ with symmetry $(S_2\times S_2)\times_s \Z_2$. The heteroclinic cycle links two saddle equilibria $P_1=(\varphi,\varphi,\varphi+\alpha,\varphi+\alpha)$ and $P_2=(\varphi,\varphi,\varphi+2\pi-\alpha,\varphi+2\pi-\alpha)$ with $S_2\times S_2$ isotropy. The robust connections $G_1$ and $G_2$ shown in red lie within two dimensional invariant subspaces with isotropy $S_2$ while the equilibria $S$ have isotropy $S_3$. This structure is an attractor for parameters in the region indicated in Figure~\ref{Fig:fouroscbifs}.
\label{Fig:fouroschet}
}
\end{center}
\end{figure}

\begin{figure}[htbp]
\begin{center}
\includegraphics[width=4.5in]{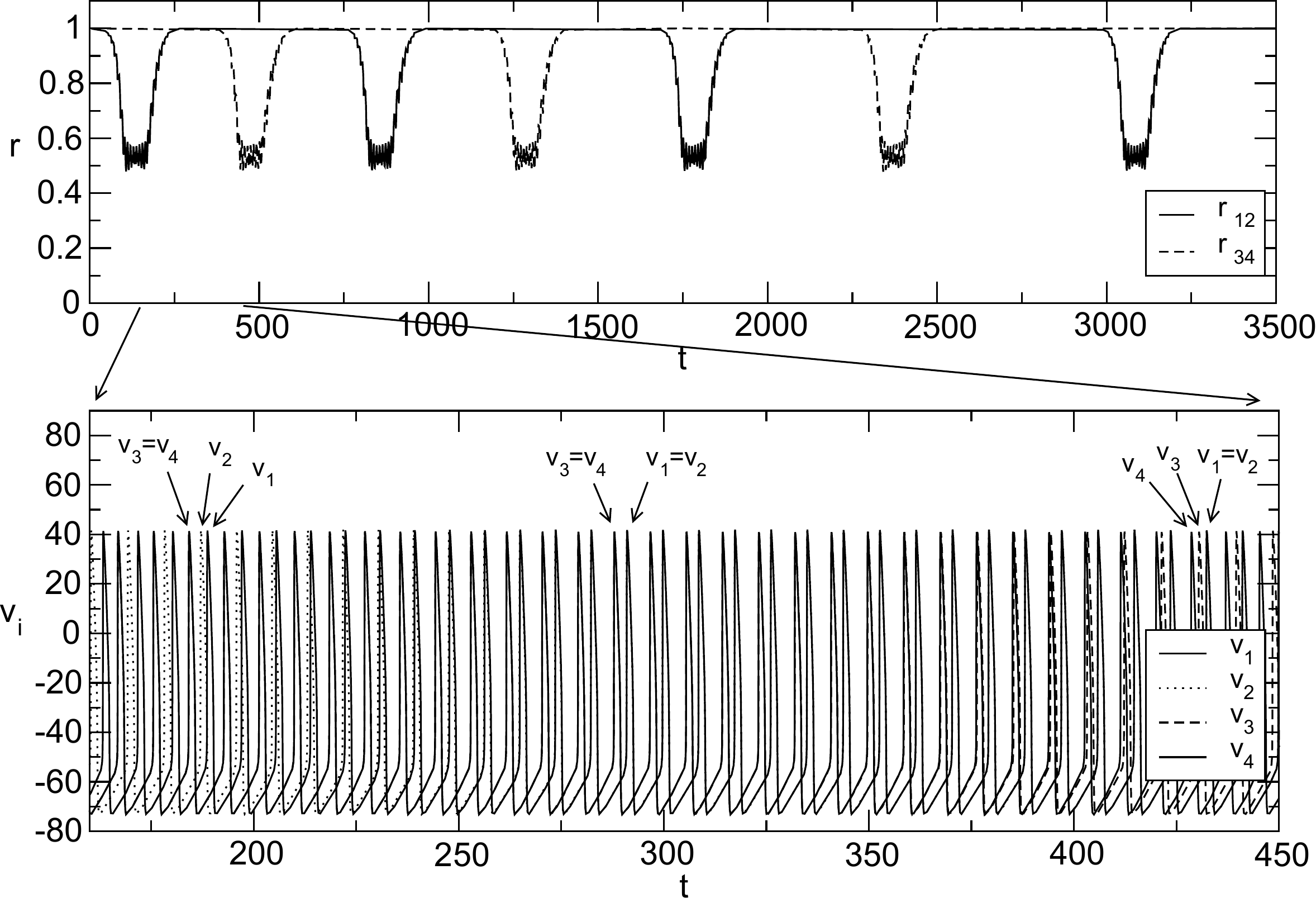}
\caption{
Robust heteroclinic cycle attractor for all-to-all coupled oscillatory Hodgkin-Huxley systems with delay synaptic coupling. The top panel shows synchronisation indices $r_{ij}$ that only equal one when the systems $i$ and $j$ are synchronised while the bottom panel shows the action potentials $v_i$ for the four oscillators; see \cite{AshKar2011} for more details. Observe that the mechanism of symmetry breaking and re-synchronisation of pairs is the same as in Fig.~\ref{Fig:fouroschet}.
\label{Fig:22cycle}
}
\end{center}
\end{figure}

The bifurcations that give rise to heteroclinic attractors in this system on varying $(r, \alpha)$ is quite complex even for small $N$. As discussed in \cite{Ashwin2008} one can only find attracting robust heteroclinic attractors in (\ref{phasenetwork}), (\ref{eq:HMMcoupling}) for $N\geq 4$: in this case Fig.~\ref{Fig:fouroscbifs} shows a region where robust heteroclinic attractors of the type illustrated in Fig.~\ref{Fig:fouroschet} appear. A trajectory approaching such a network will spend much of its time near a cluster state with two groups of two oscillators each. Each time there is a connection between the states, one of the groups will break clustering for a short time, and over a longer period the clusters will alternate between breaking up and keeping together. Qualitatively similar heteroclinic attractors can be found for example in coupled Hodgkin-Huxley type limit cycle oscillators with delayed synaptic coupling as detailed in \cite{AshKar2011} and illustrated in Fig.~\ref{Fig:22cycle}.

\begin{figure}[htbp]
\begin{center}
\includegraphics[width=4in]{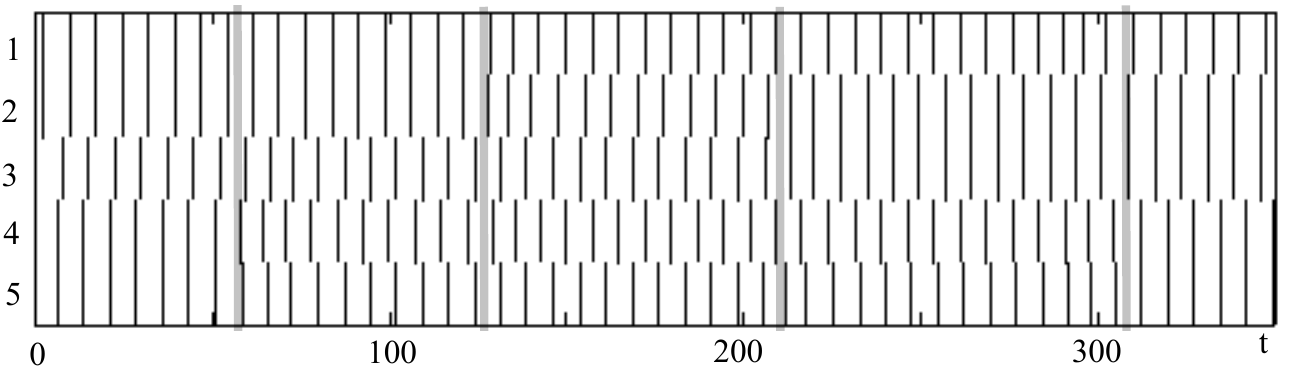}
\caption{
Raster plot showing a robust heteroclinic attractor in a system of five globally coupled phase oscillators (\ref{phasenetwork}) with phase interaction function (\ref{eq:GHMMcoupling}) and a particular choice of parameters (see text). The vertical dark lines mark times ($t$ in horizontal axis) when the oscillator represented by $\theta_k(t)$ ($k$ in vertical axis) passes through zero. Observe that most of the time there are three clusters. Occasionally the clustering splits and reforms different three clusters reforms, at times indicated by the grey bars, approximately every 70 time units. [Adapted from \cite{Ashwin07}]
\label{fig:fiveoschetraster}
}
\end{center}
\end{figure}

\begin{figure}[htbp]
\begin{center}
\includegraphics[width=3.8in]{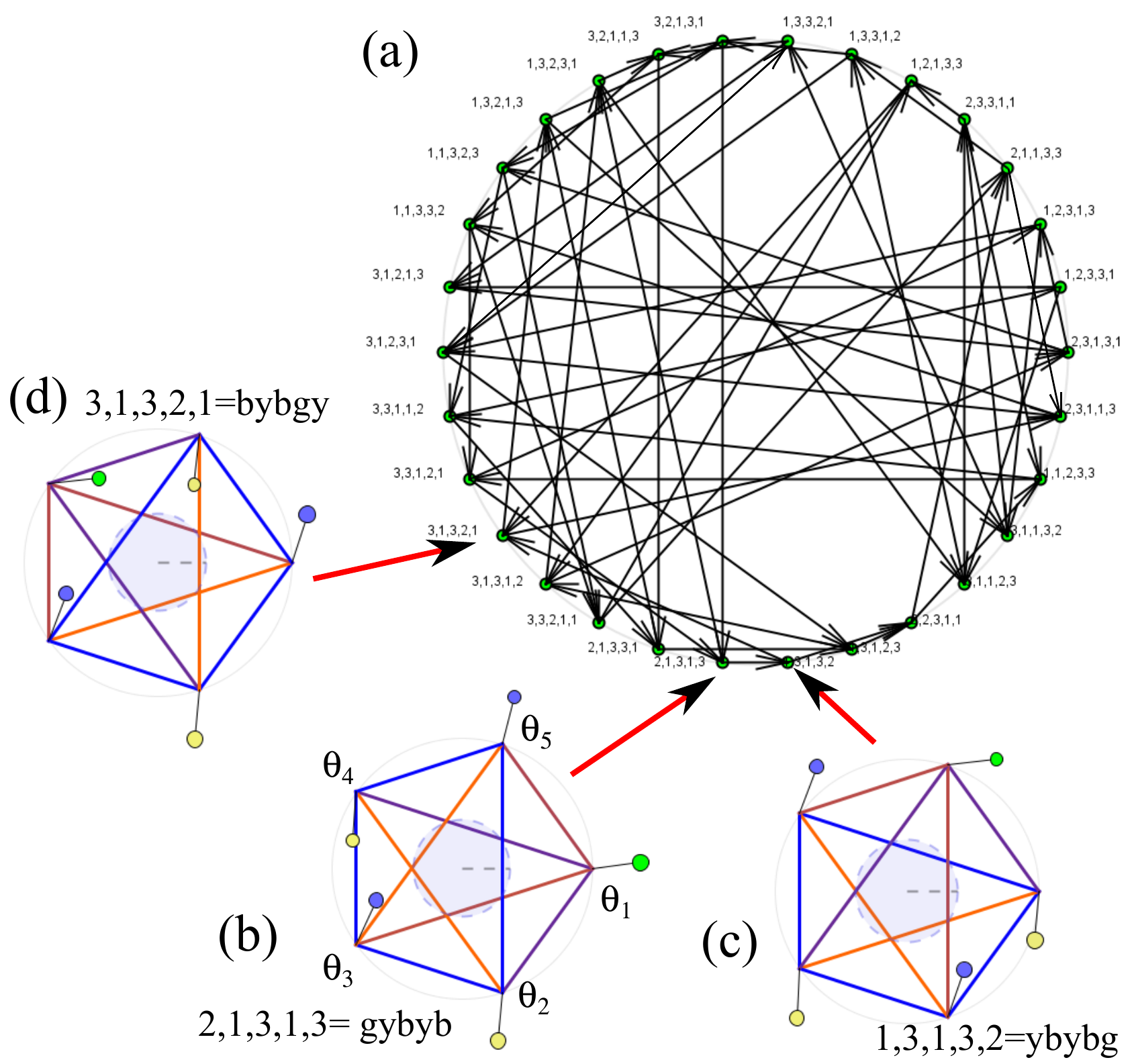}
\caption{
(a) Graph of heteroclinic connections between three cluster states for a robust heteroclinic attractor in a system of N=5 globally coupled phase oscillators. Phase interaction function and parameters as in Fig.~\ref{fig:fiveoschetraster}; see \cite{Ashwin07} for more details of the dynamics.  Each vertex represents a particular saddle periodic cluster state that is a permutation of the states shown in (b-d). Note that there are precisely two incoming and two outgoing connections at each vertex.
(b-d) show the relative phases of the five oscillators, indicated by the angles of the ``pendula" at the vertices of a regular pentagon, for a sequence of three consecutive three saddle cluster states visited on a longer trajectory that randomly explores graph (a) in the presence of noise. Inequivalent clusters are characterised by different coloured ``pendula" and numbers where yellow  corresponds to 1, green to 2 and blue to 3. The yellow cluster is stable to cluster-splitting perturbations while the blue cluster is unstable to such perturbations - observe that after the connection the yellow cluster becomes the blue cluster while the blue cluster splits up in one of two possible way.
\label{Fig:fiveoschetnetwork}}
\end{center}
\end{figure}

The heteroclinic attractors that appear for $N>4$ can have more complex structures. For $N=5$ this is investigated in \cite{AshBor2004,AshBor2005} for (\ref{phasenetwork}), (\ref{eq:HMMcoupling}) and in \cite{Ashwin07} for more general phase interaction function
\begin{equation}
H(\varphi)=-\sin(\varphi+\alpha)+r\sin (2\varphi+\beta) ,
\label{eq:GHMMcoupling}
\end{equation}
where $\alpha$, $\beta$ and $r$ are parameters. Figure~\ref{fig:fiveoschetraster} illustrates a trajectory of (\ref{phasenetwork}) with global coupling and phase interaction function (\ref{eq:GHMMcoupling}) as a raster plot for five phase oscillators with parameters $r=0.2$, $\alpha=1.8$, $\beta=-2.0$ and $\omega=1.0$ and with addition of noise of amplitude $10^{-5}$.  Observe there is a dynamic change of clustering over the course of the time-series with a number of switches taking place between cluster states of the type
$$
(\theta_1,\cdots,\theta_5)=\Omega t (1,1,1,1,1)+(y,y,g,b,b)
$$
where $y$, $g$ and $b$ represent different relative phases that are coloured ``yellow'', ``green'' or ``blue'' in Fig.~\ref{Fig:fiveoschetnetwork}
to other symmetrically related cluster states. One can check that the group orbit of states with this clustering gives 30 symmetrically related cluster states for the system; details of the connections are shown in Fig.~\ref{Fig:fiveoschetnetwork}. All cluster states connect together to form a single large heteroclinic network that is an attractor for typical initial conditions \cite{Ashwin07}. Fig.~\ref{fig:fiveoschetraster} illustrates the possible switchings between phase differences near one particular cluster state for the heteroclinic network in this case. The dynamics of this system can be used for encoding a variety of inputs, as discussed in \cite{WorAsh2008} where it is shown that a constant heterogeneity of the natural frequencies between oscillators in this system leads to a spatio-temporal encoding of the heterogeneities. The sequences produced by the system can be used by a similar system with adaptive dynamics to learn a spatio-temporal encoded state \cite{Oroszetal2009}.

Robust heteroclinic attractors also appear in a range of coupled phase oscillator models where the coupling is not global (all-to-all) but such that it still preserves enough invariant subspaces for the connections to remain robust. For example, \cite{Karabacak2009} study the dynamics of a network ``motif'' of four coupled phase oscillators and find heteroclinic attractors that are ``ratchets'', i.e. they are robust heteroclinic networks that wind preferentially around the phase space in one direction - this means that under the influence of small perturbations, phase slips in only one direction can appear.

\subsection{Winnerless competition and other types of heteroclinic attractor}

Heteroclinic connections play an key role in describing winnerless competition dynamics  \cite{Rabinovich2006a,Rabinovich2010}. This dynamics is usually associated with firing rate model of coupled neurons of ``Lotka-Volterra'' form
\begin{equation}
\frac{\d}{\d t} {x}_i = x_i (\lambda_i + \sum_{j=1}^N A_{ij} x_j) ,
\label{eq:lotkavolterra}
\end{equation}
where $x_i(t)$ is an instantaneous firing rate of a neuron or group of neurons, $\lambda_i$ is a self-excitation term and $A_{ij}$ represents coupling between the $x_i$, though it is also possible to consider generalisations \cite{Rabinovich2014}. These models were originally developed for predator-prey interaction of species in an ecosystem. Having found wide use in population ecology, they have more recently be applied to evolutionary game dynamics \cite{Hofbauer2003}, including applications in economics. Since the seminal paper of May \cite{May1975} it has been recognised that (\ref{eq:lotkavolterra}) can have robust heteroclinic attractors for an open set of parameter choices $\lambda_i$ and $A_{ij}$, as long as at least three species are involved. Indeed, the system can show a wide range of dynamics such as ``winner-takes-all'' equilibria where there is a stable equilibrium with $x_i>0$ and $x_j=0$ for $j\neq i$ ($i$ is the ``winner''), ``cooperative'' equilibria where several of the $x_i$ are non-zero as well as non-trivial periodic or chaotic dynamics. The particular case of ``winnerless competition'' gives attractors consist of chains  or networks of saddles joined by connections that are robust because the absence of a species (or in our case, the lack of firing of one neuron or group of neurons) $x_i=0$ is preserved by the dynamics of the model (\ref{eq:lotkavolterra}).

The simplest case of this appears in a ring of $N=3$ coupled neurons with dynamics
\begin{align*}
\frac{\d}{\d t} {x}_1 & = x_1(1 -x_1 -\alpha x_2-\beta x_3 ),\\
\frac{\d}{\d t} {x}_2 & = x_2(1 -x_2 -\alpha x_3-\beta x_1 ),\\
\frac{\d}{\d t} {x}_3 & = x_3(1 -x_3 -\alpha x_1-\beta x_2 ),
\end{align*}
for $\alpha+\beta>2$ and $0<\alpha<1$ \cite{May1975}, corresponding to cyclic inhibition of the neurons in one direction around the ring and cyclic excitation in the other direction. This ``rock-paper-scissors'' type of dynamics leads to winnerless competition that has been applied in a variety of more complex models of neural systems. The local behaviour near connections in such heteroclinic attractors has been called ``stable heteroclinic channels'' \cite{Rabinovich2012} and used to model a variety of low-level and high-level neural behaviours including random sequence generation, information dynamics, encoding of odours and working memory. We refer to the reviews \cite{Rabinovich2006,Rabinovich2012,Rabinovich2014}.

Analogous behaviour has been found in a range of other coupled systems, for example \cite{Aguiaretal2011} or delayed pulse-coupled oscillators \cite{KoriKuramoto2003,Timme2003,AshTim2005,Kirst2008,Broer2008,NevesTimme2012}. Recent work has also considered an explicit constructive approach to heteroclinic networks to realise arbitrary directed networks as a heteroclinic attractor of a coupled cell system \cite{AshPos2013}.

More complex but related dynamical behaviour has been studied under the names of ``chaotic itinerancy'' (see for example \cite{KanekoTsuda2003}), ``cycling chaos'' \cite{Dellnitzetal1995}, ``networks of Milnor attractors'' \cite{Kaneko1998322} and ``heteroclinic cycles for chaotic oscillators'' \cite{KuznetsovKurths2002}. It has been is suggested that these and similar models are useful for modelling of the functional behaviour of neural systems \cite{Tsuda2009}.

Because heteroclinic attractors are quite singular in their dynamical behaviour (averages of observables need not converge, there is a great deal of sensitivity of the long-term dynamics to noise and system heterogeneity), it is important to consider the effect of noise and/or heterogeneities in the dynamics. This leads to a finite average transition time between states determined by the level of noise and/or heterogeneity (which may be due to inputs to the system) and the local dynamics - see for example \cite{StoneHolmes1990}. Another useful feature of heteroclinic attractors is that they allow one to model ``input-output'' response of the system to a variety of inputs.

\section{Stochastic oscillator models}
\label{sec:stochastic}

Noise is well known to play a constructive role in the neural encoding of natural stimuli, leading to increased reliability or regularity of neuronal firing in single neurons \cite{Mainen95,Taillefumier2014} and across populations \cite{Ermentrout2008a}.
From a mathematical perspective it is natural to consider how noise may affect the reduction to a phase-oscillator description.  Naively one may simply consider the addition of noise to a deterministic phase-oscillator model to generate a stochastic differential equation. Indeed models of this type have been studied extensively at the network level to understand noise-induced first- and second-order phase transitions, and new phenomenon such as noise-induced synchrony \cite{Goldobin2005,Nakao2007,Lai2013} or asynchrony \cite{Kawamura2014}, and noise-induced turbulence \cite{Kawamura2007}.  We refer the reader to the review by Lindner \cite{Lindner2004} for a comprehensive discussion.  More recently Schwabedal and Pikovsky have extended the foundations of deterministic phase descriptions to irregular, noisy oscillators (based on the constancy of the mean first return times) \cite{Schwabedal2013}, Ly and Ermentrout \cite{Ly2011} and Nakao \textit{et al}. \cite{Nakao2010} have built analytical techniques for studying weak noise forcing, and Moehlis has developed techniques to understand the effect of white noise on the period of an oscillator \cite{Moehlis2014}.

At the network level (global coupling) a classic paper examining the role of external noise in IF populations, using a phase description, is that of Kuramoto \cite{Kuramoto91}, who analysed the onset of collective oscillations.
Without recourse to a phase-reduction it is well to mention that Medvedev has been pioneering a phase-amplitude approach to studying the effects of noise on the synchronisation of coupled stochastic limit cycle oscillators \cite{Medvedev2010,Medvedev2012}, and that Newhall \textit{et al}. have developed a Fokker-Planck approach to understanding cascade-induced synchrony in stochastically driven IF networks with pulsatile coupling and Poisson spike-train external driven \cite{Newhall2010}.  More recent work on pairwise synchrony in network of heterogeneous coupled noisy phase oscillators receiving correlated and independent noise can be found in \cite{Ly2014}.  However, note that even in the absence of synaptic coupling, two or more neural oscillators may become synchronised by virtue of the statistical correlations in their noisy input streams \cite{Abouzeid2011,Bressloff2011,Burton2012}.

\subsection{Phase reduction of a planar system with state-dependent Gaussian white noise}

For clarity of exposition let us consider the phase reduction of a planar system described by $\ID{x}=F(x)+p(x) \xi(t)$, where $\xi(t)$ is white Gaussian noise such that $\langle \xi(t) \rangle=0$ and
$\langle \xi(t) \xi(s) \rangle = 2D \delta(t-s)$, where $\langle \cdot \rangle$ denotes averaging over the realisation of $\xi$, and $D>0$ scales the noise intensity.
We employ an Stratonovich interpretation of the noise (such that the chain rule of ordinary calculus holds).  In the absence of noise we shall assume that the system has a stable $T$-periodic limit-cycle solution, with a phase that satisfies $\d{\vartheta}/{\d t} = 1$.  For weak noise perturbations the state point will leave the cycle, though will stay in some neighbourhood, which we denote by $U$.  To extend the notion of a phase off the cycle we let $\vartheta$ be a smooth function of $x$ such that $\langle \nabla_x \vartheta , F(x) \rangle=1$ holds for any point $x \in \mathcal{U}$.  We shall denote the other coordinate in $\mathcal{U}$ by $\rho$, and assume that there exists a smooth coordinate transformation $x \rightarrow (\vartheta, \rho)$.  For a noise perturbed oscillator, in the new coordinates we have
\begin{equation}
\frac{\d}{\d t} {\vartheta} = 1 + h(\vartheta,\rho) \xi(t), \qquad \frac{\d}{\d t} {\rho}= f(\vartheta,\rho)+ g(\vartheta,\rho) \xi(t) ,
\label{thetarho}
\end{equation}
where we have introduced $h=\langle \nabla_x \vartheta , p \rangle$, $f=\langle \nabla_x \rho , F \rangle$ and $g=\langle \nabla_x \rho , p \rangle$.
Note that the full coordinate transformation $(\vartheta,\rho)=(\vartheta(x),\rho(x))$ is not prescribed here, though it is common to use one such that $\rho$ can be interpreted as some distance from cycle.  Thus although results may be formally developed for the system (\ref{thetarho}) they cannot be directly interpreted in terms of a given model until the full coordinate transformation taking one from $x \in\RSet^2$ to $(\vartheta,\rho)$ is given.

One can transform (\ref{thetarho}) into a stochastic phase-amplitude equation in the It\={o} sense, where it reads
\begin{align}
\frac{\d}{\d t} {\vartheta} &= 1 + D[h_\vartheta(\vartheta,\rho) h(\vartheta,\rho)+ h_\rho(\vartheta,\rho) g(\vartheta,r) ]+ h(\vartheta,\rho) \xi(t), \nonumber \\
\frac{\d}{\d t} {\rho} &= f(\vartheta,\rho) + D[g_\vartheta(\vartheta,\rho) h(\vartheta,\rho)+ g_\rho(\vartheta,\rho) g(\vartheta,r) ] +g(\vartheta,\rho)\xi(t) ,
\nonumber
\end{align}
where the subscripts $\vartheta$ and $\rho$ denote partial derivatives with respect to $\vartheta$ and $\rho$ respectively.
Using the It\={o} form we may construct a Fokker-Planck equation for the time-dependent probability distribution $Q(t,\vartheta,\rho)$ as
\begin{align}
\PD{}{t} Q &= -\PD{}{\vartheta} \left [
\{
1 + D(h_\vartheta h +h_\rho g)
\}Q
\right ]
+ D \PD{^2 [h^2 Q]}{\vartheta^2} \nonumber \\
&-\PD{}{\rho} \left [
\{
f + D(g_\vartheta h +g_\rho g)
\}Q
\right ]
+ 2 D \PD{^2 [h g Q]}{\vartheta \partial \rho}
+D \PD{^2 [g^2 Q]}{\rho^2} ,
\label{FP}
\end{align}
with periodic boundary condition $Q(t,0,\rho)=Q(t,2 \pi,\rho)$.

When $D=0$ the steady state distribution is given by $Q_0(\vartheta,\rho)=\delta(\rho)/(2 \pi)$.
For small noise amplitude $D$ we expect $Q_0$ to still localise near $\rho=0$ \cite{Yoshimura08}, say over a region $-\rho_c < \rho < \rho_c$.  In this case it is natural to make the approximation that for large $t$,  $Q=0=\partial Q /\partial \rho$ at $\rho=\pm \rho_c$.  Now introduce the marginal distribution
\begin{equation}
P(t,\vartheta) \equiv \int_{-\rho_c}^{\rho_c} Q(t,\vartheta,\rho) \d \rho .
\nonumber
\end{equation}
Following \cite{Yoshimura08} we can integrate (\ref{FP}) over the interval $I=[-\rho_c,\rho_c]$ and generate a Fokker-Planck equation for $P$.
The last three terms in (\ref{FP}) vanish after integration due to the boundary conditions, so that we are left with
\begin{equation}
\PD{}{t} P = -\PD{}{\vartheta} \int_I D (h_\vartheta h + h_\rho g) Q \d \rho + D \PD{^2 }{\vartheta^2} \int_I h^2 Q \d \rho .
\nonumber
\end{equation}
We now expand about $\rho=0$ to give $h(\vartheta,\rho)=Z(\vartheta)+\rho h_\rho(\vartheta,0)+\ldots$ and
$g(\vartheta,\rho)=g(\vartheta,0) +\rho g_\rho(\vartheta,0)+\ldots$, where $Z(\vartheta)$ is identified as the infinitesimal phase response $\left.\langle \nabla_x \vartheta , p \right |_{\rho=0} \rangle$.
In the limit of small $D$ where $Q_0 \simeq \delta(\rho)/(2 \pi)$ we note that, for an arbitrary function $R(\vartheta)$,
\begin{equation}
\lim_{D \rightarrow 0} \PD{^n}{\vartheta^n} \int_I \rho R(\vartheta) Q \d \rho
= \PD{^n}{\vartheta^n}\lim_{D \rightarrow 0} \int_I \rho R(\vartheta) Q \d \rho =0 .
\nonumber
\end{equation}
Making use of the above gives the Fokker-Planck equation as
\begin{equation}
\PD{}{t} P = -\PD{}{\vartheta} \left [
\{ 1 + D (Z Z' + Y )\} P
\right ]
+D \PD{^2[Z^2P]}{\vartheta^2},
\label{FPphase}
\end{equation}
where $Y(\vartheta) = h_\rho(\vartheta,0)g(\vartheta,0)$.
Hence, the corresponding It\={o} equation is
\begin{equation}
\frac{\d}{\d t} {\vartheta} = 1 + D[Z(\vartheta)Z'(\vartheta)+ Y(\vartheta) ]+ Z(\vartheta) \xi(t),
\label{Itophase}
\end{equation}
while the Stratonovich version is
\begin{equation}
\frac{\d}{\d t} {\vartheta} = 1 + DY(\vartheta)+ Z(\vartheta) \xi(t).
\label{phasestrat}
\end{equation}
Equations (\ref{Itophase}) and (\ref{phasestrat}) are the stochastic phase oscillator descriptions for a limit cycle driven by weak white noise.  These make it clear that naively adding noise to the phase description misses  not only a multiplication by the iPRC but also the addition of a further term $Y(\vartheta)$ that contains information about the \textit{amplitude} response of the underlying limit cycle oscillator.

We are now in a position to calculate the steady state probability distribution $P_0(\vartheta)$ and use this to calculate the moments of the phase dynamics.  Consider (\ref{FPphase}) with the boundary condition $P(t,0)=P(t,2 \pi)$, and set $P_t=0$.
Adopting a Fourier representation for $P_0$, $Z$ and $Y$ as $P_0(\vartheta) = \sum_n P_n \e^{in \vartheta}$, $Z(\vartheta) = \sum_n Z_n \e^{in \vartheta}$, $Y(\vartheta) = \sum_n Y_n \e^{in \vartheta}$, allows us to obtain a set of equations for the unknown amplitudes $P_l$ as
\begin{equation}
-P_l + D \sum_{n,m \in \ZSet} Z_n Z_m  i(l-n) P_{l-(n+m)} -D \sum_{n \in \ZSet} Y_n P_{l-n} = K \delta_{l,0}, \qquad l \in \ZSet,
\label{amplitudes}
\end{equation}
for some constant $K$.
For $D=0$ we have that $P_0=K$, and after enforcing normalisation we may set $K=1/(2 \pi)$.  For small $D$ we may then substitute $P_0$ into (\ref{amplitudes}) and work to next order in $D$ to obtain an approximation for the remaining amplitudes,
$l \neq 0$, in the form
\begin{equation}
P_l =\frac{D}{2 \pi} \left \{
\sum_{\{n,m~|n+m=l \}} Z_n Z_m i m -Y_l
\right \} .
\nonumber
\end{equation}
Using this we may reconstruct the distribution $P(\vartheta)$, for small D, from (\ref{Fourier}) as
\begin{equation}
P(\vartheta) = \frac{1}{2\pi} + \frac{D}{2 \pi} \left (
Z(\vartheta) Z'(\vartheta) -Y(\vartheta) +Y_0
\right ), \qquad Y_0 = \frac{1}{2 \pi} \int_0^{2 \pi} Y(\vartheta) \d \vartheta .
\label{P}
\end{equation}

The mean frequency of the oscillator is defined by ${\omega} = \lim_{T \rightarrow \infty} T^{-1} \int_0^T\frac{\d}{\d t} {\vartheta}(t) \d t$.  This can be calculated by replacing the time average with the ensemble average.  For an arbitrary $2 \pi$-periodic function $R$ we set $\lim_{T \rightarrow \infty} T^{-1} \int_0^T R(t) \d t = \int_0^{2 \pi} R(\vartheta) P_0(\vartheta) \d \vartheta$.
Using (\ref{P}) and (\ref{Itophase}) we obtain
\begin{equation}
{\omega} = 1 + D Y_0 + O(D) ,
\nonumber
\end{equation}
where we have used the fact that $\langle Z(\vartheta) \xi(t) \rangle = \langle Z(\vartheta) \rangle \langle\xi(t) \rangle=0$.
We may also calculate the phase-diffusion $\widetilde{D}$ as
\begin{align}
\widetilde{D}&=\int_{-\infty}^\infty
\left \langle \left [\frac{\d}{\d t} {\vartheta}(t+\tau) - \left \langle \frac{\d}{\d t} {\vartheta} \right \rangle \right ]\left [\frac{\d}{\d t} {\vartheta}(t) - \left \langle \frac{\d}{\d t} {\vartheta} \right \rangle \right]
\right \rangle  \d \tau \nonumber \\
&=\frac{D}{\pi} \int_0^{2 \pi}  Z^2(\vartheta) \d \vartheta +O(D^2) ,
\nonumber
\end{align}
where we use the fact that $\langle \xi(t) \rangle=0$ and $\langle \xi(t) \xi(s) \rangle = 2D \delta(t-s)$.
This recovers a well known result of Kuramoto \cite{Kuramoto84}.

\subsection{Phase reduction for noise with temporal correlation}

A recent paper by Teramae \textit{et al}. \cite{Teramae09} shows that when one considers noise described by an Ornstein-Uhlenbeck (OU) process with a finite correlation time then this can interact with the attraction time-scale of the limit cycle and give fundamentally different results when compared to Gaussian white noise (which has a zero correlation time).  This observation has also been independently made in \cite{Yoshimura10}.  Both approaches assume weak noise, though \cite{Yoshimura10} makes no assumptions about relative time-scales, and is thus a slightly more general approach than that of \cite{Teramae09}.  Related work by Goldobin \textit{et al}. \cite{Goldobin10}
for noise $\eta(t)$ with zero-mean $\langle \eta (t) \rangle = 0$ and prescribed auto-correlation function $C(\tau)=\langle \eta(\tau) \eta(0) \rangle$, yields
the reduced Stratonovich phase equation
\begin{equation}
\frac{\d}{\d t} {\vartheta} = 1 + D \widetilde{Y}(\vartheta) +Z(\vartheta) \eta(t) ,
\label{phasegeneral}
\end{equation}
where
\begin{equation}
\widetilde{Y}(\vartheta) = \frac{1}{2 D} h_\rho(\vartheta,0) \int_0^\infty g(\vartheta-\psi,0) C(\psi) \e^{-\lambda \psi} \d \psi ,
\label{Ytilde}
\end{equation}
where $\lambda$ is the average rate of attraction to the limit cycle.
Note that for $C(\tau)=2 D \delta(\tau)$, $\widetilde{Y}(\vartheta)  = {Y}(\vartheta)$ and (\ref{phasegeneral}) reduces to (\ref{phasestrat}) as expected.
To lowest order in the noise strength the steady state probability distribution will simply be $P_0(\vartheta)=1/(2 \pi)$.
Therefore to lowest noise order the mean frequency is determined from an ensemble average as
\begin{equation}
\widetilde{w}=w+D \widetilde{Y}_0 + \frac{1}{4 \pi } \int_0^{2 \pi}\d \vartheta Z'(\vartheta)  \int_0^\infty \d \psi  Z(\vartheta-\psi) C(\psi) ,
\nonumber
\end{equation}
where the last term comes from using the It\={o} form of (\ref{phasegeneral}) and the subscript $0$ notation is defined as in (\ref{P}).
The phase-diffusion coefficient at lowest noise order is given by
\begin{equation}
\widetilde{D} = \frac{1}{2\pi} \int_0^{2 \pi} \d \vartheta \int_{-\infty}^\infty \d \tau Z(\vartheta) Z(\vartheta + \tau) C(\tau) .
\nonumber
\end{equation}

Let us now consider the example of OU noise so that $C(\tau)=D \gamma \exp(-\gamma |\tau|)$.  Furthermore let us take the simultaneous limit $\gamma \rightarrow \infty$ (zero correlation timescale) and $\lambda \rightarrow \infty$ (infinitely fast attraction), such that the ratio $\alpha=\lambda/\gamma$ is constant.  In this case we have from (\ref{Ytilde}) that
\begin{equation}
\widetilde{Y}(\vartheta) = \frac{Y(\vartheta)}{1+\alpha} .
\end{equation}
Hence, when the correlation time of the noise is much smaller than the decay time constant $\alpha=0$ and we recover the result for white Gaussian noise.  In the other extreme when $\alpha \rightarrow \infty$, where the amplitude of the limit cycle decays much faster than the correlation time of the noise, then $\widetilde{Y}$ vanishes and the reduced phase equation is simply $\ID {\vartheta}=1 + Z(\vartheta) \eta(t)$, as would be obtained using the standard phase reduction technique without paying attention to the stochastic nature of the perturbation.

\section{Low dimensional macroscropic dynamics and chimera states}
\label{sec:macroscopic}

The self-organisation of large networks of coupled neurons into macroscopic coherent states, such as observed in phase-locking, has inspired a search for equivalent low-dimensional dynamical descriptions.  However, the mathematical step from microscopic to macroscopic dynamics has proved elusive in all but a few  special cases.  For example, neural mass models of the type described in \S~\ref{subsec:neuralmass} only track mean activity levels and not the higher order correlations of an underlying spiking model.  Only in the thermodynamic limit of a large number of neurons firing asynchronously (producing null correlations) are such rate models expected to provide a reduction of the microscopic dynamics. Moreover, even here the link from spike to rate is often phenomenological rather than rigorous.  Unfortunately only in some rare instances has it been possible to analyse spiking networks directly (usually under some restrictive assumption such as global coupling) as in the spike-density approach \cite{Nykamp01}, which makes heavy use of the numerical solution of coupled PDEs.  Recently however, exact results for globally pulse-coupled oscillators described by the Winfree model \cite{Winfree67} have been obtained by Paz\'{o} and Montbri\'{o} \cite{Pazo2014}.  This makes use of the Ott-Antonsen (OA) ansatz, which was originally used to find solutions on a reduced invariant manifold of the Kuramoto model \cite{Kuramoto91}. The major difference between the two phase-oscillator models being that the former has interactions described by a phase product structure and the latter a phase difference structure.

\subsection{Ott-Antonsen reduction for the Winfree model}
\label{sec:Ott-Antonsen}

The Winfree model is described in \S~\ref{sec:weakcoupling} as a model for weakly globally pulse-coupled biological oscillators, and can support incoherence, frequency locking, and oscillator death when $P(\theta)=1+\cos \theta$ and $R(\theta)=-\sin\theta$ \cite{Ariaratnam2001}.
We note however that the same model is \textit{exact} when describing nonlinear IF models described by a single voltage equation, and that in this case we do not have to restrict attention to weak-coupling.  Indeed the OA ansatz has proved equally successful in describing both the Winfree network with a sinusoidal PRC \cite{Pazo2014} and a network of QIF neurons \cite{Luke2013}.  This is perhaps not surprising since the PRC of a QIF neutron can be computed using (\ref{RIF}), and for the case described by (\ref{eq:one}) with $\tau=1$ and $f(v)=v^2$ and infinite threshold and reset then $R(\theta)=a (1-\cos \theta)$ with $a=1/\sqrt{I}$ for $\theta \in [0,2 \pi)$ (which is the shape expected for an oscillator near a SNIC bifurcation).
We shall now focus on this choice of PRC and a pulsatile coupling that we write in the form
\begin{equation}
P(\theta) = 2 \pi \sum_{n \in \ZSet} \delta (\theta - 2 \pi n) \equiv \sum_{m \in \ZSet} \e^{i m \theta} ,
\nonumber
\end{equation}
where we have introduced a convenient Fourier representation for the periodic function $P$.
We now consider the large $N$ limit in (\ref{Winfree}) and let $\rho(\theta|\omega,t) \d \theta$ be the fraction of oscillators with phases between $\theta$ and $\theta + \d \theta$ and natural frequency $\omega$ at time $t$.  The dynamics of the density $\rho$ is governed by the continuity equation (expressing the conservation of oscillators)
\begin{equation}
\PD{\rho}{t} + \PD{(\rho v)}{\theta} = 0 , \qquad v(\theta,t)= \omega + \epsilon a h(t)[1 -  (\e^{i \theta}+\e^{-i\theta})/2] ,
\label{continuity}
\end{equation}
where the mean-field drive is $h(t)= \lim_{N \rightarrow \infty} \sum_j N^{-1}P(\theta_j)$.
Boundary conditions are periodic in the probability flux, namely $\rho(0|\omega,t) v (0,t) =  \lim_{\theta \rightarrow 2 \pi} \rho(\theta|\omega,t) v(\theta,t)$.
A further reduction in dimensionality is obtained for the choice that the distribution of frequencies is described by a Lorentzian $g(\omega)$ with
\begin{equation}
g(\omega) = \frac{1}{\pi}\frac{\Delta}{(\omega-\omega_0)^2+\Delta^2},
\label{Lorentzian}
\nonumber
\end{equation}
for fixed $\Delta$ and $\omega_0$ (controlling the width and mean of the distribution respectively), which has simple poles at $\omega_\pm =\omega_0 \pm i \Delta$.

A generalised set of order parameters is defined as
\begin{equation}
Z_m(t) = \int_{-\infty}^\infty \d \omega g(\omega) \int_0^{2 \pi}\d \theta \rho(\theta | \omega, t) \e^{i m \theta},\qquad m \in \NSet .
\nonumber
\end{equation}
The integration over $\omega$ can be done using a contour in the lower half complex plane so that $Z_m(t) =
\langle \e^{-i m \theta}, \rho(\theta | \omega_-, t) \rangle$, where we have introduced the inner product
$\langle u(\theta), v (\theta) \rangle = (2 \pi)^{-1}\int_0^{2 \pi}\d \theta  \overline{u(\theta)} v(\theta) \d \theta$.
The OA ansatz assumes that the density $\rho$ can be written in a restricted Fourier representation as
\begin{equation}
2 \pi \rho(\theta|\omega,t) = 1 + \sum_{m=1}^\infty \alpha(\omega,t)^m \e^{i m \theta} + \text{cc},
\label{OA}
\end{equation}
where cc stands for complex conjugate.  Substitution into the continuity equation (\ref{continuity}) and balancing terms in $\e^{i m \theta}$ shows that $\alpha$ must obey
\begin{equation}
\PD{}{t} \alpha = -i(\omega + \epsilon a h) \alpha +i \epsilon a \frac{h}{2} (1+\alpha^2) .
\label{alpha}
\end{equation}
Moreover, using the inner product structure of $Z_m$ we easily see that $Z_m(t) = [\alpha(\omega_-,t)]^m$.  Thus the Kuramoto order parameter $Z_1 \equiv R \e^{-i \Psi}$ is governed by (\ref{alpha}) with $\omega=\omega_-$ yielding:
\begin{equation}
\frac{\d}{\d t} {R} = -\Delta R - \epsilon a \frac{h}{2} (1-R^2) \sin \Psi, \qquad
\frac{\d}{\d t} {\Psi} = \omega_0 + \epsilon a h \left [ 1 - \frac{1+R^2}{2 R} \cos \Psi \right ].
\label{RPsi}
\end{equation}
To calculate the mean-field drive $h$ we note that it can be written as
\begin{align}
h &= \int_0^{2 \pi} \d \theta \rho(\theta| \omega_-,t) P(\theta) = \sum_m Z_m  \nonumber \\
&= 1+ \sum_{m=1}^\infty (Z_1)^m +(\overline{Z_1})^m = 1 + \frac{Z_1}{1-Z_1}+\frac{\overline{Z_1}}{1-\overline{Z_1}}, \qquad |Z_1|<1. \nonumber
\end{align}
Hence, we have explicitly that $h=h(R,\Psi)$ with
\begin{equation}
h(R,\Psi) =  \frac{1-R^2}{1-2 R \cos \Psi + R^2} , \qquad 0 \leq R < 1 .
\label{h}
\end{equation}
%
The planar system of equations defined by (\ref{RPsi}) and (\ref{h}) can be readily analysed using numerical bifurcation analysis.

\begin{figure}[htbp]
\begin{center}
\includegraphics[width=3in]{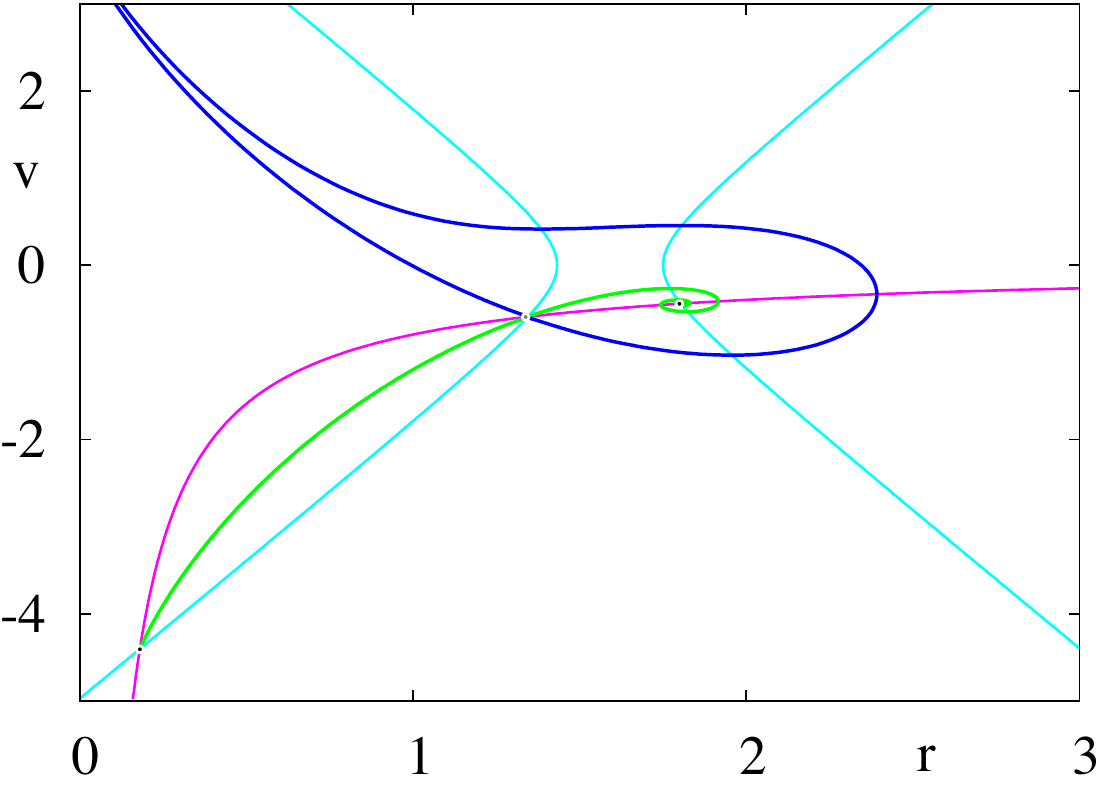}
\caption{A phase portrait for the macroscopic `$V$' dynamics showing the rate and average $(r,v)$ plane with three fixed points for $\omega_0=1$ and $\Delta=10=\epsilon$.  For this parameter set the system is bistable and the stable (blue) and unstable (green) manifolds of the saddle are plotted.  The remaining curves are nullclines.
\label{fig:Roxin}
}
\end{center}
\end{figure}

We note that the OA density (\ref{OA}) can be written in the succinct form
\begin{equation}
2 \pi \rho(\theta|\omega,t) = \frac{1}{2} \left [ \frac{1+\alpha \e^{i \theta}}{1-\alpha \e^{i \theta}} + \text{cc} \right ] .
\nonumber
\end{equation}
The form of this equation suggests recasting the density in terms of new variables $W= (1+\alpha)/(1-\alpha)$
and $V=i(1+\e^{i \theta})/(1-\e^{i \theta})=-\cot(\theta/2)$ \cite{Roxin2014}.  In this case we find that $\rho(V|\omega,t)$ has a Lorentzian shape given explicitly by
\begin{equation}
\rho(V|\omega,t) = \frac{1}{\pi} \frac{u}{(V-v)^2 +u^2}, \qquad W = u+i v ,
\nonumber
\end{equation}
with $\ID {V} = I+V^2 + \epsilon h$, where we have set $\omega=2 \sqrt{I}$ and re-scaled $V$ by a factor of $\sqrt{I}$.
Using the continuity equation for $\rho(V|\omega,t)$ and then integrating over $V$ gives the evolution of $W$ as
\begin{equation}
\PD{W}{t} = i \left [I + \epsilon h -W^2 \right ], \qquad v = \int_\infty^\infty \d V \rho(V|\omega,t) V ,
\label{W}
\end{equation}
where we identify $v$ as the average of $V$.
The firing probabilities for arbitrary cells at time $t$ are equal to the passage rate of the probability density through the spike phase, which gives the population firing rate as
\begin{equation}
r(\omega,t) = \lim_{V \rightarrow\infty} \rho(V|\omega,t) \frac{\d}{\d t} {V} = \frac{1}{\pi} u(\omega,t) .
\nonumber
\end{equation}
Thus we may use (\ref{W}) to determine the evolution of the coupled rate and average network activity by integrating over the distribution of parameters $I=\omega^2/4$.  The evolution of $(r(t),v(t))=\int \d \omega g(\omega) (r(\omega,t),v(\omega,t)) = (r(\omega_0-i \Delta,t),v(\omega_0-i \Delta,t))$ is then found as
\begin{equation}
\frac{\d}{\d t} {r} = \frac{\omega_0\Delta}{2 \pi} + 2 r v , \qquad \frac{\d}{\d t} {v} = v^2 + \frac{\omega_0^2-\Delta^2}{4} + \epsilon \pi r - \pi^2 r^2,
\label{MFOA}
\end{equation}
where we have used the result that $h=(W+\overline{W})/2=\pi r$.
Exploiting the fact that the order parameters in the `$\theta$' and `$V$' descriptions of macroscopic dynamics are related by a conformal mapping, namely $Z_1=(\overline{W}-1)/(\overline{W}+1)$, we may explore both the firing rate and degree of synchrony of the network.
Thus (\ref{MFOA}) is a mean field reduction of a population of spiking QIF neurons, and unlike a phenomenological neural mass model it is able to track a measure of within population synchrony.
The system (\ref{MFOA}) is capable of supporting bistable fixed point behaviour as shown in Fig.~\ref{fig:Roxin}, where we also plot the basin boundary (stable manifold of a saddle).
The OA ansatz has also proved remarkably useful in understanding non-trivial solutions such as chimera states (where a sub-population of oscillators synchronises in an otherwise incoherent sea).

\subsection{Chimera states}
\label{sec:chimera}

Phase or cluster synchronised states in systems of identical coupled oscillators have distinct limitations as descriptions of neural systems where not just phase but also frequency clearly play a part in the processing, computation and output of information. Indeed, one might expect that for any coupled oscillator system that is homogeneous (in the sense that any oscillators can be essentially replaced by any other by a suitable permutation of the oscillators), the only possible dynamical states are homogeneous in the sense that the oscillators behave in either a coherent or an incoherent way. This expectation however is not justified - there can be many dynamical states that cannot be easily classified as coherent or incoherent, but that seem to have a mixture of coherent and incoherent regions. Such states have been given the name ``Chimera state'' by Abrams and Strogatz \cite{Abrams2004,Abrams2008} and have been the subject of intensive research over the past five years. For reviews of chimera state dynamics we refer the reader to \cite{Laing2009a,Panaggio2015}.

\begin{figure}[htbp]
\begin{center}
\includegraphics[width=8cm]{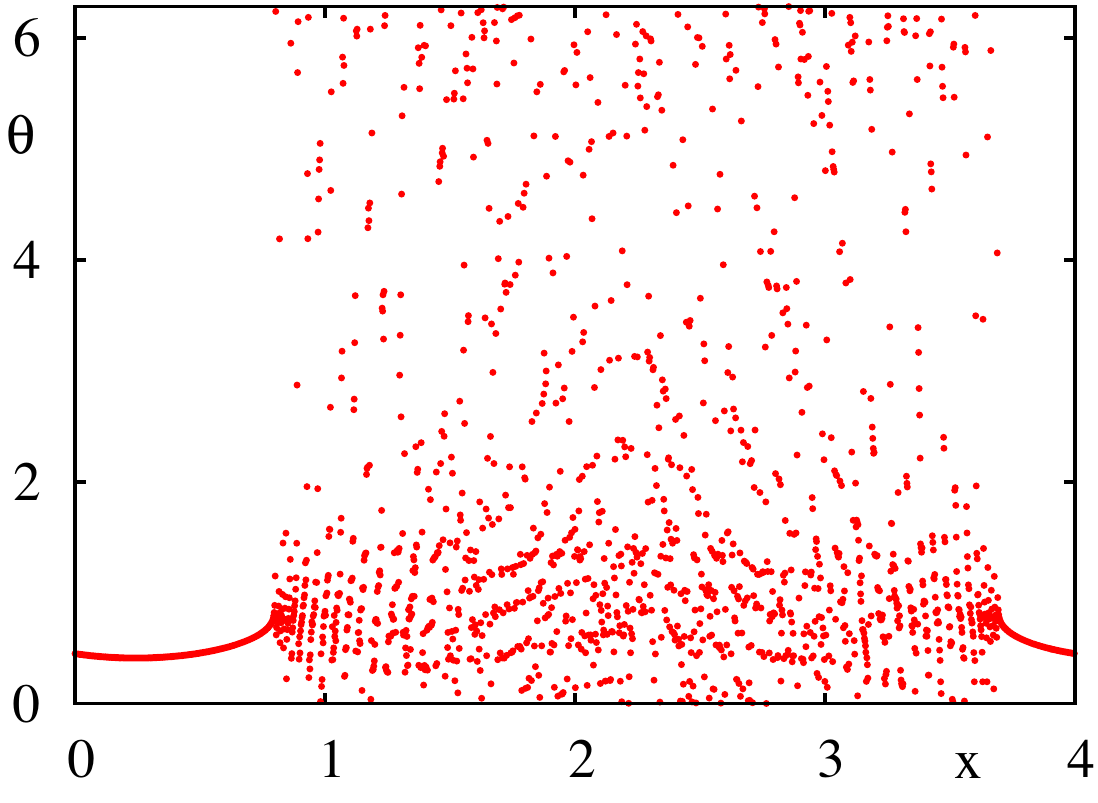}
\caption{A snapshot of a Chimera state for the model (\ref{eq:continuumnonlocal}) in a system of length $4$ using $2^9$ numerical grid points and periodic boundary conditions.  Here $\omega=0$, $\epsilon=0.1$, $\alpha=1.45$ and $\kappa=1$.
\label{Fig:Chimera}
}
\end{center}
\end{figure}

Kuramoto and Battogtokh \cite{kuramoto-battogtokh-2002,kuramoto-2003} investigated continuum systems of oscillators of the form
\begin{equation}
\frac{\partial }{\partial t}\theta(x,t) = \omega - \epsilon \int_{D} G(x-x') \sin(\theta(x,t)-\theta(x',t) +\alpha)\, \d x' ,
\label{eq:continuumnonlocal}
\end{equation}
where $\theta$ represent phases at locations $x \in D \subseteq \RSet$, the kernel $G(u)=\kappa\exp(-\kappa |u|)/2$ represents a non-local coupling and $\omega,\alpha,\kappa$ are constants.
Interestingly this model is precisely in the form presented in \S~\ref{sec:waves} as equation (\ref{phasecontinuum}) for an oscillatory model of cortex, although here there are no space-dependent delays and the interaction function is $H(\theta) = \sin(\theta -\alpha)$. Kuramoto and Battogtokh found for a range of parameters near $\alpha=\pi/2$, and carefully selected initial conditions, that the oscillators can split into two regions in $x$, one region which is frequency synchronised (or coherent) while the other region shows a nontrivial dependence of frequency on location $x$. An example of a Chimera state is shown in Fig.~\ref{Fig:Chimera}.

Note that a discretisation of (\ref{eq:continuumnonlocal}) to a finite set of $N$ coupled oscillators is
\begin{equation}
\FD{}{t}\theta(x_i,t) = \omega - \epsilon \sum_{k=1}^{N} K_{ij} \sin(\theta_i-\theta_j +\alpha)
\label{eq:discretenonlocal}
\end{equation}
where $\theta_i\in[0,2\pi)$ represents the phase at location $i=1,\ldots,N$ and the coupling matrix $K_{ij}= G(|i-j|/N)/N$ is the discretised interaction kernel (assuming a domain of length 1). Using different kernels, $G(u) = \exp(-\kappa\cos(2\pi u))$ and an approximation $G(u) = 1 -\kappa\cos(2\pi u)$ for small $\kappa$, Abrams and Strogatz \cite{Abrams2004} identified similar behaviour and \cite{Abrams2008} discussed a limiting case of parameters such that the continuum system provably has chimera solutions. The OA reduction discussed in \S~\ref{sec:Ott-Antonsen} allows an exact reduction of oscillator networks of this form and in the continuum limit this can give a solvable low-order system whose solutions include a variety of chimera states \cite{Laing2009a}.
It is useful to note that when $\alpha=\pi/2$, pure cosine coupling results in an integrable Hamiltonian system \cite{Watanabe1994}, such that disordered initial states will remain disordered. Thus $\alpha$ determines a balance between spontaneous order and permanent disorder.

However, it seems that chimera states are much more ``slippery'' in finite oscillator systems than in the continuum limit. In particular, Wolfrum and Omel'chenko \cite{wolfrum-omelchenko-2011} note that for finite approximations of the ring (\ref{eq:continuumnonlocal}) by $N$ oscillators, with a mixture of local and nearest $rN$-neighbour coupling corresponding to (\ref{eq:discretenonlocal}) with a particular choice of coupling matrix $K_{ij}$, chimera states apparently only exist as transients. However, the lifetime of the typical transient apparently grows exponentially with $N$. Thus, at least for some systems of the form (\ref{eq:discretenonlocal}), chimeras appear to be a type of chaotic saddle. This corresponds to the fact that the boundaries between the regions of coherent and incoherent oscillation fluctuate apparently randomly over a long timescale. These fluctuations lead to wandering of the incoherent region as well as change in size of the region. Eventually these fluctuations appear to result in typical collapse to either fully coherent or fully incoherent oscillation \cite{wolfrum-omelchenko-2011}.

Although this appears to be the case for chimeras for (\ref{eq:discretenonlocal}), there are networks such as coupled groups of oscillators; \cite{Martens2010} or two dimensional lattices \cite{ShimaKuramoto2004} where chimera attractors can appear. It is not clear what will cause a chimera to be transient or not, or indeed exactly what types of chimera-like states can appear in finite oscillator networks. A suggestion of \cite{ashwin-burylko-2015} is that robust neutrally stable chimeras may be due to the special type of single-harmonic phase interaction function used in (\ref{eq:continuumnonlocal},\ref{eq:discretenonlocal}).

More recent work includes investigations of chimeras (or chimera-like states) in chemical \cite{tinsley-etal-2012} or mechanical oscillator networks \cite{martens-etal-2013}; chimeras in systems of coupled oscillators other than phase oscillators have been investigated in many papers; for example in Stuart-Landau oscillators \cite{kuramoto-battogtokh-2002,omelchenko-etal-2012,sethia-sen-2014}, Winfree oscillators \cite{Pazo2014} and models with inertia \cite{Olmi2014}. Other recent work includes discussion of feedback control to stabilise chimeras \cite{sieber-etal-2014}, investigations of chimeras with multiple patches of incoherence \cite{maistrenko-etal-2014} and multicluster and traveling chimera states \cite{Xie2014}.

In a neural context chimeras have also been found in pulse-coupled LIF networks \cite{Olmi2010}, and hypothesised to underly coordinated oscillations in unihemispheric slow-wave sleep, whereby one brain hemisphere appears to be inactive while the other remains active \cite{Ma2010}.

\section{Applications}
\label{sec:Applications}

We briefly review a few examples where mathematical frameworks are being applied to neural modelling questions.  These cover functional and structural connectivity in neuroimaging, central pattern generators (CPGs) and perceptual rivalry. There are many other applications we do not review, for example to deep brain stimulation protocols \cite{Lysyansky2010} or to modelling of epileptic seizures where network structures play a key role \cite{Terry2012}.

\subsection{Functional and structural connectivity in neuroimaging}
\label{subsec:FunStruct}

\begin{figure}[htbp]
\begin{center}
\includegraphics[height=2.5in]{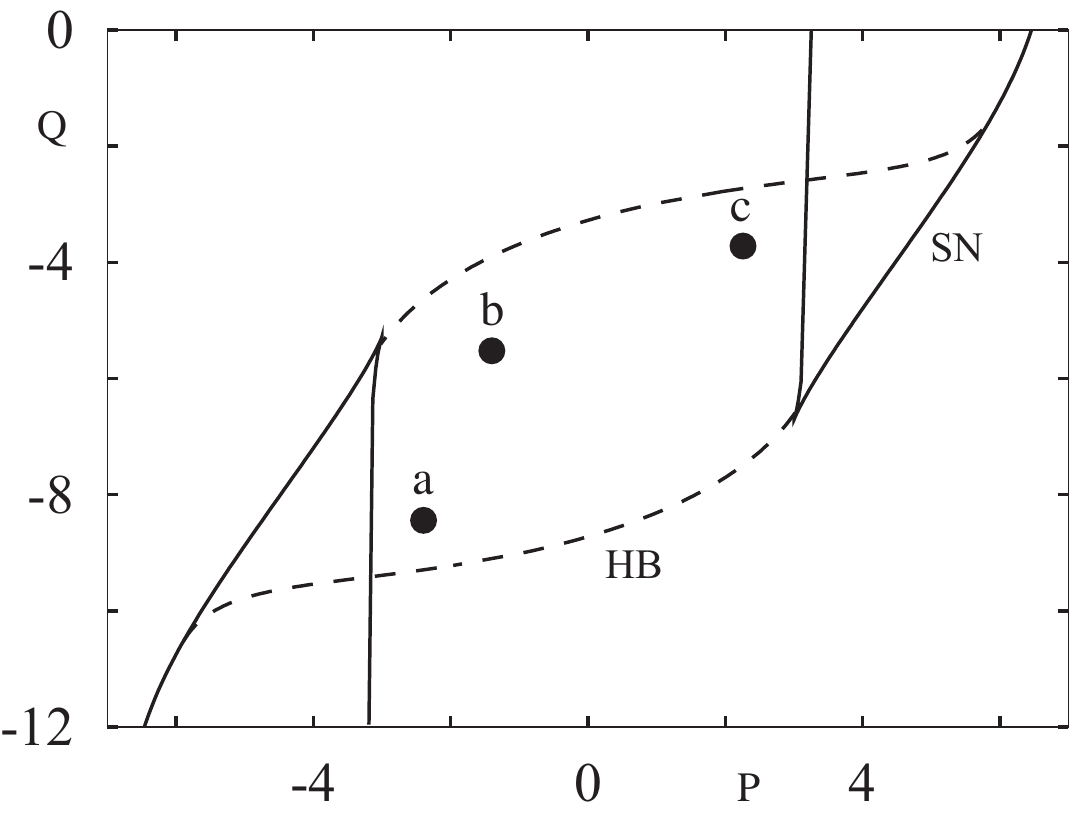}
\caption{
Left:  Bifurcation diagram for an isolated Wilson-Cowan node in the $(P,Q)$ plane. Here HB denotes Hopf bifurcation and SN a saddle node of fixed-points bifurcation. At points a and c we find $H'(0) < 0$ and at point b $H'(0) > 0$.  A breakdown of FC (loss of global synchrony) within a globally coupled network is predicted at points a and c.
\label{fig:WC}
}
\end{center}
\end{figure}

Functional connectivity (FC) refers to the temporal synchronisation of neural activity in spatially remote areas. It is widely believed to be significant for the integrative processes in brain function.
Anatomical or structural connectivity (SC), is widely believed to play an important role in determining the observed spatial patterns of FC. However, there is clearly a role to be played by the dynamics of the neural tissue.  Even in a globally connected network we would expect this to be the case, given our understanding of how synchronised solutions can lose stability for weak coupling.  Thus it becomes useful to study models of brain like systems built from neural mass models (such as the Jansen-Rit model of \S~\ref{subsec:neuralmass}), and ascertain how the local behaviour of the oscillatory node dynamics can contribute to global patterns of activity.  For simplicity, consider a network of $N$ globally coupled identical Wilson-Cowan \cite{Wilson72} neural mass models:
\begin{align}
\frac{\d}{\d t} {x}_i &= -x_i+P+c_1 f(x_i)-c_2 f(y_i) +\frac{\epsilon}{N} \sum_{j=1}^N f(x_j) ,\nonumber \\
\frac{\d}{\d t} {y}_i &= -y_i+Q+c_3 f(x_i)-c_4 f(y_i) , \nonumber
\end{align}
for $i=1,\ldots, N$.
Here $(x_i,y_i) \in \RSet^2$ represents the activity in each of a pair of coupled neuronal population models, $f$ is a sigmoidal firing rate given by $f(x) = 1/(1+\e^{-x})$ and $(P,Q)$ represent external drives.  The strength of connections within a local population is prescribed by the co-efficients $c_1,\ldots c_4$, which we choose as $c_1 = c_2 = c_3 = 10$ and $c_4 = -2$ as in \cite{Hoppensteadt97}.  For $\epsilon=0$ it is straight forward to analyse the dynamics of a local node and find the bifurcation diagram in the $(P,Q)$ plane as shown in Fig.~\ref{fig:WC}.  Moreover, for $\epsilon \ll 1$ we may invoke weak coupling theory to describe the dynamics of the full network within the oscillatory regime bounded by the two Hopf curves shown in Fig.~\ref{fig:WC} left.  From the theory of \S~\ref{sec:synchrony} we would expect the synchronous solution to be stable if $\epsilon H'(0) >0$.  Taking $\epsilon >0$ we can consider $H'(0)$ as a proxy for the robustness of synchrony.  The numerical construction of this quantity, as in \cite{Hlinka2012}, predicts that there will be regions in the $(P,Q)$ plane associated with a breakdown of FC (where $H'(0) <0$), as indicated by points a and b in Fig.~\ref{fig:WC}.  This highlights the role that local node dynamics can have on emergent network dynamics.  Moreover, we see that simply by tuning the local dynamics to be deeper within the existence region for oscillatory solutions we can, at least for this model, enhance the degree of FC.
It would be interesting to explore this simple model further for more realistic brain like connectivities, along the lines described in \cite{Arsiwalla2015}.  Moreover, given that this would preclude the existence of the synchronous state by default (since we would neither have that $H(0)=0$ or $\sum_j w_{ij}$ would be independent of $i$) then it would be opportune to explore the use recent ideas in \cite{Nicosia2013,Pecora2014} to understand how the system could organise into a regime of remote synchronisation whereby pairs of nodes with the same network symmetry could synchronise.
For related work on Wilson-Cowan networks with some small dynamic noise see \cite{Daffertshofer2011}, though here the authors construct a phase-oscillator network by linearising around an unstable fixed point, rather than use the notion of phase response.

\subsection{Central Pattern Generators}
\label{subsec:CPG}

CPGs are (real or notional) neural subsystems that are implicated in the generation of spatio-temporal patterns of activity \cite{Nadim2014}, in particular for driving the relatively autonomous activities such as as locomotion \cite{Collins1994,Sherwood2010,Sakuraietal2011} or for driving involuntary activities such as heartbeat, respiration or digestion \cite{Baletal1988}. These systems are assumed to be behind the creation of the range of walking or running patterns (gaits) that appear in different animals \cite{Cohen1988}.  The analysis of phase-locking provides a basis for understanding the behaviour of many CPGs, and for a nice overview see the review articles by Marder and Bucher \cite{Marder2001} and Hooper \cite{Hooper2001}.

In some cases, such as the Leech ({\em Hirudo medicinalis}) Heart or {\em Caenorhabditis elegans} locomotion, the neural circuitry is well studied. For more complex neural systems and in more general cases CPGs are still a powerful conceptual tool to construct notional minimal neural circuitry needed to undertake a simple task. In this notional sense they have extensively been investigated to design control circuits for actuators for robots; see for example the review \cite{Ijspeert2008}. Recent work in this area includes robots that can reproduce salamander walking and swimming patterns \cite{Crespi2013}.  Since the control of motion of autonomous ``legged'' robots is still a very challenging problem in real-time control, one hope of this research is that nature's solutions (for example, how to walk stably on two legs) will help inspire robotic ways of doing this.

\begin{figure}
\centerline{
\includegraphics[width=10cm]{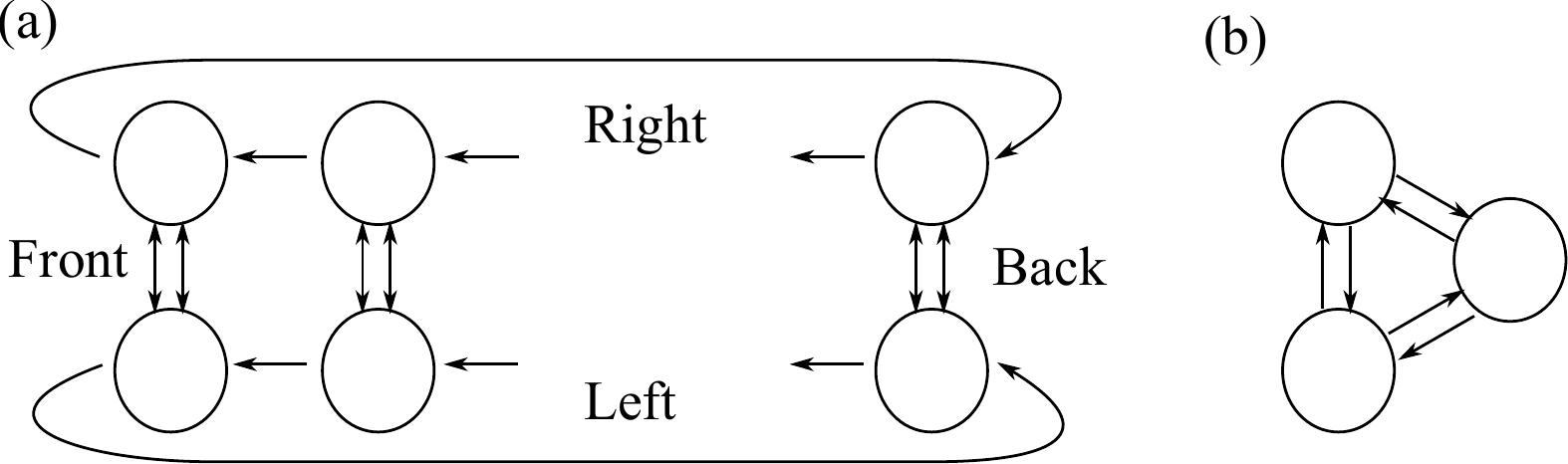}
}
\caption{Schematic diagram showing central pattern generators (a) with $4n$ coupled cells that is used to model gait patterns in animals with $2n$ legs \cite{Golubitskyetal1999} and (b) a three cell motif of bursters with varying coupling strengths, as considered in \cite{Shilnikov2008,Shilnikov2014}.}
\label{fig:GolCPG}
\end{figure}

CPGs are called upon to produce one or more rhythmic patterns of actuation; in the particular problem of locomotion, a likely CPG is one that will produce the range of observed rhythms of muscle actuation, and ideally the observed transitions between then. For an early discussion of design principles for modelling CPGs, see \cite{Kopell1988}. This is an area of modelling where consideration of symmetries as in \S~\ref{subsec:symmetry} has been usefully applied to constrain the models. For example \cite{CollinsStewart1993} examine models for generating the gaits in a range of vertebrate animals, from those with two legs (humans) through those with four (quadrupeds such as horses have a wide range of gaits - walk, trot, pace, canter, gallop - they may use) or larger numbers of legs (myriapods such as centipedes). Insects make use of six legs for locomotion while other invertebrates such as centipedes and millipedes have a large number of legs that are to some extend independently actuated. As an example, \cite{Golubitskyetal1999} consider a schematic CPG of $2n$ oscillators for animals with $n$ legs, as shown in Fig.~\ref{fig:GolCPG}(a). The authors use symmetry arguments and Theorem~\ref{thm:honk} to draw a number of model-independent conclusions from the CPG structure.

One can also view CPGs as a window into more fundamental problems of how small groups of neurons coordinate to produce a range of spatio-temporal patterns. In particular, it is interesting to see how the observable structure of the connections influences the range and type of dynamical patterns that can be produced. For example, \cite{Shilnikov2008} consider a simple three-cell ``motif'' networks of bursters and classify a range of emergent spatio-temporal patterns in terms of the coupling parameters. Detailed studies \cite{Shilnikov2014} investigate properties such multistability and bifurcation of different patterns and the influence of inhomogeneities in the system. This is done by investigating return maps for the burst timings relative to each other.

The approach mentioned in \S~\ref{subsec:groupoid} and outlined in \cite{Stewartetal2003,golubitsky-stewart2006} seems to be a promising way of providing a context in which view CPGs where there is a constrained connection structure suggested by either the neurobiology or from conceptual arguments, but this structure is not purely related to symmetries of the network. For example, \cite{Golubitsky2012} use that formalism to understand possible spatio-temporal patterns that arise in lattices or \cite{Kamei2009} that relates synchrony properties of small motif networks to spectral properties of the adjacency matrix.

\subsection{Perceptual rivalry}

Many neural systems process information - they need to produce outputs that depend on inputs. If the system effectively has no internal degrees of freedom then this will give a functional relationship between output and input so that any temporal variation in the output corresponds to a temporal variation of the input. However, this is not the case for all but the simplest systems and often outputs can vary temporally unrelated to the input. A particularly important and well-studied system that is a model for autonomous temporal output is {\em perceptual rivalry}, where conflicting information input to a neural system results, not in a rejection or merging of the information, but in an apparently random ``flipping'' between possible ``rival'' states (or percepts) of perception. This nontrivial temporal dynamics of the perception appears even in the absence of a temporally varying input. The best studied example of this type is {\em binocular rivalry}, where conflicting inputs are simultaneously made to each eye. It is widely reported by subjects that perception switches from one eye to the other, with a frequency that depends on a number of factors such as the contrast of the image \cite{Brascamp201520}. More general perceptual rivalry, often used in ``optical illusions'' such as ambiguous figures - the Rubin vase, the Necker cube - show similar behaviour with percepts shifting temporally between possible interpretations.

Various approaches  \cite{Shpiro2007} have been made to construct nonlinear dynamical models of the generation of a temporal shifting between possible percepts such as competition models \cite{Blake1989145}, bifurcation models, ones based on neural circuitry \cite{LaingChow2002}, or conceptual ones \cite{Wilson2007} based on network structures\cite{Diekman2014} or on heteroclinic attractors \cite{AshwinLavric2010}.

\section{Discussion}
\label{sec:discussion}

As with any review we have had to leave out many topics that will be of interest to the reader. In particular we have confined ourselves to ``cell'' and ``system-level'' dynamics rather that ``sub-cellular'' behaviour of neurons. We briefly mention some other active areas of mathematical research relevant to the science of rhythmic neural networks.  Perhaps the most obvious area that we have not covered in any depth is that of single unit (cell or population) forcing, which itself is a rather natural starting point for gaining insights into network behaviour and how best to develop mathematical tools for understanding response \cite{Mintchev2009,Lanford2015}.
For a general perspective on mode-locked responses to periodic forcing see \cite{Glass88} and \cite{Pikovsky01}.

For  a more recent discussion of the importance of mode-locking in auditory neuroscience see \cite{Laudanski2010,Lerud2014} and in motor systems see \cite{Daffertshofer2012}.  However, it is well to note that not much is known about nonlinear systems with three or more interacting frequencies \cite{Cartwright1999}, as opposed to periodically forced systems where the notions of Farey tree and the devil's staircase have proven especially useful.  We have also painted the notion of synchrony with a broad mathematical brush, and not discussed more subtle notions of envelope locking that may arise between coupled bursting neurons (where the within burst patterns may desynchronise) \cite{Azad2010}.  This is especially relevant to studies of  synchronised bursting \cite{Segev2001} and the emergence of chaotic phenomena \cite{Nakada2008}. Indeed, we have said very little about coupling between systems that are chaotic, such as described in \cite{Zhou2002}, the emergence of chaos in networks \cite{So2011,So2014} or chaos in symmetric networks \cite{Bick2011}. 

The issue of chaos is also relevant to notions of reliability, where one is interested in the stability of spike trains against fluctuations.
This has often been discussed in relation to stochastic oscillator forcing rather than those arising deterministically in a high dimensional setting \cite{Tiesinga2002,Goldobin2006,Ermentrout2008a,Lin2009}.  Of course, given the sparsity of firing in cortex means that it may not even be appropriate to treat neurons as oscillators.  However, some of the ideas developed for oscillators can be extended to excitable systems, as described in \cite{Ichinose98,Rabinovitch99}.  As well as this it is important to point out that neurons are not point processors, and have an extensive dendritic tree, which can also contribute significantly to emergent rhythms as described in \cite{Bressloff97,Schwemmer2012}, as well as couple strongly to glial cells. Although the latter do not fire spikes, they do show oscillations of membrane potential \cite{Amzica2000}.  At the macroscopic level it is also important to acknowledge that the amplitude of different brain waves can also be significantly affected by neuromodulation \cite{Lee2012}, say through release of norepinephrine, serotonin and acetylcholine, and the latter is also thought to be able to modulate the PRC of a single neuron \cite{Stiefel2009}.

This review has focused mainly on the embedding of weakly coupled oscillator theory within a slightly wider framework.  This is useful in setting out some of the neuroscience driven challenges for the mathematical community in establishing inroads into a more general theory of coupled oscillators.  Heterogeneity is one issue that we have mainly side-stepped, and remember that the weakly coupled oscillator approach requires frequencies of individual oscillators to be close.  This can have a strong effect on emergent network dynamics \cite{Vladimirski2008}, and it is highly likely that even a theory with heterogeneous phase response curves \cite{Tsubo2007} will have little bearing on real networks.  The \textit{equation-free} coarse-graining approach may have merit in this regard, though is a numerically intensive technique \cite{Moon2006}.

We suggest a good project for the future is to develop a theory of strongly coupled heterogeneous networks based upon the phase-amplitude coordinate system described in \S~\ref{subsec:phase-amplitude}, with the challenge to develop a reduced network description in terms of a set of phase-amplitude interaction functions, and an emphasis on understanding the new and generic phenomena associated with nontrivial amplitude dynamics (such as clustered phase-amplitude chaos and multiple attractors).  To achieve this one might further tap into recent ideas for classifying emergent dynamics based upon the group of structural symmetries of the network.  This can be computed as the group of automorphisms for the graph describing the network.  For many real-world networks, this can be decomposed into direct and wreath products of symmetric groups \cite{MacArthur2008}.  This would allow the use of tools from computational group theory \cite{Holt2005} to be used, and open up a way to classify the generic forms of behaviour that a given network may exhibit using the techniques of equivariant bifurcation theory.

\section*{Appendix A}
The Hodgkin-Huxley description of nerve tissue is completed with:
\begin{align}
\alpha_m(V) &= \frac{0.1(V+40)}{1-\exp [-0.1(V+40)]}, & \alpha_h(V) &= 0.07
\exp [-0.05(V+65)], \nonumber \\
\alpha_n(V) &= \frac{0.01(V+55)}{1-\exp[-0.1(V+55)]}, & \beta_m(V) &= 4.0
\exp[-0.0556(V+65)], \nonumber \\
\beta_h(V) &= \frac{1}{1+ \exp[-0.1(V+35)]}, & \beta_n(V) &= 0.125
\exp[-0.0125(V+65)] ,\nonumber
\end{align}
and
$C=1 \mu$F cm$^{-2}$, $g_L = 0.3$mmho cm$^{-2}$, $g_K=36$mmho cm$^{-2}$, $g_{Na}=120$mmho
cm$^{-2}$, $V_L=-54.402$mV,
$V_K=-77$mV and $V_{Na}=50$mV.  (All potentials are measured in mV, all times in ms and all currents in
$\mu$A per cm$^2$).

\section*{Glossary}

We give a brief list of some of the abbreviations used within the review.

\begin{description}
\item[DDE] Delay differential equation
\item[IF] Integrate and fire (model for neural oscillator)
\item[iPRC] Infinitesimal phase response curve
\item[ISI] Inter-spike interval
\item[FHN] Fitzhugh-Nagumo equation (model for neural oscillator)
\item[HH] Hodgkin-Huxley equation (model for neural oscillator)
\item[LIF] Leaky integrate and fire (model for neural oscillator)
\item[ML] Morris-Lecar equation (model for neural oscillator)
\item[MSF] Master stability function
\item[ODE] Ordinary differential equation
\item[PDE] Partial differential equation
\item[PRC] Phase response curve
\item[QIF] Quadratic integrate and fire (model for neural oscillator)
\item[SDE] Stochastic differential equation
\item[SHC] Stable heteroclinic channel
\item[SNIC] Saddle-node on an invariant circle (bifurcation)
\item[WLC] Winnerless competition
\end{description}



\section*{Competing interests}
  The authors declare that they have no competing interests.

\section*{Author's contributions}
    PA, SC and RN contributed equally. All authors read and approved the final manuscript.

\section*{Acknowledgements}
	We would like to thank Kyle Wedgwood and \'Aine Byrne for useful comments made on draft versions of this manuscript.  SC was supported by the European Commission through the FP7 Marie Curie Initial Training Network 289146, NETT: Neural Engineering Transformative Technologies.


\bibliographystyle{plain}


\end{document}